\documentclass[authoryear,12pt,1p,preprint]{elsarticle}
\usepackage{snapshot}
\usepackage{arydshln}
\usepackage{url}
\newcommand{\pkg}[1]{\textbf{\texttt{#1}}}
\newcommand{\proglang}[1]{\textbf{\texttt{#1}}}
\newcommand{\code}[1]{\texttt{#1}}

\usepackage{amsmath,amssymb}
\usepackage{bm}
\usepackage{natbib}
\DeclareMathOperator{\arctanh}{arctanh}

\DeclareMathOperator{\logit}{logit}

\DeclareMathOperator{\tr}{tr}

\DeclareMathOperator{\rank}{rank}

\usepackage{xspace}
\newcommand{\bpnd}{\ensuremath{\text{BP}_{ND}}\xspace}
\newcommand{\htt}{SERT\xspace}

\newcommand{\var}{\ensuremath{\mathbb{V}\text{ar}}}
\newcommand{\cov}{\ensuremath{\mathbb{C}\text{ov}}}

\newcommand{\E}{\ensuremath{\mathbb{E}}}
\newcommand{\pr}{\ensuremath{\mathbb{P}}}
\newcommand{\one}{\bm{1}}
\newcommand{\R}{\ensuremath{\mathbb{R}}}

\newcommand{\N}{\ensuremath{\mathbb{N}}}

\newcommand{\abs}[1]{\ensuremath{\left\vert#1\right\vert}}

\renewcommand{\subset}{\subseteq} 

\newcommand{\argmax}{\operatorname{arg\,\max}}

\newcommand{\mvec}{\operatorname{vec}}

\newcommand{\independenT}[2]{
  \mathrel{\setbox0\hbox{$#1#2$}%
    \copy0\kern-\wd0\mkern4mu\box0}}

\newcommand{\bias}{\operatorname{bias}}

\newcommand{\fun}[2]{#1(#2)}

\newcommand{\ft}[2]{\ensuremath{{#1}_{#2}}}
\newcommand{\ftb}[1]{\ft{\bm{#1}}{\bm{\theta}}}
\newcommand{\SigmaT}{\ftb{\Sigma}}
\newcommand{\SigmaF}{\ftb{\Omega}}
\newcommand{\IA}{\ftb{G}}
\newcommand{\IAi}{(\bm{1}-\ftb{A})^{-1}}
\newcommand{\AT}{\ftb{A}}
\newcommand{\PT}{\ftb{P}}
\newcommand{\wmu}{\widehat{\bm{\mu}}}
\newcommand{\wS}{\widehat{\bm{\Sigma}}}
\newcommand{\dif}{\operatorname{d}\!\,}

\usepackage{ifthen}
\newboolean{notesBoolean}
\setboolean{notesBoolean}{true} 
\newcommand{\mynote}[1]{%
  \ifthenelse{\boolean{notesBoolean}}{
    \begingroup
    \hbadness 20000
    \marginpar{\tiny\textsf{#1}}
    \endgroup
  }{}
}

\usepackage{Sweave}
\usepackage{calc}
\newcounter{origsize}
\newcommand{\figsize}[2]{\setkeys{Gin}{width=#1\textwidth/10,
    height=#2\textwidth/10}}
\newcommand{\origfigsize}{\figsize{\value{origsize}}{\value{origsize}}}
\origfigsize
\newcounter{cursize}
\newcommand{\relsize}[1]{
  \setcounter{cursize}{\value{origsize}}
  \addtocounter{cursize}{#1}
  \figsize{\value{cursize}}{\value{cursize}}
}

\renewenvironment{Schunk}{
\small
}
{
\normalsize
}

\begin{document}

\begin{frontmatter}
  \journal{}
  \title{Linear Latent Variable Models: The \pkg{lava}-package}
  \author[biostat]{Klaus K\"{a}hler Holst}
  \ead{k.k.holst@biostat.ku.dk}
  \author[biostat]{Esben Budtz-J{\o}rgensen}
  \ead{e.budtz-joergensen@biostat.ku.dk}
  \address[biostat]{University of Copenhagen, Department of
    Biostatistics}
  \begin{abstract}
    An R package for specifying and estimating linear latent variable
models is presented. The philosophy of the implementation is to
separate the model specification from the actual data, which leads to
a dynamic and easy way of modeling complex hierarchical structures.
Several advanced features are implemented including robust standard
errors for clustered correlated data, multigroup analyses, non-linear
parameter constraints, inference with incomplete data, maximum
likelihood estimation with censored and binary observations, and
instrumental variable estimators.  In addition an extensive simulation
interface covering a broad range of non-linear generalized structural
equation models is described. The model and software are demonstrated
in data of measurements of the serotonin transporter in the human
brain.


  \end{abstract}
  \begin{keyword}
    latent variable model\sep 
    maximum likelihood\sep 
    multigroup analysis\sep
    structural equation model\sep
    \proglang{R}\sep
    serotonin\sep
    \htt
  \end{keyword}
\end{frontmatter}



\section{Introduction}
Multivariate data are often modelled using random effects in order to
account for correlation between measurements in the statistical
analysis. The dominating model is the linear mixed effects model
\citep{laird82}, which is available in most standard statistical
software packages, e.g.  \proglang{SAS} \pkg{PROC MIXED} and in
\pkg{R} in the packages \pkg{nlme} \citep{pinheirobatesNLME} and
\pkg{lme4} \citep{lme4} with the latter one also offering some support
for generalized linear mixed models.

Another type of random effect model is the \emph{structural equation
  model} (SEM) \citep{MR996025}, where the terminology \emph{latent
  variable} often is used instead of random effects. While the mixed
effect model and structural equation model have many aspects in
common, the aim of a SEM analysis is typically to analyze the
association between the latent variable, representing some process
that is only partially observed, and some other variables (observed or
latent). Thus, in SEMs focus is mainly on the latent variable, where
one normally ascribes less interpretation to the random effects in a
mixed effects model, which primarily serves as a way of capturing
covariance between measurements.  Because observed variables can be
viewed as representations of underlying true variables SEMs offer a
natural framework for handling measurement errors in study variables,
and it often provides an efficient analysis of high dimensional data
\citep{budtz-jorgensen03}.  The framework was pioneered by Jöreskog
\citep{joreskog70} and since then it has been an active area of
research with focus on relaxing linearity and distributional
assumptions \citep{genmulsemrabehesketh}.

SEMs have proven to be useful in many different fields of
research. However, applications have been dominated by
\emph{covariance structure analyses}, and as residuals on the
individual level are not available in this setup, model assessment has
been based on more or less heuristic omnibus tests. The lack of
profound model diagnostics have undoubtedly lead to several poor
applications of SEMs \citep{steiger01drivinfastrever}, and thus
likely, however unjustified, giving the framework a somewhat bad
reputation among groups of statisticians.

Several properitary software solutions are available for analyzing
SEMs with some of the most popular being \proglang{LISREL},
\proglang{SAS} \pkg{PROC CALIS}, \proglang{AMOS}, \proglang{EQS},
\proglang{Stata 12} \pkg{sem}, \proglang{Stata} \pkg{gllamm}
\citep{genmulsemrabehesketh} and \proglang{Mplus} \citep{mplus5},
where the last two programs stand-out because of their general
modelling framework.  Common for all these solutions is that they are
difficult to extend and therefore possibilities for examination of new
methodological ideas are limited.  In part because details of many
features of properitary software often remains hidden from the user.
Implementations in an open-source environment such as \proglang{R}
\citep{Rmanual} directly address this problem. Currently two such
solutions are available: the \pkg{sem} package
\citep{fox06:teachcorner, fox09sem} and the \pkg{OpenMx} package
\citep{OpenMX}. While the former package is
limited to standard covariance structure analysis, \pkg{OpenMx} offers
sophisticated methods such as multiple group analysis, models for
ordinal data and mixture models. The predecessor package \proglang{Mx}
has been a popular for analyzing family data in epidemiological
genetics and \pkg{OpenMx} will undoubtedly become an important tool in
that field of research.

This paper presents the \pkg{lava}-package for statistical analysis in
a very general modelling framework known as the \emph{Linear Latent
  Variable Model}, which includes structural equation models and mixed
models as important special cases.  This model class also allows for
non-linear effects of covariates and non-linear parameter constraints.
The \pkg{lava}-package offers a superior user interface for
specifying, altering and visualizing the model design. Models are
specified independently of data using commands that are similar to
standard regression modeling in \proglang{R}.  In addition path diagrams can be
generated to help give the user an overview of the assumptions
specified. Further, the package gives access to an extensive simulation
procedure which covers, but is not limited to, linear latent variable
models. This tool will be extremely useful e.g. for understanding the
biases caused by different types of model misspecification. The
\pkg{lava}-package also includes sophisticated inferential methods
such as multigroup analyses, robust standard errors for clustered
correlated data, maximum likelihood based inference with data missing
at random and inference for indirect and total effects. In addition
advanced model diagnostic techniques for structural equation models
(fitted in \code{lava}) are available via the \pkg{gof}-package
\citep{holst09:gof}, and extensions to models for censored and binary
outcomes are available via the package \pkg{lava.tobit}
\citep{lavatobit} covered briefly in this article.

Modular programming has been a key concept during the software
development thus making the process of extending the program
(e.g. implementing new estimators, changing optimization routines
etc.) easy, as exemplified by the above mentioned add-on packages.

Our hope is that the package will serve as a platform for testing,
developing and sharing new ideas in the field of latent variable
models.


\section{Linear Latent Variable Models}

We will define the \emph{Linear Latent Variable Model} as the 
model defined by a \emph{Measurement part} describing the responses
$\bm{Y}_i = (Y_{i1},\ldots,Y_{ip})'$:
\begin{align}
  \begin{split}
    Y_{ij} &= \nu_j + \sum_{k=1}^l\lambda_{jk}\eta_{ik} +
    \sum_{r=1}^q\kappa_{jr}X_{ir}
    + \sum_{k=1}^l  \delta_{jk}V_{ijk}\eta_{ik} + \epsilon_{ij},
  \end{split}
\end{align}
and a \emph{structural part} describing the \emph{latent variables}
$\bm{\eta}_i = (\eta_{i1},\ldots,\eta_{il})'$:
\begin{align}
  \begin{split}
    \eta_{is} &= \alpha_{s} + \sum_{k=1}^l \beta_{sk}\eta_{ik} +
    \sum_{r=1}^q \gamma_{sr}X_{ir} + \sum_{k=1}^l
    \tau_{sk}W_{isk}\eta_{ik} + \zeta_{is},  \end{split}
\end{align}
where $i=1,\ldots,n$ is the index of the sampling unit
(e.g. individuals), $j=1,\ldots,p$ is the index of the observed
variables (measurements or within cluster observations) and
$s=1,\ldots,l$ is the index of the $l$ distinct latent variables. In a
more compact matrix notation the model can be written as
\begin{align}\label{eq:measurement}
  \bm{Y}_i = \bm{\nu} + \bm{\Lambda}\bm{\eta}_i + \bm{K}\bm{X}_i +
  (\bm{\Delta}\odot\bm{V}_i)\bm{\eta}_i +
  \bm{\epsilon}_i,
\end{align}
\begin{align}\label{eq:structural}
  \bm{\eta}_i = \bm{\alpha} + \bm{B}\bm{\eta}_i + 
  \bm{\Gamma}\bm{X}_i + (\bm{T}\odot\bm{W}_i)\bm{\eta}_i 
  + \bm{\zeta}_i,
\end{align}  
where $\bm{\nu}\in\R^{p}$ and $\bm{\alpha}\in\R^{l}$ are intercepts,
and $\bm{\Lambda},\bm{\Delta}\in\R^{p\times l}$, $\bm{K}\in\R^{p\times
  q}$, $\bm{\Gamma}\in\R^{l\times q}$, $\bm{B}, \bm{T}\in\R^{l\times
  l}$ are regression coefficient matrices (defining both fixed effects
and random slopes), and
$\bm{X}_i\in\R^q, \bm{V}_i\in\R^{p\times l},
\bm{W}_i\in\R^{l\times l}$ are covariates. The $\odot$ denotes the
Schur product (element-wise multiplication).  The residual terms
follow multivariate normal distributions, $\bm{\epsilon}_i\sim{ }$
$\mathcal{N}_p(0,\bm{\Sigma_{\epsilon}})$ and
$\bm{\zeta}_i\sim\mathcal{N}(0,\bm{\Psi})$, which typically are
assumed to be independent.


Note that the terms including the covariates $\bm{V}_i$ and $\bm{W}_i$
define random slope components as in the Laird-Ware mixed model
formulation and differentiates the LLVM from the usual SEM formulation
\citep{sanchez05}.  Many cases can however be modeled without such
terms, resulting in a more computational efficient model formulation
with constant variance between individuals.  In the following we will
therefore initially assume that the model is parameterized by some
$\bm{\theta}$ defining the matrices
\begin{align}\label{eq:parlist}
  (\bm{\nu},\bm{\alpha},\bm{\Lambda},\bm{K},\bm{B},\bm{\Gamma},\bm{\Sigma_{\epsilon}},\bm{\Psi}),
\end{align}
with some restrictions on the parameter space to guarantee
identification (obviously zeroes in the diagonal of $\bm{B}$), and
possibly non-linear constraints between the different parameters.
In Section \ref{sec:randominteractions} we will return to the
general case, and demonstrate how to write up the model in atoms
adapted to the case without any interaction terms. Note that we will
allow non-linear constraints on the parameters between any of the
elements in (\ref{eq:parlist}).

In the setup with $\bm{\Delta}=\bm{0}$ and $\bm{T}=\bm{0}$ it follows that
the mean and variance of $\bm{Y}_i$ given the covariates are 
\begin{align}
  \begin{split}\label{eq:margmean}
    \bm{\mu_i}(\bm{\theta}) = \E(\bm{Y}_i\mid\bm{X}_i) &= \bm{\nu} +
    \bm{\Lambda}(\one-\bm{B})^{-1}\bm{\alpha} \\ &\qquad +
    \left[\bm{\Lambda}(\one-\bm{B})^{-1}\bm{\Gamma} +
      \bm{K}\right]\bm{X_i},
  \end{split}
\end{align}
\begin{align}\label{eq:margvar}
  \SigmaT = \var(\bm{Y}_i\mid \bm{X}_i) &= 
  \bm{\Lambda}(\one-\bm{B})^{-1}\bm{\Psi}(\one-\bm{B})^{-1}{}'\bm{\Lambda}'
  + \bm{\Sigma_{\epsilon}},
\end{align}
where the fundamental property of the normal distribution that the
marginals also are normal is exploited.  Inference about
$\bm{\theta}$ can then be obtained by maximizing the
corresponding likelihood \citep{MR996025}
\begin{align*}
  L(\bm{\theta}; \bm{Y},\bm{X}) \propto \prod_{i=1}^n \exp\left\{-\frac{1}{2}\left(\bm{Y_i}-\mu_i(\bm{\theta})\right)'\SigmaT^{-1}\left(\bm{Y_i}-\mu_i(\bm{\theta})\right)\right\}\abs{\SigmaT}^{-\tfrac{1}{2}}.
\end{align*}

\subsection{Implementation}
From a mathematical and implementation-wise point of view it is convenient
to supplement the model formulation with the equivalent
\emph{Reticular Action Model} (RAM) formulation
\citep{mcardle:mcdlonald:1984, fox09sem}, which have a direct
connection with the underlying path diagram and also explicitly covers
path analysis models. 
Let $\bm{U}$ be the stochastic vector including the latent variables,
$\bm{\eta}$,
\begin{align}\label{eq:Udef}
  \bm{U} = (Z_1, \ldots, Z_{p+q}, \eta_1, \ldots, \eta_l)' = (\bm{Z}', \bm{\eta}')',
\end{align}
where $\bm{Z} = (Z_1,\ldots,Z_{p+q})'$ is the stochastic vector
containing all observed variables $\bm{Z} =
(Y_1,\ldots,Y_p,X_1,\ldots,X_q)'$.  The RAM formulation states
that
\begin{align}\label{eq:RAM}
  \bm{U} = \ftb{v} + \ftb{A}\bm{U} + \bm{\epsilon},
\end{align}
where $\ftb{v}$ describes the intercepts, and
$\epsilon$ is a residual term assumed to follow a zero-mean normal
distribution with
\begin{align} 
  \var(\bm{\epsilon}) = \PT.
\end{align} 
Hence the model is completely specified by the matrices $\ftb{v}$,
$\PT$ and $\AT$ where the matrices generally are sparse, and the
latter have zeros in the diagonal.  
In graph-terms the matrix $\AT$
represents the asymmetric paths whereas $\PT$ represent the symmetric
paths.

Let $k$ be the total number of variables in the model
(i.e. $\ftb{A}\in\R^{k\times k}$).  We let $\bm{J}$ be the matrix
that picks out the observed variables from $\bm{U}$ (see Section
\ref{sec:zeroone}), and define
\begin{align}
  \IA = \bm{J}(\one-\AT)^{-1}.
\end{align}
Now it follows that 
\begin{align}
  \var(\bm{Z}) = \SigmaF = \IA\PT\IA'.
\end{align}
and similarly the mean of the
observed variables is then specified by the model structure as:
\begin{align}
  \E(\bm{Z}) = \ftb{\xi} = \IA\ftb{v}.
\end{align}
The parameter $\bm{\theta}$ can then be estimated by maximizing the
log-likelihood, $\ell$, for the full data vector $\bm{Z}$
\begin{align}
  \widehat{\bm{\theta}}_{ML} = \argmax_\theta \ell(\bm{\theta}\mid
  \bm{Z} = \bm{z}).
\end{align}


At a first glance this formulation seems restrictive in the sense that
the covariates also have to be normally distributed. However, if we
split the parameter vector into $\bm{\theta} =
(\bm{\theta}_1,\bm{\theta}_2)$, where $\bm{\theta}_1$ parameterizes the
conditional distribution of $\bm{Y}$ given $\bm{X}$ and $\bm{\theta}_2$ are the mean and
variance parameters of the covariates, then by Bayes formula the
probability density for the joint distribution can be written as the
product of the conditional density multiplied by the marginal density:
\begin{align}\label{eq:ffull}
  f_{\bm{\theta}_1,\bm{\theta}_2}(\bm{y},\bm{x}) = f_{\bm{\theta}_1}(\bm{y}\mid \bm{x})f_{\bm{\theta}_2}(\bm{x}).
\end{align}
It follows that maximum likelihood inference about the parameters
$\bm{\theta}_1$ is independent of the model for the covariates, and
hence finding the MLE of the conditional likelihood is equivalent to
finding the MLE of the joint likelihood. If we fix $\bm{\theta}_2$ to
the corresponding MLE, then
the expected information agrees in the two models:
\begin{align}
  -\E\left(\frac{\partial^2\log f_{\bm{\theta}_1}(\bm{y}\mid \bm{x})}{\partial\bm{\theta}_1\partial\bm{\theta}_1'}\right) =
  -\E\left(\frac{\partial^2\log f_{\bm{\theta}_1,\widehat{\bm{\theta}}_2}(\bm{y},\bm{x})}{\partial\bm{\theta}_1\partial\bm{\theta}_1'}\right).
\end{align}
An important advantage of the RAM formulation is that the empirical
variance and mean are sufficient statistics for the parameters, thus
giving the joint formulation a computational advantage over the
conditional likelihood which requires explicit calculation of the
likelihood contribution of each individual observation. In particular given
the sufficient statistics, which are easily computed, the
computational complexity in the RAM parameterization is independent of
$n$.



\subsection{Inference - standard SEM}

In the following we  will describe inference under \emph{a priori} non-linear
constraints $\bm{\Omega}=\SigmaF$ (twice differentiable
w.r.t. $\bm{\theta}$). Letting $\wmu$ denote the empirical mean of the observed variables, we define
\begin{align}
  \ftb{W} = \left[\wmu-\ftb{\xi}\right]\left[\wmu-\ftb{\xi}\right]'
\end{align}
and
\begin{align}
  \ftb{T} = \widehat{\bm{\Sigma}} + \ftb{W},
\end{align}
where $\widehat{\bm{\Sigma}}$ is the ML covariance matrix estimate
(non-central estimate).
We will exploit that
\begin{align}
  \begin{split}
    \sum_{i=1}^n (\bm{z}_i-\ftb{\xi})(\bm{z}_i-\ftb{\xi})' &=
    \sum_{i=1}^n (\bm{z}_i-\wmu + \wmu - \ftb{\xi})(\bm{z}_i-\wmu +
    \wmu -
    \ftb{\xi})' \\
    &= n(\wS + \ftb{W}) + \\
    &\qquad n\big[ (\wmu-\wmu)(\wmu-\ftb{\xi})' +
    (\wmu-\ftb{\xi})(\wmu-\wmu)' \big] \\
    &= n\ftb{T}.
  \end{split}
\end{align}
The complete log-likelihood then is given by 
\begin{align}\label{eq:loglik1}
  \begin{split}
    \ell(\bm{\theta}\mid \bm{z}_1,\ldots,\bm{z}_n) &= \sum_{i=1}^n\Big\{ -\frac{k}{2}\log(2\pi) -
      \frac{1}{2}\log\abs{\SigmaF} - \\
      &\qquad \frac{1}{2}
      (\bm{z}_i-\ftb{\xi})'\SigmaF^{-1}(\bm{z}_i-\ftb{\xi})\Big\} \\
    &= -\frac{nk}{2}\log(2\pi) - \frac{n}{2}\log\abs{\SigmaF} -
    \frac{n}{2}\tr\{\ftb{T}\SigmaF^{-1}\},
  \end{split}
\end{align}
with score
\begin{align}
  \begin{split}
    \frac{\partial \ell(\bm{\theta})}{\partial\bm{\theta}} &=
    \frac{n}{2}\left(\frac{\partial\mvec\SigmaF}{\partial\bm{\theta}'}
    \right)'\mvec\left[\SigmaF^{-1}\ftb{T}\SigmaF^{-1}\right] \\
    &\qquad
    -\frac{n}{2}\left(\frac{\partial\mvec\SigmaF}{\partial\bm{\theta}'}
    \right)'\mvec\left[\SigmaF^{-1}\right]
    \\
    &\qquad -
    \frac{n}{2}\left(\frac{\partial\mvec\ftb{T}}{\partial\bm{\theta}'}\right)'\mvec\left(\SigmaF^{-1}\right),
  \end{split}
\end{align}
and the MLE is obtained by solving the corresponding score
equation by Fisher scoring or a similar iterative procedure.
See \ref{sec:scorehessian} for details w.r.t. expressions for
the relevant matrix derivatives and information matrix.

In certain situations we may need to calculate conditional moments,
for instance to calculate conditional residuals (model diagnostics) or the
conditional likelihood given the covariates (likelihood ratio
testing, model selection, etc.). Here we need to apply the selection matrix,
$\bm{J}_{\bm{Y}}$, that picks out the \emph{endogenous variables},
$\bm{Y}$, of $\bm{U}$, and the cancellation matrix, $\bm{p}_{\bm{X}}$,
that sets all \emph{exogenous variables}, $\bm{X}$, of $\bm{U}$ to
zero (see Section \ref{sec:zeroone}). Notice we do not put any
distributional assumptions on the exogenous variables.

Now
\begin{align}
  \bm{\mu_i}(\bm{\theta}) = \E(\bm{Y}\mid\bm{X}=\bm{x}) =
  \bm{J}_{\bm{Y}}(\one-\ftb{A})^{-1}
  (\bm{p}_{\bm{X}}\ftb{v} + \bm{v}_{\bm{x}}),
\end{align}
where $\bm{v}_{\bm{x}}$ is the $p$-vector which is zero everywhere but on the index of the
exogenous variables where it is set to $\bm{x}$, and
\begin{align}
  \SigmaT = \var(\bm{Y}\mid\bm{X}) =
  \bm{J}_{\bm{Y}}(\one-\ftb{A})^{-1}
  (\bm{p}_{\bm{X}}\ftb{P}\bm{p}_{\bm{X}}')
  (\one-\ftb{A})^{-1}{}'\bm{J}_{\bm{Y}}'.
\end{align}



\subsection{Interactions with latent
  variables}
\label{sec:randominteractions}
Including interactions between covariates and random effects in the model, i.e. with
non-zero $\bm{\Delta}$ and $\bm{T}$ in (\ref{eq:measurement}) and
(\ref{eq:structural}), we clearly loose the property of constant
variance between individuals and the empirical mean and variance are
therefore not sufficient.  With $\bm{\Delta}$ and $\bm{T}$
consisting of ones in all entries, the marginal variance of
$\bm{Y}_i$, $\var(\bm{Y}_i\mid \bm{X}_i,\bm{V}_i,\bm{W}_i)$, will
however take the same form as (\ref{eq:margvar}), exchanging $\bm{B}$
with
\begin{align}\label{eq:Bi}
  \bm{B}_i = \bm{B}+\bm{W_i},
\end{align}
and
\begin{align}
  \bm{\Lambda}_i = \bm{\Lambda}+\bm{Z_i},
\end{align}
hence the expression for the likelihood contribution and its derivatives of a single
individual will take the same form as derived in
\ref{sec:scorehessian}. 

With free parameters in $\bm{\Delta}$ or $\bm{T}$ we can still adapt
the model by adding one or more degenerate random effects. For instance, the term 
\begin{align}
  \xi_{ij} = \delta_{jk}V_{ijk}\eta_{ik} + \epsilon_{ij}
\end{align}
where $\eta_{ik}$ and $\zeta_{ijk}$ follow normal distributions, can
trivially be parameterized as the simultaneous equation
\begin{align}
  \xi_{ij} = \delta_{jk}\widetilde{\eta}_{ik} + \epsilon_{ij}, \\
  \widetilde{\eta}_{ik} = V_{ijk}\eta_{ik} + 0,
\end{align}
hence the model is expanded to include a random effect with residual
term with variance 0 with fixed slope parameter $Z_{ijk}$, and the
variance of the observed variables in such a model therefore takes the
form as described above (\ref{eq:Bi}), i.e. $\bm{\Delta}=\bm{1}$.

The possibility of including interactions with the random effects adds
important flexibility to the model class for instance when modeling
longitudinal data or to account for certain types of variance
heterogeneity.


\subsection{Non-linear effects}\label{sec:nonlinear}

Allowing non-linear parameter constraints or non-linear effects of
some covariates opens up for several interesting applications for
instance in dose-response modeling. An important extension of the model
framework is therefore to allow parametric non-linear functions
$\phi^{(j)}, j=1,\ldots,p$ and $\psi^{(s)}, s=1,\dots,l$ of the
covariates to enter the model:
\begin{align}
  \begin{split}\label{eq:summeas2}
    Y_{ij} &= \nu_j + \sum_{k=1}^l\lambda_{jk}\eta_{ik} +
    \sum_{r=1}^q\kappa_{jr}X_{ir}
    + \sum_{k=1}^l  \delta_{jk}V_{ijk}\eta_{ik} \\
    &\qquad\qquad + \phi^{(j)}_{\bm{\gamma}_1}(X_{i1},\ldots,X_{iq}) + \epsilon_{ij},
  \end{split}
\end{align}
and 
\begin{align}
  \begin{split}\label{eq:sumstruct2}
    \eta_{is} &= \alpha_{s} + \sum_{k=1}^l \beta_{sk}\eta_{ik} +
    \sum_{r=1}^q \gamma_{sr}X_{ir} + \sum_{k=1}^l
    \tau_{sk}W_{isk}\eta_{ik} \\
    &\qquad\qquad + \psi^{(s)}_{\bm{\gamma}_2}(X_{i1},\ldots,X_{iq}) + \zeta_{is}.
  \end{split}
\end{align}
Though we introduce non-linear effects in the model, we will still
denote this model a \emph{Linear Latent Variable Model}, as there are
only linear effects of latent variables and endogenous variables in
the model. Hence the observed data likelihood and its derivatives
still have a closed form solution.

\subsection{Multigroup analysis}

A useful generalization of the model framework is the \emph{multigroup
  model}, where we have several groups of data and specify a LLVM
for each group. This naturally leads to the log-likelihood
\begin{align}\label{eq:multigroup}
  \log L(\bm{\theta}\mid \bm{Y},\bm{X},\bm{V},\bm{W}) = \sum_{g\in G}
  \log L_g(\bm{\theta}_g\mid \bm{Y}_g, \bm{X}_g,\bm{V}_g,\bm{W}_g)
\end{align}
where the intersection of the parameters $(\bm{\theta}_g)_{\{g\in
  G\}}$ is not empty. Unbalanced designs and data with values missing
at random \citep{MR1925014} is naturally handled by this extension by
forming groups from the different missing data patterns. Note that the
additive structure of the log-likelihood makes the calculation of
score functions and information matrices for the multigroup model
explicitly available.



\section{Model specification}

The \pkg{lava} package aims to deliver a dynamic model specification
experience in the sense that adding or removing model elements should
be as easy as possible, and the model specification should be familiar
to users accustomed to specifying models in for example \code{glm} in
\proglang{R}. In order to achieve this we have designed a formal
system for interactively specifying the complex hierarchical
structure of a latent variable model. We believe this to be an important novel
contribution, since the difficult specification of models in other
languages often proves to be a significant obstacle.

The implementation relies on \proglang{R} \citep{Rmanual} and the following
packages all available from the Comprehensive R Archive Network (CRAN)
: \pkg{mvtnorm} \citep{mvtnorm}, \pkg{graph}
\citep{graphR}, \pkg{survival} \citep{survival},
\pkg{numDeriv} \citep{numderivR} and \pkg{gof}
\citep{holst09:gof}. The graphical system builds on \proglang{graphviz}
\citep{graphviz} and the R-package \pkg{Rgraphviz} \citep{rgraphvizR}
(available from Bioconductor \cite{bioconductor}).

\vspace*{0.5cm}


\begin{table}[ht]
  \centering
  \begin{tabular}{llp{6.5cm}}
    & \textbf{Function} & \textbf{Task} \\ \hline
    Primary functions \\
    & \code{lvm} & Constructor of new model   \\
    & \code{regression} & Add regression association to model \\
    & \code{covariance} & Add correlation between residuals terms\\
    & \code{intercept} & Add intercept parameter\\ 
    & \code{constrain} & Add non-linear covariate effects or \\
    & & parameter constraints \\ 
    Secondary functions \\    
    & \code{latent} & Define latent variables in model \\ 
    & \code{addvar} & Add variable to model \\ 
    & \code{parfix} & Define equality constraints index of parameters \\
    & \code{parameter} & Add a parameter name (for use with \code{constrain})\\ 
    & \code{cancel} & Remove previously defined associations \\
    & \code{kill} & Remove variables from model \\
    \hline
  \end{tabular}
  \caption{Model building blocks}\label{tab:modelbuilding}
\end{table}
\noindent The specification of models in the \pkg{lava}-language is primarily
achieved via the constructor function \code{lvm} and the two methods
\code{regression} and \code{covariance} (see Table \ref{tab:modelbuilding}).  
  
A new model object is initialized with the constructor
\code{lvm}\index{\code{lvm}}
\begin{Schunk}
\begin{Sinput}
> m1 <- lvm()
\end{Sinput}
\end{Schunk}
which creates an empty \code{lvm}-object. Variables (or a
multivariate regression formula as described below) can be fed to
the \code{lvm}-function as arguments in order to control the order of
entry in the graph layout.
However, this is optional as variables automatically will be added
during the process of defining the linear structure. A list of
formulas is also valid as a shortcut for successive calls to
\code{regression} (see below).

\subsection{Specifying Linear Relationships}
\label{sec:modelspec}

\begin{figure}[htbp]\centering
\includegraphics{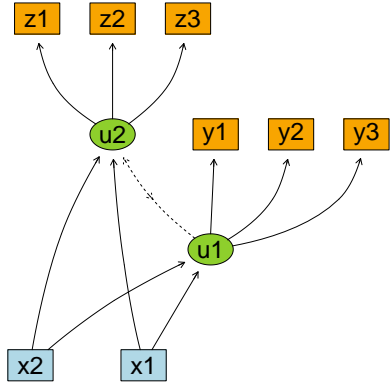}
\caption{Path diagram of the model \code{m1} with correlation between residuals of
  the latent variables. Obtained with the command
  \code{plot(m1)}. Following the convention of path
  diagrams observed variables are framed with rectangles whereas
  latent variables are framed with ellipses. Regression associations
  are depicted as one-headed arrows (parent=predictor, child=response) and
  covariance/correlations are shown as (dashed) double-headed
  arrows. For easier interpretation the following color-codes are
  used: exogenous:=light blue, endogenous:=orange, latent:=green. }
\label{fig:m1cor}
\end{figure}

Linear associations between variables are specified via the member
function
\code{regression}\index{\code{regression},\code{regression<-}} taking
the character vector arguments \code{to} and \code{from}. For
convenience a replacement function, \code{regression<-}, is also
available which in addition supports specification via the usual
\emph{formula} statements in \pkg{R}. As a simple example we will
specify the following structural equation model with two measurement models
\begin{align*}
  Y_{j} &= \mu_j + \lambda_{1j} U_1 + \epsilon_{1j},  \\
  Z_{j} &= \nu_j + \lambda_{2j} U_2 + \epsilon_{2j}, \quad j=1,\ldots,3, \\
\end{align*}
and structural model defined by
\begin{align*}
  U_{1} &= \delta_{1}X_1 + \delta_{2}X_2 + \zeta_{1}, \\
  U_{2} &= \beta_{1}X_1 + \beta_{2}X_2 + \zeta_{2},  
\end{align*}
and $\cov(\zeta_1,\zeta_2)\neq 0$.
The model is illustrated in the path diagram of Figure \ref{fig:m1cor}.

The following commands specifies a
multivariate linear regression model with two covariates \code{x1} and
\code{x2} and two outcomes \code{u1} and \code{u2}
\begin{Schunk}
\begin{Sinput}
> m1 <- regression(m1, "u1", c("x1", "x2"))
> regression(m1) <- u2 ~ x1 + x2
\end{Sinput}
\end{Schunk}

In the following we will focus on the replacement functions and the
formula specification as defined by the second line, but generally for
all the available methods a standard function is available and
arguments can be given as character vectors as above.  A more compact
call would simply be
\begin{Schunk}
\begin{Sinput}
> m1 <- lvm(c(u1, u2) ~ x1 + x2)
\end{Sinput}
\end{Schunk}

Next, we define \code{u1} and the
\code{u2} as a latent/un\-obs\-erved variables using the
\code{latent}-function\index{\code{latent},\code{latent<-}}:
\begin{Schunk}
\begin{Sinput}
> latent(m1) <- ~u1 + u2
\end{Sinput}
\end{Schunk}
Again arguments can generally be given as character vectors instead
of a formulas. To remove the latent status from a variable we simply
use the \code{cancel=TRUE} argument (\code{latent(m1,cancel=TRUE) <- ...}).

Next we define the measurement part of the model:
\begin{Schunk}
\begin{Sinput}
> regression(m1) <- c(y1, y2, y3) ~ u1
> regression(m1) <- c(z1, z2, z3) ~ u2
\end{Sinput}
\end{Schunk}
Covariance between residual terms can be specified using
the replacement function \code{covariance<-} (or the function
\code{covariance})\index{\code{covariance},\code{covariance<-}} where
the argument is a vector (or formula) of variables that are assumed to
be pairwise correlated.
In the current model specification, the residuals of the two latent
variables are assumed to
be conditionally independent given the covariates, and in order to
define correlation between the residual terms of \code{u1} and
\code{u2} we write
\begin{Schunk}
\begin{Sinput}
> covariance(m1) <- u1 ~ u2
\end{Sinput}
\end{Schunk}
which specifies $\cov(\zeta_1,\zeta_2)\neq 0$,
thus completing the specification of the model defined by the path
diagram in Figure \ref{fig:m1cor}. Note that the model is specified
independently of any data. The model is linked to data when parameters
are estimated using the \code{estimate} function (see Section
\ref{sec:inference}). Here it is important that the manifest variable
names used in the model specification corresponds to the variable
names in the data-frame.


\relsize{-2}
\origfigsize 

\begin{figure}[htbp]\centering
\includegraphics{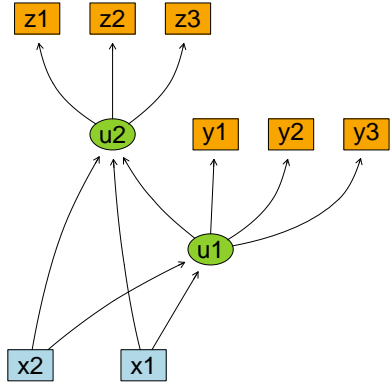}
\caption{Path diagram of the final specification of the model
  \code{m1} with a direct regression association between 
  the two latent variables.}
\label{fig:m1corfinal}
\end{figure}

Removal of associations or variables can be achieved with the
\code{cancel} (and
\code{cancel<-})\index{\code{cancel},\code{cancel<-}} function which
takes a character vector (or formula) as argument, removing any
associations between all the variables in the vector. To remove the
previously specified correlation and instead add a regression
association between \code{u1} and \code{u2}, we write
\begin{Schunk}
\begin{Sinput}
> cancel(m1) <- ~u1 + u2
> regression(m1) <- u2 ~ u1
\end{Sinput}
\end{Schunk}
Notice that the last \code{regression} call defining the association
between \code{u1} and \code{u2} does \emph{not} cancel the earlier
defined predictors of \code{u2}. Hence the current definition says
(see Figure \ref{fig:m1corfinal}) that
\begin{align}
  \E(U_2\mid U_1,X_1,X_2) = \mu + \beta_0 U_1 + \beta_1 X_1 + \beta_2 X_2
\end{align}
for some parameters $(\mu,\beta_0,\beta_1,\beta_2)$.

To completely remove one or more variables from the model we can use
the \code{kill<-}\index{\code{kill},\code{kill<-}} function.

\relsize{-2}
\origfigsize 

\subsection{Constraining Parameters}
Defining restrictions on some parameters
is usually needed in order to obtain an identifiable
model, and by default \pkg{lava} will automatically set reasonable
restrictions when model parameters are estimated (see Section \ref{sec:inference}).
Also, in situations where we need to test associations or
use a priori knowledge in the model building, constraints on some
parameters are needed. The \pkg{lava} package allows specification of completely general
non-linear constraints on all parameters.

The most common type of constraints are identity constraints where one
or more parameters are fixed to either a specific numerical value or
to be equal to a common free parameter (equality constraints). For
covariance and regression parameters (slope-parameters) the
\code{regression} and \code{covariance} function can be used, and for
the intercepts $\bm{\alpha}$ and $\bm{\nu}$ in
(\ref{eq:measurement}-\ref{eq:structural}), the \code{intercept<-} and
\code{intercept} \index{\code{intercept},\code{intercept<-}} functions
are used. 

As an continuing example we will specify a new multivariate regression
model with three outcomes, $(Y_1,Y_2,Y_3)$, and two predictors, $(X_1,X_2)$:
\begin{align*}
  Y_i = \beta_i X  + \gamma_i Z + \epsilon_i, i=1,\ldots,3
\end{align*}

\begin{Schunk}
\begin{Sinput}
> mregr <- lvm(c(y1, y2, y3) ~ x + z)
\end{Sinput}
\end{Schunk}

\subsubsection{Constraining regression parameters}

The restrictions of slope parameters can be accomplished
with the \texttt{regression} function. 
For example fixing the slopes of $Y_1$, $Y_2$
and $Y_3$ on $X$ to be identical, $b_1$, and defining $Z$ as an offset
(i.e. slope 1), can be achieved with the calls 
\begin{Schunk}
\begin{Sinput}
> regression(mregr, c(y1, y2, y3) ~ x) <- "b1"
> regression(mregr, c(y1, y2, y3) ~ z) <- 1
\end{Sinput}
\end{Schunk}
To simultaneously define several different constraints a list can be
given as the right-hand side argument
\begin{Schunk}
\begin{Sinput}
> regression(mregr, c(y1, y2) ~ x + z) <- list(1, "a", 2, "b")
\end{Sinput}
\end{Schunk}
All parameters for the first response on the covariates are given
first, then for the second response, and so on. Hence in the above
example we have
\begin{align}
  Y_1 = X + a Z + \cdots \,\quad\text{ and }\quad 
  Y_2 = 2 X + b Z + \cdots
\end{align}

When defining constraints (intercepts, covariance or regression
constraints) any missing associations will automatically be added to
the model object. Hence the following call will add an extra level to the
model, with a top-level response $W$ with identical effects of
$Y_1$, $Y_2$ and $Y_3$ (see Figure \ref{fig:mregr}):
\begin{Schunk}
\begin{Sinput}
> regression(mregr, w ~ y1 + y2 + y3) <- "beta"
\end{Sinput}
\end{Schunk}

To remove the constraints again (but not removing the associations) we simply
fix to the logical constant \code{NA}:
\begin{Schunk}
\begin{Sinput}
> regression(mregr, w ~ y1 + y2 + y3) <- NA
\end{Sinput}
\end{Schunk}

\subsubsection{Constraining covariance parameters}

Constraints on the covariance between residual terms are set
with the \code{covariance} function, with a similar syntax.
For example, to fix the covariance of the residuals
terms of $Y_1$ and $Y_2$ to $0.5$, and the variance of the residual term
of $Y_1$ to a parameter \code{v1}:
\begin{Schunk}
\begin{Sinput}
> covariance(mregr, y1 ~ y1 + y2) <- list("v1", 0.5)
\end{Sinput}
\end{Schunk}
If we only need to constrain the variance parameters (i.e. not
covariances) then the following syntax can be used
\begin{Schunk}
\begin{Sinput}
> covariance(mregr, ~y1 + y2) <- "v"
\end{Sinput}
\end{Schunk}
here setting the residual variance of $Y_1$ and $Y_2$ to be identical.

To fix the variance of the variables to different values, we simply
give a list of the correct length as argument, e.g.
\begin{Schunk}
\begin{Sinput}
> covariance(mregr, ~y1 + y2) <- list("v", 0.3)
\end{Sinput}
\end{Schunk}
If we are interested in fixing only the covariance parameters (and not
the diagonal) we can add the \code{pairwise=TRUE} argument. For
instance to specify that the covariances between $Y_1$,$Y_2$, and $Y_3$
are the same, we can call
\begin{Schunk}
\begin{Sinput}
> covariance(mregr, ~y1 + y2 + y3, pairwise = TRUE) <- "r1"
\end{Sinput}
\end{Schunk}
thus specifying a compound-symmetry structure.

\begin{figure}[!ht]
  \centering 
\includegraphics{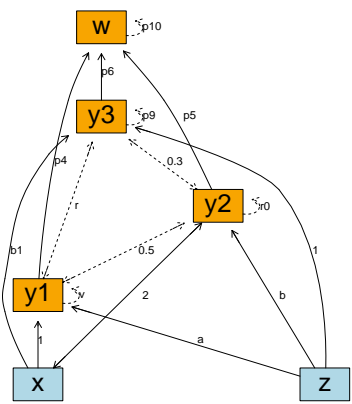}
\caption{\code{plot(mregr, labels=TRUE, diag=TRUE)}}
\label{fig:mregr}
\end{figure}

Finally a syntax like the one used by the \code{regression}-function
can be used, such that
\begin{Schunk}
\begin{Sinput}
> covariance(mregr, c(y1, y2) ~ y2 + y3) <- list(0.5, "r", "r0", 
+     0.3)
\end{Sinput}
\end{Schunk}
defines the following covariance structure between residual terms
$\varepsilon_1, \varepsilon_2, \varepsilon_3$ corresponding to $Y_1,
Y_2, Y_3$ (with the
$\cdot$ denoting elements that are not affected by the call):
\begin{align}
  \var((\varepsilon_1,\varepsilon_2,\varepsilon_3)') =
  \begin{pmatrix}
    \cdot & 0.5 & r \\
    0.5 & r_0 & 0.3 \\
    r & 0.3 & \cdot
  \end{pmatrix}
\end{align}
As with intercept and slope parameters, we can remove covariance
constraints by fixing parameters to the value \code{NA}.  We end up
with a model as defined by the path-diagram in Figure \ref{fig:mregr}.



\relsize{-2}
\origfigsize

\subsubsection{Constraining intercepts}
To fix the intercepts of the three outcomes to
be identical, we can write
\begin{Schunk}
\begin{Sinput}
> intercept(mregr, ~y1 + y2 + y3) <- "mu"
\end{Sinput}
\end{Schunk}
Instead of a parameter name, \code{mu}, we could have chosen a
numerical, say 0.  

Notice that a character vector could also have been given instead of
the formula. The value on the right can be given as a list, hence to
fix the intercept of $Y_1$ and $Y_2$ to be identical and
$Y_3$ to zero, we could call
\begin{Schunk}
\begin{Sinput}
> intercept(mregr, ~y1 + y2 + y3) <- list("mu", "mu", 0)
\end{Sinput}
\end{Schunk}


\subsection{Simultaneously specifying constraints on intercepts,
  slopes and variances}

Using the formula syntax with the \code{regression} method it is
possible to simultaneously specify constraints on intercept, regression
and covariance parameters. A special function, \code{f}, can be used
within the formula to specify the slope parameters, and further a pair of
square brackets can be appended to each variable in the
formula. Inside the square bracket the intercept of the variable can
be defined, or alternatively both the intercept and residual variance
separated by a colon. As an example we can specify
\begin{align*}
  Y_1 &= a + b U + \epsilon_1, \quad \epsilon_1\sim\mathcal{N}(0,1), \\
  Y_2 &= a + b U + \epsilon_2, \quad \epsilon_2\sim\mathcal{N}(0,v), \\
  U &= b_2 X + \zeta,
\end{align*}
using the square-bracket syntax:
\begin{Schunk}
\begin{Sinput}
> m2 <- lvm()
> regression(m2) <- c(y1[a:1], y2[a:v]) ~ f(u[0], b)
> regression(m2) <- u ~ f(x, b2)
\end{Sinput}
\end{Schunk}
An equivalently even more compact model specification can be obtained
using a list of formulas with the model initializer \code{lvm}, hence
an equivalent way of specifying \code{m2} would be
\begin{Schunk}
\begin{Sinput}
> m2 <- lvm(list(c(y1[a:1], y2[a:v]) ~ f(u[0], b), u ~ f(x, b2)))
\end{Sinput}
\end{Schunk}

If the index of the parameter is known (see the \code{coef}
method below) the \code{parfix} method can also be used to
simultaneous constrain parameters. For example to fix
the parameters of a model \code{m}, at positions 1, 4 and 5 to the
values \code{a}, \code{a} and \code{1}, we can call
\begin{Schunk}
\begin{Sinput}
> parfix(m, c(1, 4, 5)) <- list("a", "a", 1)
\end{Sinput}
\end{Schunk}

\subsection{Random slopes}
Random slope effects (i.e. the matrices $\bm{V}_i$ and $\bm{W}_i$) can
be defined by constraining the slope parameter of a latent variable to
the name of a covariate. The covariate does not necessarily need to be
added to the model explicitly, as the slope parameters are matched to
the column names of the \code{data.frame} that is used during
estimation (see Section \ref{sec:inference}).

\begin{figure}[htbp]
  \centering
  \includegraphics[height=8cm,keepaspectratio=true]{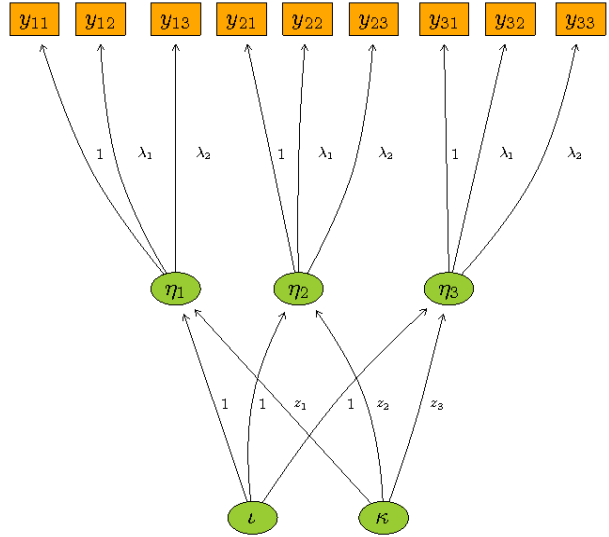}
  \caption{Longitudinal analysis with measurement models. Here
    illustrated with three time points where 
    $\eta_1$, $\eta_2$, $\eta_3$ are modeled by a random
    intercept, $\iota$, and a random slope defined by the covariate
    $z_i$, $i=1,2,3$ and the random effect $\kappa_i$:
    $\eta_i = \iota + \kappa_i z_i + \zeta_i$. As the $\eta$'s are only
    indirectly observed a measurement model is employed at each time point.    
  }
  \label{fig:longitudinal}
\end{figure}

In a standard structural equation model covariates enter the model in
order to describe differences in the mean structure. Another question
might be whether covariates can influence the variation in an
outcome. In a situation where the variance depends on a covariate, the
usual assumptions of variance homogeneity of the latent variables are
not met and the model is not a SEM. In the LLVM this situation can be
handled by defining a regression on the variance component using the
random slope specification. Another important example is to use the
random slopes to describe variation in longitudinal studies, and in
combination with measurement models this allows us to formulate random
regression models taking measurement error into account (see Figure
\ref{fig:longitudinal}).

\subsection{Non-linear constraints and effects}

Non-linear parameter constraints are defined using the
\code{constrain<-}\index{\code{constrain},\code{constrain<-}} function.
The syntax is 
\begin{Schunk}
\begin{Sinput}
> constrain(m, formula) <- function(x) ...
\end{Sinput}
\end{Schunk}
where \code{m} is the \code{lvm}-object, and the left-hand side in the
\code{formula} specifies the parameter that is a (non-linear) function
of the parameters or covariates defined by the right-hand side, and the
\code{function} defines this association. The result of the function
can optionally be given the attributes \code{grad}, defining the
analytic gradient, and \code{inv}, defining a monotonic
transformation (typically the inverse function) used in conjunction
with confidence limit calculations (see Section \ref{sec:inference} on
Statistical Inference).

As an example we will define the model
\begin{align}
  Y_{ij} = \mu + b_j X_{i} + \epsilon_{ij}, \quad j=1,2
\end{align}
with $\cov(\varepsilon_{i1},\varepsilon_{i2})=0, \var(\epsilon_{ij})=v$:
\begin{Schunk}
\begin{Sinput}
> mconstr <- lvm()
> regression(mconstr, c(y1, y2) ~ x) <- list("b1", "b2")
> intercept(mconstr, ~y1 + y2) <- "mu"
> covariance(mconstr, ~y1 + y2) <- "v"
\end{Sinput}
\end{Schunk}
To restrict the variance-parameter to live on the positive real axis,
we add the constraint
\begin{align}
  v = \exp(\alpha)
\end{align}
with
\begin{Schunk}
\begin{Sinput}
> constrain(mconstr, v ~ alpha) <- exp
\end{Sinput}
\end{Schunk}
using a log-link, and the parameter \code{alpha} (log-variance) is
added to list of model parameters. In this example we do not set any
attributes and numerical derivatives of the constraint will therefore be used
(based on the package \pkg{numDeriv}) during estimation. 
Notice that constraints on the variance
parameters can be set automatically by the
\code{estimate}-function (see Section \ref{sec:inference}). For
general domain constraints the function \code{range.lvm} can be used
as the right-hand side argument, e.g. \code{range.lvm(a=1,b=Inf)} will
bound the parameter to the interval $(1,\infty)$.

Continuing the example, we could define the intercept, $\mu$, to be the
product of the two slope-parameters,
\begin{Schunk}
\begin{Sinput}
> constrain(mconstr, mu ~ b1 + b2) <- prod
\end{Sinput}
\end{Schunk}

Constraints can be removed by letting the RHS be \code{NULL} (or \code{NA})
\begin{Schunk}
\begin{Sinput}
> constrain(mconstr, "mu") <- NULL
\end{Sinput}
\end{Schunk}

If we instead wish to add a non-linear effect of $x$ on $y_1$ and
$y_2$:
\begin{align}
  Y_{ij} = \alpha + \Phi(\beta X_{ij}) + b_j X_{ij} + \epsilon_{ij}, \quad i=1,2
\end{align}
with $\Phi$ denoting the standard normal cumulative distribution
function, we can make the call
\begin{Schunk}
\begin{Sinput}
> constrain(mconstr, mu ~ alpha + beta + x) <- function(x) x[1] + 
+     pnorm(x[2] * x[3])
\end{Sinput}
\end{Schunk}

\subsection{Complex models with feedback or co-existance of regression associations
  and covariance between residuals}

The linear latent variable model framework in principle allows pathways
going in both direction between two variables, i.e. feedback,
and also simultaneous presence of both a regression association and
covariance between residual terms of the same two variables.
Note that both cases are different from a simple correlation between the
residuals.



\begin{figure}[htbp]\centering
\includegraphics{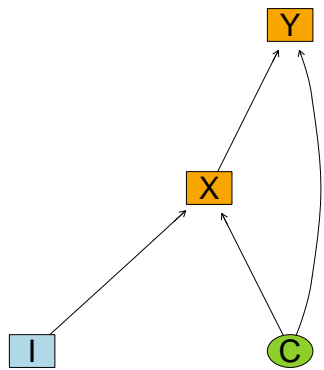}
\caption{Path diagram showing the association between the variables $Y$ and
$X$ with unmeasured confounder $C$ and instrumental variable $I$.}
\label{fig:unmeasconf}
\end{figure}

A simple example of an identified model is a illustrated in Figure
\ref{fig:unmeasconf}, where the aim is to estimate the effect of $Y$
on $X$ while taking the unmeasured confounder $C$ into account.  This
can be achieved using an \emph{instrumental variable}, $I$, which by
assumption must be (strongly) correlated with $X$, independent with
the confounder, and conditionally independent with the $Y$
given $X$. We can model the inflation in covariance between
$Y$ and $X$ using a correlation between their residual terms, which
can be implemented as
\begin{Schunk}
\begin{Sinput}
> m <- lvm()
> regression(m) <- y ~ x
> regression(m) <- x ~ i
> covariance(m) <- x ~ y
\end{Sinput}
\end{Schunk}

\relsize{-2}
\origfigsize

\section{Inspecting the model assumptions}

\begin{table}[ht]
  \centering
  \begin{tabular}{ll}
    \textbf{Function} & \textbf{Task} \\ \hline
    \code{plot} & Plot the graph of the model \\
    \code{regression} & Display parameter restrictions \\ 
    \code{intercept} & Display parameter restrictions \\ 
    \code{covariance} & Display parameter restrictions  \\ 
    \code{exogenous} & Extract (or define) exogenous variables
    (predictors) \\ 
    \code{endogenous} & Extract set of endogenous variables
    (responses) \\ 
    \code{latent} & Extract set of latent variables\\     
    \code{manifest} & Extract set of manifest (observed) variables \\     
    \code{vars} & Extract all variables \\     
    \code{children} & Extract children of a node \\ 
    \code{parents} & Extract parents of a node \\ 
    \code{coef} & Get list of parameters \\
    \code{constrain} & Display non-linear constraints \\ 
    \code{path} & Extract direct and indirect unidirectional \\
    & pathways between nodes \\
    \code{subset} & Extract sub-model \\
    \code{merge},\code{\%+\%} & Merge models \\
   \hline
  \end{tabular}
  \caption{Model inspection functions.}\label{tab:modelinspection}
\end{table}

\noindent The philosophy of the \pkg{lava} package is to separate the model
specification from the actual data, since examination of the model
structure before actual estimation is often an important aspect of
modelling within this class of statistical models. 
The \pkg{lava} package includes several functions as an aid to obtain an
overview of the model assumptions (see Table \ref{tab:modelinspection}).
In general, the below described methods also applies for a
\code{lvmfit}-object (see Section \ref{sec:inference}).

The \code{plot}-method (see Figure \ref{fig:m1cor} and
\ref{fig:mregr}) visualizes the model using a path-diagram, i.e. a
graph structure where linear (causal) associations are shown with
directed edges and covariance between residuals are shown as
bidirectional edges. Manifest variables are shown as rectangles and
latent variables as ellipsoids. The parameter constraints can be added
as labels on the edges with the argument \code{labels=TRUE} and
variance parameters can be added with the argument \code{diag=TRUE}
(see Figure \ref{fig:mregr}). The \code{plot}-function will be
explained in more details in Section \ref{sec:graphics}.

Via the \code{summary} function, a complete overview of the model and
(identity) parameter constraints can be obtained. Returning to the
model \code{mregr} defined in Figure \ref{fig:mregr}:
\begin{Schunk}
\begin{Sinput}
> summary(mregr)
\end{Sinput}
\begin{Soutput}
Latent Variable Model 
	with: 6 variables.
Npar=11+2

Regression parameters:
      y1 y2 y3 x z w
   y1              *
   y2              *
   y3              *
   x  1  2  b1      
   z  a  b  1       
   w                
Covariance parameters:
      y1  y2  y3  w
   y1 v   0.5 r    
   y2 0.5 r0  0.3  
   y3 r   0.3 *    
   w              *
Intercept parameters:
    y1 y2 y3 w
    mu mu 0  *
\end{Soutput}
\end{Schunk}
Here the adjacency-matrix for the graph of all the unidirectional
edges of the path-diagram can be read off under the title
\emph{Regression parameters} (i.e. slopes). Similarly the covariance
structure of the residual terms and the intercept structure are
shown.  An empty element indicates that there is no direct
association. A star indicates a free parameter and all other entries
are either fixed numerical values or parameter names as defined by
identity constraints during model specification. These three matrices
can also be extracted via calls to \code{regression},
\code{covariance} or \code{intercept}.

The exogenous variables (covariates) of a \code{lvm} object can be
identified with the \code{exogenous} function. Similarly a list of
the endogenous (manifest) variables can be obtained with the function
\code{endogenous} and the subset of these variables that do not
predict other variables (top-level outcomes) can be shown by including
the argument \code{top=TRUE}.  In a similar way the observed and
latent variables can be shown with \code{manifest} and \code{latent}.
All variables of the model are listed with the \code{vars} function

\begin{Schunk}
\begin{Sinput}
> exogenous(mregr)
\end{Sinput}
\begin{Soutput}
[1] "x" "z"
\end{Soutput}
\begin{Sinput}
> endogenous(mregr, top = TRUE)
\end{Sinput}
\begin{Soutput}
[1] "w"
\end{Soutput}
\end{Schunk}
The \code{children} and \code{parents} functions extracts the children
respectively the parents of one or several nodes in the unidirectional
graph of the model, e.g.
\begin{Schunk}
\begin{Sinput}
> children(mregr, ~x + y1)
\end{Sinput}
\begin{Soutput}
[1] "y1" "y2" "y3" "w" 
\end{Soutput}
\begin{Sinput}
> parents(mregr, ~w)
\end{Sinput}
\begin{Soutput}
[1] "y1" "y2" "y3"
\end{Soutput}
\end{Schunk}
The pathways from one variable to another can be viewed with the 
\code{path} function which returns a list of character
vectors indicating the (causal) path
\begin{Schunk}
\begin{Sinput}
> path(mregr, w ~ x)
\end{Sinput}
\begin{Soutput}
[[1]]
[1] "x"  "y1" "w" 

[[2]]
[1] "x"  "y2" "w" 

[[3]]
[1] "x"  "y3" "w" 
\end{Soutput}
\end{Schunk}
The function \code{subset} can be used to extract subsets of a
model. To extract the upper level of the path analysis we call
\begin{Schunk}
\begin{Sinput}
> subset(mregr, ~y1 + y2 + y3 + w)
\end{Sinput}
\end{Schunk}
which keeps all parameter restrictions of the original
model. Conversely, \code{lvm} models can be merged with the \code{merge} method
(or using the operator syntax: \code{m\%++\%m2}). 

To examine the parameters (and in particular their order) one can call the
\code{coef}-function
\begin{Schunk}
\begin{Sinput}
> coef(mregr)
\end{Sinput}
\begin{Soutput}
       m1        m2        p1        p2        p3        p4        p5        p6 
     "y1"       "w"   "y1<-z"   "y2<-z"   "y3<-x"   "w<-y1"   "w<-y2"   "w<-y3" 
       p7        p8        p9       p10       p11 
"y1<->y1" "y2<->y2" "y3<->y3"   "w<->w" "y1<->y3" 
\end{Soutput}
\end{Schunk}
where "\verb|<-|" represents slope parameters (e.g. \code{z} on \code{y1}) and
"\verb|<->|" represents covariance (See also the \code{describecoef}
function). With the argument \code{labels=TRUE} we can get the same
vector but with all parameter labels substituted by their constraints
\begin{Schunk}
\begin{Sinput}
> coef(mregr, labels = TRUE)
\end{Sinput}
\begin{Soutput}
       m1        m2        p1        p2        p3        p4        p5        p6 
     "mu"       "w"       "a"       "b"      "b1"   "w<-y1"   "w<-y2"   "w<-y3" 
       p7        p8        p9       p10       p11 
      "v"      "r0" "y3<->y3"   "w<->w"       "r" 
\end{Soutput}
\end{Schunk}
The non-linear parameter constraints or non-linear regression
specifications can be shown with
\begin{Schunk}
\begin{Sinput}
> constrain(mconstr)
\end{Sinput}
\begin{Soutput}
$v
function (x)  .Primitive("exp")
attr(,"args")
[1] "alpha"

$mu
function (x) 
x[1] + pnorm(x[2] * x[3])
attr(,"args")
[1] "alpha" "beta"  "x"    
\end{Soutput}
\end{Schunk}


\section{Simulation}\label{sec:simulation}

\begin{table}[ht]
  \centering
  \begin{tabular}{lp{8cm}}
    \textbf{Function} & \textbf{Task} \\ \hline
    \code{sim} & Simulation method for \code{lvm}-objects \\
    \code{functional},\code{constrain} & Introduce non-linearities in simulation \\
    \code{distribution} & Change distribution and link of variables \\ 
    \code{heavytail} & Define heavy tailed distribution of a variable\\ 
    \code{normal.lvm} & Normal distribution  \\     
    \code{poisson.lvm} & Poisson distribution \\     
    \code{binomial.lvm} & Binomial distribution \\     
    \code{uniform.lvm} & Uniform distribution\\     
    \code{weibull.lvm} & Weibull accelerated failure times
    \\
    \hline
  \end{tabular}
  \caption{Simulation methods.}\label{tab:simulation}
\end{table}

Simulation is a major component of modern statistics, which allows us
to experimentally
study the properties of a statistical method under various
alternatives and to verify (or reject) preliminary ideas. The
\pkg{lava} package includes the \code{sim} method offering a convenient
tool for performing simulation studies from very general models.

As an initial example we will create a \code{data.frame} with 100 observations
from the structural equation model \code{m1} defined in Section \ref{sec:modelspec}:
\begin{Schunk}
\begin{Sinput}
> mydata <- sim(m1, 100)
\end{Sinput}
\end{Schunk}
The default parameter values are that all intercepts are 0, slope and
residual variance parameters are 1, and covariance parameters are
0.5. To change the simulation parameters one can either fix the
relevant parameters of the model to the desired numerical values as
described in the previous section, or give the parameters as the
argument \code{p} directly to the \code{sim} method.
For instance the following two calls will both simulate 1000
observations from the model \code{mregr} with the default parameter values,
except that the residual variance of $w$ is set to 2, the intercepts
of $Y_1$ and $Y_2$ ($\mu$) are set to 1, and $\beta=1.5$ (the slope of
$Y_1$ on $W$)
\begin{Schunk}
\begin{Sinput}
> d.mregr <- sim(mregr, 1000, p = c(mu = 2, beta = 1.5, `w<->w` = 2))
> d.mregr2 <- sim(mregr, 1000, p = c(y1 = 2, `w<-y1` = 1.5, `w<->w` = 2))
\end{Sinput}
\end{Schunk}
To simulate data with heavier tails than the normal distribution the
\code{heavytail} method can be used. The following defines \code{y1}
and \code{y2} to be realizations from an unstructured
bivariate normal distribution $\mathcal{N}(\mu,\Sigma)$: 
\begin{Schunk}
\begin{Sinput}
> mhtail <- lvm()
> covariance(mhtail) <- y1 ~ y2
\end{Sinput}
\end{Schunk}
We let $Y$ be a stochastic variable with this distribution, then the
following call
\begin{Schunk}
\begin{Sinput}
> heavytail(mhtail, df = 3) <- ~y1 + y2
\end{Sinput}
\end{Schunk}
will allow us to draw simulations of \code{y1} and \code{y2} from the
distribution of $\smash{Y_j(3/Q)^{0.5}}$
where $Q\sim\chi^2_{3}$, i.e., leading to a multivariate $t$-distribution with
covariance matrix $\Sigma$, mean $\mu$, and $\nu=3$
degrees of freedom, described by the density
\begin{align}
  f(\bm{x}\mid \bm{\mu},\bm{\Sigma},\nu) =
  \frac{\Gamma((k+\nu)/2)}{\nu^{k/2}\Gamma(\nu/2)\Gamma(1/2)^k}\frac{\abs{\bm{\Sigma}}^{-1/2}}{\left(1+\frac{1}{\nu}(\bm{x}-\bm{\mu})'\Sigma^{-1}(\bm{x}-\bm{\mu})\right)^{(\nu+k)/2}}
\end{align}
where $k=2$ is the dimension.
The same realization of $Q$ will be used on both \code{y1} and
\code{y2} in the above code. To make simulations where a different
realization of the $\chi^2$-distribution is used for each outcome
(leading to a star-shaped distribution), one simply has to make
separate \code{heavytail} calls as in
\begin{Schunk}
\begin{Sinput}
> heavytail(mhtail, df = 3) <- ~y1
> heavytail(mhtail, df = 3) <- ~y2
\end{Sinput}
\end{Schunk}
The method can be used with different degrees of freedom for
different variables in the model, and thus gives access to an easy way
of simulating models with various degrees of outlier contamination.

To allow simulations from quite general models, two additional replacement
functions are available, \code{functional} and \code{distribution}. 
The \code{functional} replacement function is used for defining
(nonlinear) functional relationships between variables and has the
syntax
\begin{verbatim}
  functional(x,to,from) <- value
\end{verbatim}
where \code{x} is a \code{lvm}-object, \code{from} and \code{to} are
predictor and outcome, respectively, and \code{value} is a real function
describing the functional form. In the model \code{mregr} (Figure
\ref{fig:mregr}) we have 
\begin{align}
  \E(Y_3|X,Z) = b_1 X + Z
\end{align}
With the call 
\begin{Schunk}
\begin{Sinput}
> functional(mregr, y3 ~ x) <- function(x) x^2
\end{Sinput}
\end{Schunk}
we can simulate from the model 
\begin{align}
  \E(Y_3|X,Z) = b_1 X^2 + Z
\end{align}
with the coefficient $b_1$ defined by the earlier identity constraint. To
define a more complex polynomial effect of $X$ we can make a copy of
the predictor with the \code{copy} function and apply
\code{functional} on the copy
\begin{Schunk}
\begin{Sinput}
> copy(mregr) <- x ~ x2
> regression(mregr, y3 ~ x2) <- "b2"
> functional(mregr, y3 ~ x2) <- function(x) x^3
\end{Sinput}
\end{Schunk}
leading to the mean-structure
\begin{align}
  \E(Y_3|X,Z) = b_1 X^2 + b_2 X^3 + Z
\end{align}
An alternative approach is to use the
\code{constrain} function, e.g.
\begin{Schunk}
\begin{Sinput}
> functional(mregr, y3 ~ x) <- NA
> kill(mregr) <- ~x2
> intercept(mregr, ~y3) <- "y3"
> constrain(mregr, y3 ~ b0 + b2 + b3 + x) <- function(x) x[1] + 
+     x[2] * x[4]^2 + x[3] * x[4]^3
\end{Sinput}
\end{Schunk}
would define the model
\begin{align}
  \E(Y_3|X,Z) = b_0 + b_1 X + b_2 X^2 + b_3 X^3 + Z
\end{align}
The major difference between the two methods is that \code{functional} only
has an impact on the \code{sim} method and not on the inferential
methods, whereas \code{constrain} alters the model fitted by
\code{estimate} (see Section \ref{sec:inference}).

The \code{distribution} replacement function is used for defining the
link/dist\-ribu\-tion of variables, with syntax
\begin{verbatim}
  distribution(x,variable) <- value
\end{verbatim}
where \code{x} is a \code{lvm}-object, \code{variable} is the
variable to define the distribution of, and \code{value} is a
function defining the random generator for \code{variable} taking the form
\begin{verbatim}
  function(n, mu, var)
\end{verbatim}
where $n$ defines the number of samples, \code{mu} is the mean, and
\code{var} is the variance as defined by the latent variable model.
Some of the most common distributions have been predefined in the
functions \texttt{uniform.lvm}, \texttt{normal.lvm},
\texttt{binomial.lvm}, \texttt{poisson.lvm}, \texttt{weibull.lvm}.

As an example we will define a simple hierarchical model structure (path analysis)
\begin{Schunk}
\begin{Sinput}
> msim <- lvm(t ~ y + u + z)
> regression(msim) <- y ~ u + x + z
> regression(msim) <- c(z, u) ~ x
\end{Sinput}
\end{Schunk}

\relsize{-3}
\begin{figure}[htbp]\centering

\includegraphics{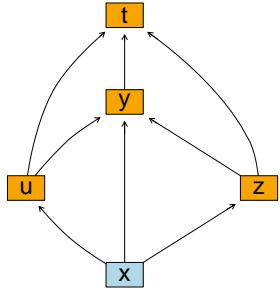}
\caption{\code{plot(msim)}}
\label{fig:msim}
\end{figure}
\origfigsize
To change the distribution of $Y$ to a Bernoulli distribution we simply call
\begin{Schunk}
\begin{Sinput}
> distribution(msim, ~y) <- binomial.lvm()
\end{Sinput}
\end{Schunk}
The default link is logit 
\begin{align}
  \pr(Y=1\mid U,X,Z) = \frac{\exp(\beta_0 + \beta_1X + \beta_2U +
    \beta_3Z)}{1+\exp(\beta_0+\beta_1X + \beta_2U+\beta_3Z)},
\end{align}
but a complementary log-log (cloglog) or
probit link can be chosen via the link argument, e.g.
\begin{Schunk}
\begin{Sinput}
> distribution(msim, ~y) <- binomial.lvm("probit")
\end{Sinput}
\end{Schunk}
Similarly we can define the conditional distribution of $Z$ as Poisson
\begin{align}
  Z \sim pois(\E(Z\mid X)), \\
  \log(\E(Z\mid X)) = \gamma_0 + \gamma_1X,
\end{align}
and the conditional distribution of $U$ as uniform
\begin{align}
  U = \lambda_0+\lambda_1X + \sqrt{\sigma_U^2}U_0, \\
  U_0 \sim unif(-1,1)\sqrt{3},
\end{align}
and let $T$ follow a proportional hazards model with Weibull 
baseline with scale parameter 1.25 and shape parameter 2 (and no
censoring)
\begin{align}
  \lambda(t) = \lambda_0^{Weibull}(1.25;2)\exp(-\alpha_0+\alpha_1U+\alpha_2Y+\alpha_3Z).
\end{align}

\begin{Schunk}
\begin{Sinput}
> distribution(msim, ~z + u + t) <- list(poisson.lvm(), uniform.lvm(), 
+     weibull.lvm(1.25, 2, cens = Inf))
\end{Sinput}
\end{Schunk}
The default simulation parameter values leads to intercepts
$\beta_0=\lambda_0=\gamma_0=\alpha_0=0$ and all the remaining
parameters are 1 (including the residual variance $\sigma_U^2$).

The default distribution of exogenous variables is the standard normal
distribution, $X \sim \mathcal{N}(0,1)$ (and independence between the
exogenous variables).
It is also possible to let a variable be deterministic by simply
assigning a list encapsulating the data:
\begin{verbatim}
  distribution(msim,~x) <- list(seq(0,1,length.out=100))
\end{verbatim}
Obviously an error will occur if this variable definition differs
in length from the number of samples to be drawn in the simulation.

Parameters can be altered in the same manner as endogenous variables
(e.g. \code{sim(msim, p=c(x=3,`x<->x`=2))}), but they can also
be fixed directly via the function \code{distribution}, e.g.
\begin{Schunk}
\begin{Sinput}
> distribution(msim, ~x) <- binomial.lvm(p = 0.4)
\end{Sinput}
\end{Schunk}
defines $X$ to be simulated from a Bernoulli-process
($\pr(X=1)=0.4$), with similar options for the other methods
(e.g. \code{poisson.lvm(lambda=2)}).

For a \code{lvmfit}-object (see Section \ref{sec:inference}) the \code{sim}
method by default simulates observations with the parameter vector \code{p}
set equal to the estimated parameter vector and with the same number
of observations as in the original data set. In this case the defaults
causes the simulations to be drawn from the
conditional distribution given the exogenous variables, hence the
exogenous variables will be fixed at their original values (can be
changed with the argument \code{xfix=FALSE}, in which case the
covariates are simulated with mean and variance parameters set to the
empirical mean and covariance).

\section{Inference}\label{sec:inference}
  \begin{table}[htbp!]
    \centering
    \begin{tabular}{ll}
      \textbf{Function} & \textbf{Task} \\ \hline    
      \code{estimate} & Estimates parameters of \code{lvm}-objects\\
      \code{effects} & Calculates direct and indirect effects\\
      \code{compare} & Likelihood ratio, Wald and Score tests\\
      \code{gof} & Goodness-of-fit measures \\
      \code{logLik} & Extracts individual Likelihood
      values \\      
      \code{information} & Extracts information matrix \\
      \code{score} & Extracts individual contribution to score\\
      \code{bootstrap} & Non-parametric bootstrap \\
      \code{confint} & Calculates confidence limits (Wald and profile)\\
      \code{constraints} & Extracts non-linear parameter constraints \\
      \code{modelsearch} & Model searching via score tests \\
      \code{equivalence} & Finds empirically equivalent models \\
      \hline
    \end{tabular}
    \caption{Inferential tools.}\label{tab:inference}    
  \end{table}
\noindent  Parameter estimation is achieved with the \code{estimate}-function
  which returns an object of class \code{lvmfit}. The syntax of the
  estimation procedure is:
  \begin{center}
\begin{verbatim}
  estimate(x, data, estimator, control=list(), 
           missing, weight, cluster, fix, ...)
\end{verbatim}
  \end{center}
with the following arguments
\begin{description}
\item[\code{x}:] A \code{lvm}-object.
  
\item[\code{data}:] \code{data.frame} with variables with the same
  names as the observed variables of the \code{lvm} object:
  \code{manifest(x)}. The order of the variables is not relevant.
  Alternatively, for models where the empirical mean and variance are
  sufficient statistics (e.g. structural equation models), a list with
  the observed covariance matrix (\code{S}), the observed mean
  \code{mu} and the number of observations \code{n} can be passed as
  arguments, e.g. \\
  \code{data=list(S=cov(mydata),mu=colMeans(mydata),n=nrow(mydata))}.
  
\item[\code{estimator}:] Choice of estimator. Default is
  \code{gaussian} which is the MLE of the model defined by
  (\ref{eq:measurement}) and (\ref{eq:structural}). See also
  Section \ref{sec:beyondmle}.
  
\item[\code{control}:] A list of parameters controlling the estimation
  and optimization procedures. See below for more details.

\item[\code{missing}:] Logical variable. If \code{FALSE} (default) a
  complete case analysis is performed and else a full information
  likelihood analysis is conducted under a MAR assumption. Note that
  observations with missing covariates are excluded. 

\item[\code{weight}:] Optional weight matrix to be used by the estimation
  procedure defined by \code{estimator} (for multigroup analyses this
  should be a list of weights).  

\item[\code{cluster}:] Optional cluster variable identifying 
  correlated groups of observations in the data set (for multigroup analyses this
  should be a list of cluster variables) to be used in the
  calculation of robust sandwich variance estimates.

\item[\code{fix}:] Logical variable (defaults to \code{TRUE}).  Care
  has to be taken when specifying a SEM in order to obtain an
  identifiable model. As a rule of thumb one regression coefficient in
  each measurement model should be fixed, e.g. to 1, and the intercept
  of one the indicator set to 0. By default the model is altered to
  fulfill these requirements unless \code{fix=FALSE}.  Other
  parameterizations can be selected by setting the option \code{param}
  via the \code{lava.options} function. The above mentioned
  parameterization is obtained with
  \code{lava.options(param="relative")} and loading parameters and
  intercepts are then interpreted as differences compared to the
  reference indicator. Setting \code{param="absolute"} will result in
  a parameterization where the variance of the latent variables are
  fixed to 1 and the intercept to 0 if not already fixed at some other
  values. \code{lava.options(param="hybrid")} alters the model such
  that the intercept of latent variables are set to 0, and one factor loading
  in each measurement model is set to 1. 
  
  Calling
  \code{lava.option(param="none")} has the same effect as
  \code{fix=FALSE} (in which case the user has to manually define parameter
  constraints that guarantee model identification).
\end{description}
The optional \code{control} argument must be a list of parameters for
the optimization procedure. The element \code{method} should be a string
pointing to a generic optimizer conforming to the syntax of
\code{stats::nlminb} (the default optimizer), i.e. accepting the
objective function (e.g. the log likelihood), the gradient, the
Hessian and control parameters. Setting \code{method="nlminb0"} will
only use the objective function during optimization, where
\code{method=} \code{"nlminb1"} also uses the gradient, and \code{method=}\code{"nlminb2"}
the Hessian as well. Defining \code{method="NR"} will use an alternative
Newton-Raphson algorithm. Variance parameters are as default modeled
using a $\log$-link. This can be disabled by setting \code{constrain=FALSE}.
Additional options like the number of
iterations (\code{iter.max}), turning trace information on
(\code{trace=1}), starter function (\code{starterfun} (a string
pointing to a function generating starting values for the
optimization), convergence tolerance (\code{tol}), reduction of
step-size of the "\code{NR}" optimizer (\code{gamma}), 
etc. can also be defined (see the \proglang{R} help page of
\code{nlminb} and \code{estimate}).  

The control parameters can be set globally via the function
\code{lava.options}. For instance to turn on the trace information of
the optimizer as default in the current session, we would submit
\code{lava.options(trace=1)}.

\vspace*{0.5cm}

We demonstrate the procedure in the
ongoing example (see Figure \ref{fig:m1corfinal}) with the data obtained from the simulation in Section
\ref{sec:simulation}, using the hybrid parameterization:
\begin{Schunk}
\begin{Sinput}
> lava.options(param = "hybrid")
> e <- estimate(m1, mydata)
\end{Sinput}
\end{Schunk}
\begin{Schunk}
\begin{Sinput}
> summary(e)
\end{Sinput}
\begin{Soutput}
Latent variables: u1 u2 
Number of rows in data=100
--------------------------------------------------
                    Estimate Std. Error  Z value  Pr(>|z|)   std.xy
Measurements:                                                      
   y1<-u1            1.00000                                0.85423
   y2<-u1            0.85609    0.07774 11.01174    <1e-12  0.85902
   y3<-u1            0.88990    0.08488 10.48415    <1e-12  0.83429
    z1<-u2           1.00000                                0.93812
    z2<-u2           1.09038    0.05153 21.16099    <1e-12  0.96411
    z3<-u2           1.02444    0.04998 20.49764    <1e-12  0.95745
Regressions:                                                       
   u1<-x1            1.11118    0.13232  8.39775    <1e-12  0.61181
   u1<-x2            0.96907    0.13392  7.23601    <1e-12  0.51159
    u2<-u1           0.94651    0.17037  5.55569 2.765e-08  0.56706
    u2<-x1           0.72832    0.21753  3.34815 0.0008135  0.24025
    u2<-x2           1.05500    0.20428  5.16434 2.413e-07  0.33367
Intercepts:                                                        
   u1                0.00000                                0.00000
   u2                0.00000                                0.00000
   y1                0.06805    0.15264  0.44582    0.6557  0.03110
   y2                0.07461    0.12917  0.57759    0.5635  0.04005
   y3               -0.08326    0.14244 -0.58453    0.5589 -0.04176
   z1                0.07929    0.17709  0.44776    0.6543  0.02384
   z2                0.07514    0.17400  0.43186    0.6658  0.02130
   z3                0.07933    0.16811  0.47185     0.637  0.02376
Residual Variances:                                                
   u1                1.02921    0.23949  4.29753            0.29462
   u2                0.87832    0.23741  3.69960            0.09024
   y1                1.29397    0.23560  5.49234            0.27029
   y2                0.90931    0.16777  5.41987            0.26209
   y3                1.20812    0.21027  5.74543            0.30396
   z1                1.32627    0.23349  5.68022            0.11993
   z2                0.87762    0.19360  4.53323            0.07050
   z3                0.92797    0.18768  4.94448            0.08328
--------------------------------------------------
Estimator: gaussian 
--------------------------------------------------
Number of observations = 100 
 Log-Likelihood = -1010.407 
 BIC = 2167.944 
 AIC = 2066.814 
 log-Likelihood of model = -1010.407 
 log-Likelihood of saturated model = -1003.438 
 Chi-squared statistic: q = 13.93726 , df = 16 , P(Q>q) = 0.6033881 
--------------------------------------------------
\end{Soutput}
\end{Schunk}
The parameter estimates in the output are divided into slope
parameters belonging to the measurements part of the model
(Measurements) (i.e. factor loadings), the structural part
(Regressions) and the intercepts and residual (co)variances of the
model. The \code{summary} method outputs all parameters of the model
including the parameters that were fixed in contrast to the
\code{print} method that only outputs the canonical parameters of the
model.  Notice that a regression parameter in each of the measurement
models has been fixed to one in order to identify the model, and the
slope parameter of $u_2$ on $u_1$ is therefore interpreted on the
scale of $Y_{11}$ and $Y_{21}$. The standardized coefficients in the
last column are interpreted as the change in standard deviation of the
outcome when increasing the predictor one standard deviation. These
parameters can be used to compare effects of predictor variables
measured on different scales. If non-linear constraints were defined
then the relevant estimates and approximate standard errors will be
shown in the last part of the summary output. These effects can also
be extracted with the \code{constraints} function.

The p-values for the variance parameters are deliberately omitted from
the output as the asymptotic distribution under the null is
non-standard (derived in special-cases such as the random intercept
model as an equal mixture of a $\chi^2_1$-distribution and the Dirac
measure in $0$. Hence the naive p-values based on a
$\chi^2_1$-distribution will tend to be conservative).  Different
versions of the information matrix can be used, via the argument
\code{type}$\in\{$\code{"E"},\code{"hessian"},\code{"outer"},\code{"robust"}$\}$
(expected information (default), minus the second derivative of
the log-likelihood, outer product of the score, and the
robust/sandwich variant \citep{white82mlemis}), to calculate the
asymptotic standard errors of the parameters (see also the
\code{information} function).

The last part of the output includes some fit criteria (Akaike and
Bayesian) and the omnibus goodness-of-fit $\chi^2$-test which is a
likelihood ratio test of the current model against the saturated model
structure.  In the conditional model formulation, the least
restrictive model allows covariance between residuals of all
endogenous variables and the mean-vector $\bm{\mu}$ to be a general
linear combination of all the covariates. In the unconditional model
formulation (\ref{eq:ffull}), this corresponds to a completely free
mean structure and a covariance $\bm{\Sigma}$ that can be any
symmetric positive definite matrix. Clearly the maximum likelihood is
attained at the sample mean and non-central empirical covariance
matrix, $\bm{S}$. Hence the log-likelihood of the saturated linear
model is
\begin{align}\label{eq:lll}
   -\frac{n}{2}\left(k\log(2\pi) + \log(\abs{\bm{S}}) +
    \frac{n-1}{n}(p+q)\right).
\end{align}
This part of the output can also be obtained with the call
\begin{Schunk}
\begin{Sinput}
> gof(e, chisq = TRUE)
\end{Sinput}
\begin{Soutput}
Number of observations = 100 
 Log-Likelihood = -1010.407 
 BIC = 2167.944 
 AIC = 2066.814 
 log-Likelihood of model = -1010.407 
 log-Likelihood of saturated model = -1003.438 
 Chi-squared statistic: q = 13.93726 , df = 16 , P(Q>q) = 0.6033881 
 RMSEA (90% CI): 0 (0;0.08127)
rank(Information) = 23 (p=23)
condition(Information) = 0.007602741
||score||^2 = 1.306582e-08 
\end{Soutput}
\end{Schunk}
Here the omnibus test gives a p-value of 0.60, thus indicating a
reasonable agreement between the model implied and empirical
covariance structure.

Byproducts of the maximum likelihood estimation such as the score,
information and log-likelihood can be obtained with the functions
\code{score}, \code{information} (see also \code{vcov}), and
\code{logLik}. The individual contribution to score and log-likelihood
are calculated with the argument \code{indiv=TRUE}. The methods also
allow altering the parameter, data, weights and type of estimator
(arguments \code{p}, \code{data}, \code{weight} and
\code{estimator}). Predictions can be obtained via the \code{predict}
method (and conditional residuals via \code{residuals}), where the
latent variables are predicted by the conditional expectation given all
manifest variables (the prediction can be based on conditioning on a
subset of the manifest variables defined by the second argument of
\code{predict}). For instance to estimate $\E(U_1\mid
Y_{1},Y_{2},Y_{3},X_1,X_2)$ we can call
\begin{Schunk}
\begin{Sinput}
> u1hat <- predict(e, ~y1 + y2 + y3)[, "u1"]
\end{Sinput}
\end{Schunk}
Prediction of the residual terms can be obtained by
setting the argument \code{residual=TRUE}.
\begin{Schunk}
\begin{Sinput}
> zeta1hat <- predict(e, ~y1 + y2 + y3, residual = TRUE)[, "u1"]
\end{Sinput}
\end{Schunk}

Assessment of linearity and distributional assumptions can be based on
examination of different residuals and their association with for
example covariates in the model.  We refer to the \pkg{gof} package
\citep{holst09:gof} for residual based goodness-of-fit methods for
structural equation models fitted with \code{lava}.

\subsection{Direct and indirect effects}

One of the strengths of the structural equation model framework is the
possibility of decomposing the effects of a predictor into direct and
indirect effects. In the model \code{m1} (Figure \ref{fig:m1corfinal})
we have the following relations
\begin{align}
  Z_{3} &= \lambda_{23}U_2 + \epsilon_{23}, \label{eq:line1}\\
  U_2 &= \beta_{1} X_1 + \beta_{2}X_2 + \gamma U_1 + \xi_{2}, \label{eq:line2}\\
  U_1 &= \delta_1 X_1 + \delta_2 X_2 + \xi_{1}, \label{eq:line3}
\end{align}
and we wish to quantify the effect of $X_1$ on $Z_{3}$. The direct
effect is zero as $X_1$ is not present in (\ref{eq:line1}). By
substituting (\ref{eq:line2}) and (\ref{eq:line3}) into
(\ref{eq:line1}), it follows that
\begin{align}    
  Z_{3} &= \beta_1\lambda_{23} X_1 + \beta_2\lambda_{23}X_2 + 
  \beta_2\lambda_{23}X_2 + \delta_1\gamma\lambda_{23} X_1 \\
  &\qquad +  \delta_2\gamma\lambda_{23} X_2 + \gamma\lambda_{23}\xi_1 +
  \lambda_{23}\xi_2 + \epsilon_{23}.
\end{align}
Hence the total effect of $X_1$ is the sum of the two
indirect effects $\beta_1\lambda_{23}$ and
$\delta_1\gamma\lambda_{23}$. The estimation uncertainty of this
effect can be approximated by
the delta method. In general the distribution of a product of
estimators can be approximated in the following way. Let
\begin{align}
  f(\widehat{\bm{\beta}}) = f(\widehat{\beta}_1,\ldots,\widehat{\beta}_k) = \prod_{i=1}^k\widehat{\beta}_i,
\end{align}
where the estimated parameters $\widehat{\bm{\beta}}$ are
asymptotically normally distributed with covariance matrix
$\bm{\Sigma}_{\bm{\beta}}$. Now
\begin{align}
  \nabla f(\bm{\beta}) =
  \begin{pmatrix}
    \prod_{i\neq 1}\beta_i \\
    \vdots \\
    \prod_{i\neq k}\beta_i
  \end{pmatrix}
\end{align}
and
\begin{align}
  \sqrt{n}\Big(f(\widehat{\bm{\beta}})-f(\bm{\beta}_0)\Big)
  \overset{\mathcal{D}}{\longrightarrow} \mathcal{N}\Big(0, \, \nabla
  f(\widehat{\bm{\beta}})'\bm{\Sigma}_{\widehat{\bm{\beta}}}\nabla
  f(\widehat{\bm{\beta}})\Big)
\end{align}
The approximate distribution of linear combinations of products is obtained by
straightforward calculations (i.e. $\nabla f_1 + \nabla f_2$).

The \code{effects} function can be used to estimate the (direct and indirect) effect of one variable on another together with approximate
standard errors, e.g.
\begin{Schunk}
\begin{Sinput}
> (f <- effects(e, z3 ~ x1))
\end{Sinput}
\end{Schunk}
\begin{Schunk}
\begin{Soutput}
Total effect of 'x1' on 'z3':
		1.823569 (Approx. Std.Err = 0.1524713)
Direct effect of 'x1' on 'z3':
		0 (Approx. Std.Err = NA)
Indirect effects:
	Effect of 'x1' via x1->u1->u2->z3:
		1.077448 (Approx. Std.Err = 0.2170743)
	Effect of 'x1' via x1->u2->z3:
		0.7461208 (Approx. Std.Err = 0.222262)
\end{Soutput}
\end{Schunk}
\begin{Schunk}
\begin{Sinput}
> coef(f)
\end{Sinput}
\begin{Soutput}
                Estimate   Std.Err   z value     Pr(>|z|)
Total          1.8235689 0.1524713 11.960083 0.000000e+00
Direct         0.0000000        NA        NA           NA
z3<-u2<-u1<-x1 1.0774481 0.2170743  4.963498 6.923458e-07
z3<-u2<-x1     0.7461208 0.2222620  3.356943 7.880941e-04
\end{Soutput}
\end{Schunk}

\subsection{Hypothesis testing}

Next we estimate the parameters of two nested models where we have
restricted all factor loadings to 1 and in
addition removed the effect from $U_1$ to
$U_2$
\begin{Schunk}
\begin{Sinput}
> m1a <- m1
> regression(m1a, c(z1, z2, z3) ~ u2) <- 1
> regression(m1a, c(y1, y2, y3) ~ u1) <- 1
> m1b <- m1a
> cancel(m1b) <- u2 ~ u1
> ea <- estimate(m1a, mydata)
> eb <- estimate(m1b, mydata)
\end{Sinput}
\end{Schunk}

\subsubsection{Likelihood Ratio Test}

For nested models, $\mathcal{M}_1\subset\mathcal{M}_2$, the natural test is the
\emph{likelihood ratio test} (LRT) 
\begin{align}
  -2\left[\log L_1(\widehat{\bm{\theta}}_1)-\log L_2(\widehat{\bm{\theta}}_2)\right] \overset{\text{approx.}}{\sim} \chi^2_{\Delta df}.
\end{align}
For non-nested models, one choice is the Akaike's Information Criterion
(AIC) favoring models with low values of
\begin{align}
  AIC = -2\log(L) + 2 n_{\text{par}},
\end{align}
where $n_{\text{par}}$ is the number of parameters in the model
(implemented in the \code{AIC} function), or the Bayesian Information
Criterion
\begin{align}
  BIC = -2\log(L) + n_{\text{par}}\log(N),
\end{align}
where $N$ denotes the \emph{total} number of observations
\citep{raftery93:_bayes_model_selec_struc_equat_model}, i.e. the
number of endogenous variables times the number of individuals.

Successive LRT between nested models can be calculated with
\begin{Schunk}
\begin{Sinput}
> (LRT1 <- compare(e, ea, eb))
\end{Sinput}
\end{Schunk}
\begin{Schunk}
\begin{Soutput}
[[1]]

	Likelihood ratio test

data:  
chisq = 6.6542, df = 4, p-value = 0.1553
sample estimates:
log likelihood (model 1) log likelihood (model 2) 
               -1010.407                -1013.734 


[[2]]

	Likelihood ratio test

data:  
chisq = 33.9868, df = 1, p-value = 5.549e-09
sample estimates:
log likelihood (model 1) log likelihood (model 2) 
               -1013.734                -1030.728 
\end{Soutput}
\end{Schunk}
Hence we accept the hypothesis that all factor loadings are equal (the
model \code{m1a}) but reject the hypothesis that the two latent
variables $U_1$ and $U_2$ are conditional independent given the covariates.

\subsubsection{Wald Test}
The \code{compare} method can also deal with hypothesis testing via
Wald or Score tests. The hypothesis
\begin{align}
  H_0\colon \bm{\beta} = \bm{\beta}_0 
\end{align}
for a subset, $\bm{\beta}$, of all the parameters,
can be tested with a Wald test using the \code{par} and \code{null}
arguments (the latter defaults to 0), for instance to test if all
loading parameters are 1 (equivalent to the LRT of \code{m1} against
\code{m1a}), we can write
\begin{Schunk}
\begin{Sinput}
> (W1 <- compare(e, par = c("y2<-u1", "y3<-u1", "z2<-u2", "z3<-u2"), 
+     null = rep(1, 4)))
\end{Sinput}
\end{Schunk}
\begin{Schunk}
\begin{Soutput}
	Wald test

data:  
chisq = 7.2687, df = 4, p-value = 0.1223
\end{Soutput}
\end{Schunk}
For a general estimable contrast $\bm{C}$, we can also test the hypothesis
\begin{align}
  H_0\colon \bm{C}\bm{\beta} = \widetilde{\bm{\beta}}_0,
\end{align}
where $\bm{C}$ is matrix (or vector) with the same number of columns
as the number of parameters, or alternatively a sub-matrix with column
names given by parameter names (implicitly assuming that omitted
columns are zero), leading to the test statistic
\begin{align}
  (\bm{C}\widehat{\bm{\beta}}-\widetilde{\bm{\beta}}_0)'(\bm{C}\bm{\Sigma}_{\widehat{\bm{\beta}}}\bm{C}')^{-1}
  (\bm{C}\widehat{\bm{\beta}}-\widetilde{\bm{\beta}}_0) \sim \chi^2_{\rank(\bm{C)}}
\end{align}
The covariance matrix $\bm{\Sigma}_{\widehat{\bm{\beta}}}$ will by
default be the variance matrix as defined by the chosen estimator (for
the linear gaussian models, \texttt{estimator="gaussian"}, this is the
inverse of the expected information), but it can
optionally be given as the argument \texttt{Sigma}, e.g. to use a robust
variance estimate (see \texttt{information} method).

To test whether all the intercepts of the outcomes sum to zero, we can
write
\begin{Schunk}
\begin{Sinput}
> C <- rep(1, 6)
> names(C) <- endogenous(m1)
> (W2 <- compare(e, contrast = C))
\end{Sinput}
\end{Schunk}
\begin{Schunk}
\begin{Soutput}
	Wald test

data:  
chisq = 0.1769, df = 1, p-value = 0.6741
\end{Soutput}
\end{Schunk}
hence we accept the hypothesis of equal intercepts.

\subsubsection{Score Test}
With the \code{scoretest} argument we can conduct Score tests. 
Letting $\widetilde{\bm{\theta}}$ be the parameter belonging to the
less restrictive model $\mathcal{M}_2$, which is equal to
$\widehat{\bm{\theta}}_1$, the MLE of the restrictive model
$\mathcal{M}_1$, for all the parameters shared by
$\mathcal{M}_2$ and zero elsewhere. The test statistic is
then given by
\begin{align}
  S =
  \mathcal{S}_2(\widetilde{\bm{\theta}})'\mathcal{I}_2^{-1}(\widetilde{\bm{\theta}})\mathcal{S}_2(\widetilde{\bm{\theta}})  
\end{align}
with approximate $\smash{\chi^2_{\Delta df}}$-distribution under the null,
where $\mathcal{S}_2$ and $\mathcal{I}_2$ are the score and
information matrix of model $\mathcal{M}_2$.

We will test whether adding correlation between the residuals terms of
$Z_{3}$ and $Z_{2}$ significantly improves the model fit:
\begin{Schunk}
\begin{Sinput}
> (S1 <- compare(e, scoretest = z3 ~ z2))
\end{Sinput}
\end{Schunk}
\begin{Schunk}
\begin{Soutput}
	Score test

data:  z3 ~ z2 
chisq = 0.1186, df = 1, p-value = 0.7306
\end{Soutput}
\end{Schunk}
which does not indicate evidence against the conditional
independence assumption.

Similarly we can test the statistical significance of simultaneously
adding two extra correlation parameters:
\begin{Schunk}
\begin{Sinput}
> (S2 <- compare(e, scoretest = c(z3 ~ z2, z1 ~ z2)))
\end{Sinput}
\end{Schunk}
\begin{Schunk}
\begin{Soutput}
	Score test

data:  z3 ~ z2 z1 ~ z2 
chisq = 2.03, df = 2, p-value = 0.3624
\end{Soutput}
\end{Schunk}

\subsubsection{Model searching with the Score test}
An advantage of the Score test over the LRT is that the MLE is only
needed in the more restrictive model making it an ideal instrument for
model searching, in order to check that important aspects of the
covariance structure has not been neglected in the model
specification. The \code{modelsearch} function can be used to examine
all one-parameter extensions of the model. The following call give the
5 most significant one-parameter extensions
\begin{Schunk}
\begin{Sinput}
> print(ms <- modelsearch(e), tail = 5)
\end{Sinput}
\end{Schunk}
\begin{Schunk}
\begin{Soutput}
 Score: S P(S>s)  Index   holm BH    
 2.024    0.1549  u2<->y1 1    0.9031
 2.124    0.145   y1<->z1 1    0.9031
 2.283    0.1308  y1<->y2 1    0.9031
 2.892    0.08904 y2<->z2 1    0.9031
 4.207    0.04025 y2<->z1 1    0.9031
\end{Soutput}
\end{Schunk}
As expected we do not see any significant improvements of the model
among the 5 most significant Score tests (with the first and second
column being the test statistic and corresponding p-value, and the
last two columns being the p-values adjusted by the Bonferroni-Holm
procedure to control the overall Type I error \citep{holm79}, and
q-values of the Benjamini-Hochberg procedure controlling the FDR).
Similarly, the most important $k$-parameter extensions to the model,
can be examined with the argument \code{k}, but the number of models
to search through will increase dramatically with $k$.

\subsection{Model equivalence}
A challenge in multivariate modeling is the problem of equivalent
models, where two different parameterizations leads to identical model
fit (likelihood) for all data sets. Hence without strong a priori
knowledge of the model structure, e.g. based on other scientific
evidence, the interpretation of model parameters must be made
cautiously.  Formal proofs of model equivalence can be difficult
\citep{MR996025}. To identify candidates of equivalent models we
suggest using the Score test. The idea is to study all one-parameter
extensions of a given model using the score test. Two models are said
to be empirically equivalent  if the score tests agree. 
This can be achieved with the \code{equivalence} function. Two
variables of the model are chosen, which not necessarily are defined
as being directly related in the model structure. The score function
for the model including covariance between the residuals of the two
selected variables is then compared with score functions of models
omitting this association, but with the same number of parameters.

\begin{figure}[htbp]
\centering
\includegraphics{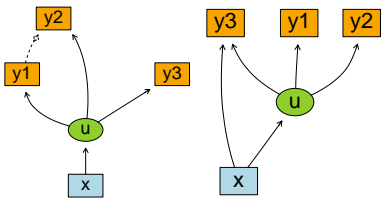}
\caption{Example of two equivalent models (\code{mEq}).}
\label{fig:mEq}
\end{figure}

As a simple example we will investigate the structural equation model
in the path diagram of the left panel of Figure \ref{fig:mEq}
\begin{Schunk}
\begin{Sinput}
> mEq <- lvm(list(c(y1, y2, y3) ~ u, u ~ x))
> latent(mEq) <- ~u
> covariance(mEq) <- y1 ~ y2
> dEq <- sim(mEq, 100)
> est.mEq <- estimate(mEq, dEq)
\end{Sinput}
\end{Schunk}

Below we are examining whether the inclusion of a residual correlation
between $Y_1$ and $Y_2$ has any equivalent formulations
\begin{Schunk}
\begin{Sinput}
> (Eq <- equivalence(est.mEq, y1 ~ y2))
\end{Sinput}
\end{Schunk}
In fact, an equivalent model is defined by instead adding a direct
effect of $X$ on $Y_3$ (see
Figure \ref{fig:mEq}).
\begin{Schunk}
\begin{Soutput}
  0)	 y1<->y2  (10.31)
Empirical equivalent models:
  1)	 y3<->x  (10.31)
Candidates for model improvement:
	 none
\end{Soutput}
\end{Schunk}
\figsize{4}{2}
\origfigsize

\subsection{Confidence limits}
Wald confidence limits can be created using the method
\code{confint}. However, for some parameters better coverage can be
achieved with alternative methods.
One method is the non-parametric bootstrap which can be calculated
with the function \code{bootstrap}. The bootstrap is a computational
intensive method, and parallel computation can be done by registering
a \code{foreach} \citep{foreach} parallel adaptor. In this example we
will compute the bootstrap in parallel using the \pkg{parallel} and
\pkg{doParallel} packages distributing the bootstrap computations across the
available CPU cores
\begin{Schunk}
\begin{Sinput}
> library(doParallel)
> registerDoParallel()
> (B <- bootstrap(e, 500))
\end{Sinput}
\end{Schunk}
\begin{Schunk}
\begin{Soutput}
Non-parametric bootstrap statistics (R=500):

           Estimate          Bias    Std.Err      2.5 %    97.5 %
y1       0.08467677  1.662669e-02 0.15922250 -0.2417240 0.3787349
y2       0.07691911  2.312410e-03 0.12146566 -0.1457868 0.3101638
y3      -0.07196788  1.129167e-02 0.13941055 -0.3536971 0.2112750
z1       0.08407469  4.780799e-03 0.18185690 -0.2799668 0.4421753
z2       0.08527725  1.013330e-02 0.17283489 -0.2966441 0.3995207
z3       0.09493044  1.560519e-02 0.17017286 -0.2081544 0.4407723
u1<-x1   1.11860351  7.421090e-03 0.15172036  0.8141602 1.4084714
u1<-x2   0.96908130  1.303164e-05 0.15096572  0.6741721 1.2717508
u2<-u1   0.96947448  2.296893e-02 0.16712619  0.6754198 1.3478912
u2<-x1   0.69630433 -3.201448e-02 0.23181926  0.1494670 1.0941781
u2<-x2   1.02390307 -3.109418e-02 0.20826885  0.5773626 1.4170756
y2<-u1   0.85971870  3.632407e-03 0.10370758  0.6779335 1.0642511
y3<-u1   0.89440961  4.508961e-03 0.08414864  0.7388379 1.0566335
z2<-u2   1.09661468  6.232894e-03 0.05341720  0.9971030 1.2131320
z3<-u2   1.02856111  4.118107e-03 0.04733460  0.9419200 1.1232677
u1<->u1  0.99249288 -3.671648e-02 0.27738180  0.5390826 1.5590308
u2<->u2  0.81896688 -5.935355e-02 0.26481969  0.3523745 1.3589482
y1<->y1  1.28947917 -4.494201e-03 0.26570998  0.8302451 1.8369823
y2<->y2  0.89753510 -1.177998e-02 0.17311848  0.5861767 1.2511752
y3<->y3  1.18530385 -2.281194e-02 0.18883960  0.8092845 1.5525326
z1<->z1  1.30673170 -1.953404e-02 0.21030950  0.9245536 1.7315556
z2<->z2  0.84766364 -2.995799e-02 0.17431201  0.4971759 1.1894273
z3<->z3  0.92206325 -5.906032e-03 0.18632891  0.5572348 1.3095068
\end{Soutput}
\end{Schunk}
To bootstrap other statistics a user-defined function can be
supplied as the argument \code{fun}. A parametric bootstrap can be
computed setting the argument \code{parametric=TRUE} and setting the
argument \code{p} to parameter values of the null model from which to
simulate from. The parallel computation functionality can be disabled
via the call \texttt{lava.options(parallel=FALSE)}.

As an alternative to the resample-based approach, we can also
calculate the confidence limits based on the profile likelihood:
\begin{Schunk}
\begin{Sinput}
> (ci <- confint(e, profile = TRUE, parm = "u2<-u1", level = 0.95))
\end{Sinput}
\end{Schunk}
\begin{Schunk}
\begin{Soutput}
           2.5 %   97.5 %
u2<-u1 0.6392854 1.332559
\end{Soutput}
\end{Schunk}
where the parameter of interest can be given as the index or label
name.

A third option is to use a variance stabilizing transform of the
parameter. As an example we will calculate the confidence limits of
the \emph{partial
  correlation} (or conditional correlation) between two outcomes $Y_1$
and $Y_2$ given covariates $\bm{X}$. Hence we assume that
\begin{align}
  Y_i = \bm{\beta}_i' \bm{X} + \varepsilon_i, \quad i=1,2
\end{align}
and aim to estimate the correlation, $\rho$, between $\varepsilon_1$ and
$\varepsilon_2$. 

We define this model with a single covariate and simulate some
observations and find the corresponding MLE:
\begin{Schunk}
\begin{Sinput}
> m <- lvm(c(y1, y2) ~ x)
> covariance(m, y1 ~ y2) <- "C"
> covariance(m, ~y1 + y2) <- list("v1", "v2")
> d <- sim(m, 100)
> e.pcor <- estimate(m, d)
\end{Sinput}
\end{Schunk}
Note with the default parameter values the correlation between
$\epsilon_1$ and $\epsilon_2$ is 0.5.
Next we define the correlation parameter using a non-linear parameter
constraint and obtain the estimate with confidence limits based on the
delta method 
\begin{Schunk}
\begin{Sinput}
> constrain(e.pcor, rho ~ C + v1 + v2) <- function(x) x[1]/(x[2] * 
+     x[3])^0.5
> constraints(e.pcor)
\end{Sinput}
\begin{Soutput}
     Estimate Std. Error  Z value     Pr(>|z|)      2.5%     97.5%
rho 0.4449733 0.08019988 5.548303 2.884553e-08 0.2877844 0.6021621
\end{Soutput}
\end{Schunk}
Near the boundary of the parameter space these limits will tend to
perform poorly and a better approach is to apply the variance
stabilizing $\arctanh$ transform (Fishers z-transform):
\begin{align}
  \mathcal{Z}\colon \rho {\mapsto} \frac{1}{2}\log\left(\frac{1+\rho}{1-\rho}\right)
\end{align}
Here we also supply the analytical gradient (optional) calculated with the
chain-rule and in addition we set the attribute \code{inv} defining
the inverse transformation, thus giving us the confidence limits on
original correlation scale:
\begin{Schunk}
\begin{Sinput}
> constrain(e.pcor, z ~ C + v1 + v2) <- function(x) {
+     f <- function(p) p[1]/sqrt(p[2] * p[3])
+     res <- atanh(f(x))
+     df <- function(p) c(1/sqrt(p[2] * p[3]), -f(p)/(2 * p[2]), 
+         -f(p)/(2 * p[3]))
+     datanh <- function(r) 1/(1 - r^2)
+     attributes(res)$grad <- function(p) datanh(f(p)) * df(p)
+     attributes(res)$inv <- tanh
+     return(res)
+ }
> constraints(e.pcor)
\end{Sinput}
\begin{Soutput}
        Estimate Std. Error  Z value     Pr(>|z|)      2.5%     97.5%
rho    0.4449733 0.08019988 5.548303 2.884553e-08 0.2877844 0.6021621
z      0.4784149 0.10000000 4.784149 1.717133e-06 0.2824185 0.6744113
inv(z) 0.4449733         NA       NA           NA 0.2751420 0.5878741
\end{Soutput}
\end{Schunk}
In fact $\sqrt{n} \mathcal{Z}(\widehat{\rho}_n)
\overset{\mathcal{D}}{\rightarrow} \mathcal{N}(0,1)$ as $n\to\infty$
\citep{MR2135927}. Note that the \code{correlation} method calculates
the correlation coefficients of a \code{lvmfit} object in a more
elegant way and with a slightly more precise variance estimate
\citep{hotelling;correlation} (see also the \code{partialcor}
function).

\section{Multigroup models}\label{sec:multigroup}
Multigroup analysis (\ref{eq:multigroup}) can be used to combine
different models linked via some shared parameters. Among other things
this extension can be useful in testing general hypotheses of linear
interactions, and the \code{lava} package supports this
generalization via the \code{estimate}-function taking a list of
\code{lvm}-objects and a list of \code{data.frame}'s as arguments and
returning an object of class \code{multigroupfit}:
\begin{Schunk}
\begin{Sinput}
> estimate(list(m1, m2, m3, ...), list(d1, d2, d3, ...), ...)
\end{Sinput}
\end{Schunk}
The list of \code{lvm} objects can optionally be named, as in the
example below, to enhance the output. Parameters that are shared across
the models \code{m1},\code{m2},\code{m3},$\ldots$ will be also be
shared in the multigroup analysis, whereas all other parameters will be
estimated independently between the groups.  In many applications the
first argument will therefore be repetitions of the same
\code{lvm}-object.
Note, that when the different datasets are defined from a single
\code{data.frame} using a grouping variable, the function \code{split}
can be applied to define the second argument. A typical multigroup
analysis call will therefore resemble 
\begin{Schunk}
\begin{Sinput}
> estimate(rep(m, n), split(d, d$x))
\end{Sinput}
\end{Schunk}
where the data-frame \code{d} here is split into a list defined from
the variable \code{d\$x} (with, in this case, \code{n} distinct values).

As an example we will create two nearly identical lvm-objects
describing simple factor models (see Figure \ref{fig:mclustmarg}):
\begin{Schunk}
\begin{Sinput}
> mg1 <- lvm()
> regression(mg1, Y1 ~ H) <- 1
> intercept(mg1, ~Y1) <- 0
> regression(mg1) <- c(Y2, Y3) ~ H
> regression(mg1) <- H ~ E
> latent(mg1) <- ~H
> mg1 <- baptize(mg1)
> covariance(mg1, endogenous(mg1)) <- NA
> mg2 <- mg1
> intercept(mg2, ~Y2 + Y3) <- 0
\end{Sinput}
\end{Schunk}

The \code{baptize} function labels all free parameters of the model,
giving the parameter the names as defined by the \code{coef} function
(\code{"Y1<-H"}, \code{"H<->H"} etc.). An optional argument
\code{labels} can be given to define custom labels. 

In the above example the restrictions of the variances of the
residuals of the endogenous variables are removed, and 
hence the two models \code{mg1} and \code{mg2} share all parameters
except for these variance parameters, and the intercepts which are
identical in \code{mg2}.

Next we simulate two datasets from model 1 (thus in fact only a single
group):
\begin{Schunk}
\begin{Sinput}
> data1 <- sim(mg1, 200)[, manifest(mg1)]
> data2 <- sim(mg1, 200)[, manifest(mg1)]
\end{Sinput}
\end{Schunk}



To estimate parameters using MLE we simply type
\begin{Schunk}
\begin{Sinput}
> (e.mg <- estimate(list(`Arm 1` = mg1, `Arm 2` = mg2), list(data1, 
+     data2)))
\end{Sinput}
\end{Schunk}
\begin{Schunk}
\begin{Soutput}
--------------------------------------------------
Group 1: Arm 1 (n=200)
                    Estimate Std. Error  Z value Pr(>|z|)
Measurements:                                            
   Y2<-H             0.99827    0.06189 16.13044   <1e-12
   Y3<-H             1.04265    0.06527 15.97469   <1e-12
Regressions:                                             
   H<-E              0.87361    0.06612 13.21214   <1e-12
Intercepts:                                              
   H                -0.09120    0.06334 -1.43978   0.1499
   Y2               -0.06491    0.09168 -0.70803   0.4789
   Y3               -0.06794    0.10117 -0.67155   0.5019
Residual Variances:                                      
   Y1                0.93968    0.13726  6.84580         
   H                 1.03203    0.12038  8.57301         
   Y2                1.00505    0.14252  7.05186         
   Y3                1.30951    0.17326  7.55823         
--------------------------------------------------
Group 2: Arm 2 (n=200)
                    Estimate Std. Error  Z value Pr(>|z|)
Measurements:                                            
   Y2<-H             0.99827    0.06189 16.13044   <1e-12
   Y3<-H             1.04265    0.06527 15.97469   <1e-12
Regressions:                                             
   H<-E              0.87361    0.06612 13.21214   <1e-12
Intercepts:                                              
   H                -0.09120    0.06334 -1.43978   0.1499
Residual Variances:                                      
   Y1                0.82723    0.12448  6.64570         
   H                 1.03203    0.12038  8.57301         
   Y2                1.03718    0.14115  7.34796         
   Y3                1.08498    0.15032  7.21790         
\end{Soutput}
\end{Schunk}

Comparisons of multigroup structures can be conducted using a LRT. As
an example we fit the single group LLVM and perform a LRT to test
whether the residual variances are the same in both groups and the
intercepts are zero
\begin{Schunk}
\begin{Sinput}
> e0 <- estimate(mg2, rbind(data1, data2))
\end{Sinput}
\end{Schunk}
\begin{Schunk}
\begin{Sinput}
> compare(e0, e.mg)
\end{Sinput}
\begin{Soutput}
	Likelihood ratio test

data:  
chisq = 2.3744, df = 5, p-value = 0.7953
sample estimates:
log likelihood (model 1) log likelihood (model 2) 
               -2001.323                -2000.136 
\end{Soutput}
\end{Schunk}

\section{Data with missing values}
\label{sec:missing}
Missing data are common in studies with multivariate outcomes and complete case
analyses can in these settings become quite inefficient and are
further only consistent when data are missing completely at
random (MCAR).

Under the more general assumption that data are missing at random
(MAR), i.e. that the missing data mechanism depends only on the
observed variables \citep{MR1925014}, then inference can be based on the
marginal likelihood, where the missing values has been integrated out
\begin{align}
  f(y_{obs}; \theta) = \int f(y_{obs},y_{mis}; \theta)\,dy_{mis}.
\end{align}
Here $f(y_{obs},y_{mis}; \theta)$ is the full likelihood of both the observed
data ($y_{obs}$) and the missing part ($y_{mis}$), parameterized by $\theta$.

In \pkg{lava}, MLE under the MAR assumption can be obtained by adding
the \code{missing=TRUE} argument to \code{estimate} (both for
\code{lvm} and \code{multigroup} objects) using the multigroup
framework on the different missing patterns of the data.

To demonstrate this, we will imitate a MCAR missing data mechanism on
the first of the datasets simulated from \code{mg1} (see Section \ref{sec:multigroup}),
with a massive 30\% probability of missingness on each outcome
\begin{Schunk}
\begin{Sinput}
> d0 <- makemissing(data1, p = 0.3, cols = endogenous(mg1))
\end{Sinput}
\end{Schunk}

and the full-information maximum likelihood estimates can then be achieved
with the call:
\begin{Schunk}
\begin{Sinput}
> e.mis <- estimate(mg1, d0, missing = TRUE)
\end{Sinput}
\end{Schunk}
\begin{Schunk}
\begin{Sinput}
> summary(e.mis, std = NULL, labels = FALSE)
\end{Sinput}
\begin{Soutput}
Latent variables: H 
Number of rows in data=199 (73 complete cases, 7 groups)
--------------------------------------------------
                    Estimate Std. Error  Z value Pr(>|z|)
Measurements:                                            
   Y1<-H             1.00000                             
   Y2<-H             0.90265    0.10861  8.31067   <1e-12
   Y3<-H             0.95524    0.11812  8.08673   <1e-12
Regressions:                                             
   H<-E              0.83840    0.11018  7.60938   <1e-12
Intercepts:                                              
   Y1                0.00000                             
   H                -0.14287    0.11042 -1.29388   0.1957
   Y2                0.05932    0.11132  0.53288   0.5941
   Y3               -0.01611    0.12958 -0.12433   0.9011
Residual Variances:                                      
   Y1                0.68217    0.17881  3.81499         
   H                 1.20761    0.21942  5.50375         
   Y2                1.02999    0.19284  5.34120         
   Y3                1.37374    0.23948  5.73625         
--------------------------------------------------
Estimator: gaussian 
--------------------------------------------------
 Log-Likelihood = -730.9461 
 BIC = 1525.811 
 AIC = 1481.892 
--------------------------------------------------
\end{Soutput}
\end{Schunk}
with standard errors based on the observed information \citep{MR1665713}.

\section{Beyond the standard linear Gaussian case}\label{sec:beyondmle}

While linear Gaussian models cover many useful situations there are
clearly cases where these models are no longer adequate. In this section we
will briefly describe extensions of \pkg{lava} that covers some of
these cases.

\subsection{Clustered correlated data}

Models with very complex hierarchical structures can be estimated in
\proglang{lava}. However, the full specification of such a model can
be challenging and perhaps more importantly, as the lowest level in a
such a model is often not of primary interest, it\ can be more
natural to relax the model assumptions for this part of the model.

\begin{figure}[htbp]\centering
\includegraphics{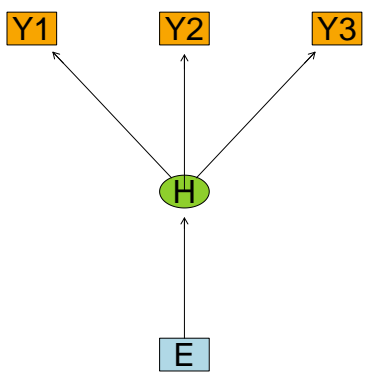}
\caption{\code{plot(mclust): Marginal model for the noise-health example.}}
\label{fig:mclustmarg}
\end{figure}

As a hypothetical example we can imagine that the aim of a study is to
estimate the association between noise levels and health. In practice
this is done by measuring the average noise level, $E$, in different
neighborhoods. We assume that the health status is measured indirectly
for each subject by three proxy measures, $Y_{1i}, Y_{2i}, Y_{3i}$
(e.g. blood pressure and cholesterol levels), and that the with-in
subject correlation between these measurements can be described by a
single latent variable, $H_i$ (see Figure \ref{fig:mclustmarg}).  The
effect of noise on health is quantified as the linear association
between $H_i$ and $E$.  The study is complicated by the fact that
measurements within neighborhoods are correlated beyond what $H_i$ is
capturing (air pollution, crime levels, traffic and other factors that
could affect stress levels) and disregarding this with-in cluster
correlation will generally lead to too optimistic standard errors of
the noise effect.

Inference can instead be based on the i.i.d. decomposition of the score
leading to a sandwich estimator (GEE-type) of the variance
\citep{williams00:_note_robus_varian_estim_clust_correl_data}
\begin{align}
  \left(\frac{\partial\mathcal{S}(\bm{\theta})}{\partial\bm{\theta}}\right)^{-1}
  \left(\frac{K}{K-1}\sum_{c=1}^K\mathcal{S}_{(c)}(\bm{\theta})^{\otimes
      2}\right)
  \left(\frac{\partial\mathcal{S}(\bm{\theta})}{\partial\bm{\theta}}\right)^{-1},
\end{align}
where $\mathcal{S}$ is the total score and
$\mathcal{S}_{(c)}$ is the sum of the scores within cluster 
$c$, and $K$ denotes the total number of clusters. Simulation studies
\citep{pmid15027075,paik88} indicate that the sandwich estimator
works well with $K>50$.

With 5 individuals sampled from each cluster/neighborhood, the above
model could specified with
\begin{Schunk}
\begin{Sinput}
> K <- 5
> mclust1 <- lvm()
> for (i in 1:K) {
+     xyz <- c("Y1", "Y2", "Y3") %+% i
+     h <- "H" %+% i
+     regression(mclust1, to = c(xyz), from = h) <- list(1, "l1", 
+         "l2")
+     regression(mclust1, to = h, from = c("U", "E")) <- list(1, 
+         "b")
+     intercept(mclust1, c(xyz)) <- list("mx", "my", "mz")
+     covariance(mclust1, c(xyz, h)) <- list("vx", "vy", "vz", 
+         "v")
+ }
> latent(mclust1) <- c("H" %+% 1:K, "U")
> intercept(mclust1, latent(mclust1)) <- 0
\end{Sinput}
\end{Schunk}
We simulate data from 250 clusters and obtain the MLE
\begin{Schunk}
\begin{Sinput}
> dclust <- sim(mclust1, 250, p = c(b = 0.3))[, manifest(mclust1)]
> eclust <- estimate(mclust1, dclust)
\end{Sinput}
\end{Schunk}
and we wish to compare this with the marginal model (see Figure \ref{fig:mclustmarg}):
\begin{Schunk}
\begin{Sinput}
> dclustWide <- reshape(dclust, direction = "long", varying = lapply(list("Y1", 
+     "Y2", "Y3"), function(x) x %+% 1:K), v.names = c("Y1", "Y2", 
+     "Y3"))
> dclustWide <- dclustWide[order(dclustWide$id), ]
> mclust <- lvm(list(c(Y1, Y2, Y3) ~ H, H ~ E))
> latent(mclust) <- ~H
> eclust0 <- estimate(mclust, dclustWide)
\end{Sinput}
\end{Schunk}

\relsize{-2}
\origfigsize

The marginal estimates with robust standard errors 
are obtained easily by giving the name of the column in
the \code{data.frame} that specifies the clusters (or an actual vector)
as argument to the \code{estimate} function
\begin{Schunk}
\begin{Sinput}
> eclust1 <- estimate(mclust, data = dclustWide, cluster = "id")
\end{Sinput}
\end{Schunk}

In this example we see a substantial under-estimation of the standard
errors of the pollution effect estimate in the na\"ive approach (with
covariates varying within clusters the bias could go in the opposite
direction as well). In contrast, the results of the marginal approach
is close to those the full model.
\begin{Schunk}
\begin{Sinput}
> res <- rbind(coef(eclust, 2)["b", ], coef(eclust1, 2)["H<-E", 
+     ], coef(eclust0, 2)["H<-E", ])
> rownames(res) <- c("full MLE", "Marg.robust", "Marg.naive")
> res
\end{Sinput}
\begin{Soutput}
             Estimate Std. Error  Z value     Pr(>|z|)
full MLE    0.3683574 0.07816049 4.712835 2.442944e-06
Marg.robust 0.3702012 0.07658741 4.833708 1.340131e-06
Marg.naive  0.3702012 0.04843684 7.642967 2.131628e-14
\end{Soutput}
\end{Schunk}

Typically the loss of power in this marginal approach is modest,
and is countered by circumventing the need
for explicit (mis)specification of the distribution of the cluster random
effect. 


\subsection{Mixture models}
\label{sec:mixture}
The normal distribution is a central assumption in
(\ref{eq:measurement}) and (\ref{eq:structural}), and one way to relax
this assumption while still avoiding the need for computational
intense numerical approximations of the likelihood function of the observed data,
is to allow mixtures
of normal distributions in the model.  Applications include
pattern recognition and cluster analysis (machine learning), outlier
analysis and modeling of heterogeneity, e.g. adjusting for unknown
subpopulations in a sample.

In general we will allow models to be described
by the convex combination
\begin{align}
  f_\theta(\bm{y}\mid \bm{z}) = \sum_{j=1}^K \pi_j f_{\theta_j}^{(j)}(\bm{y}\mid \bm{z}),
  \quad \sum_{j=1}^K\pi_j=1, \quad \pi_j\in]0,1[
\end{align}
where $f_{j,\theta_j}$ is a probability density of a LLVM with responses
$\bm{y}$ and covariates $\bm{z} = (\bm{x}',\bm{v}',\bm{w}')'$, and $\theta$
is the parameter-vector $(\cup_j \theta_j, \pi_1, \ldots, \pi_{K-1})$,
noting that the $\theta_j$'s need not to be disjoint. We denote the
number classes $K$.

The likelihood for the mixture model is therefore
\begin{align}
  L(\theta | \bm{y},\bm{z}) = \prod_{i=1}^n\sum_{j=1}^K \pi_j
  f^{(j)}_{\theta_j}(\bm{y}_i\mid \bm{z}_i)
\end{align}
To solve the corresponding score equation, the EM algorithm is
typically applied \citep{dempsterlairdrubin77}. We introduce latent indicator variables
$\xi_{ij}$, $i=1,\ldots,n$, describing the class
membership of the observation $(\bm{y}_i,\bm{z}_i)$.
\begin{align}
  \xi_{ij} = \one_{\{y_{ij} \text{ belongs to class } j\}}
\end{align}
and hence $\E \xi_{ij}=\pr(y_{ij} \text{ belongs to class } j) =
\pi_{j}$. We can then treat the mixture analysis as a missing data
problem, $\bm{v} = (\bm{y},\bm{z},\bm{\xi})$,
and complete-data log-likelihood:
\begin{align}
  \log L_{\mathcal{C}}(\theta | \bm{v}) &= 
  \sum_{i=1}^n\sum_{j=1}^K \xi_{ij}\log\left(
    f_{\theta_j}^{(j)}(\bm{y}_i\mid \bm{z}_i)\right)
\end{align}

While the EM-algorithm is generally slower than Newton-Raphson
(sub-linear vs. quadratic convergence), this disadvantage is
compensated by the fact that the EM-algorithm tends to be less
sensitive to choice of starting values as the algorithm guarantees a
non-decreasing likelihood in each iteration. In contrast, NR can
behave erratically for poor choices of starting values. To address
possible problems with convergence to a local maximum, it is still
advisable to start the algorithm at several different starting points in the
parameter space.  The EM algorithm also implicitly constraints the
probability vector to the correct parameter space, as the E-step at
iteration $l$ leads to a simple expression of the posterior class
probabilities
\begin{align}\label{eq:pihat}
  \widehat{\pi}_{ji}^{(l)}  =
  \frac{\widehat{\pi}_{j}^{(l)}f_{j,\widehat{\theta}^{(l)}}(\bm{y}_i\mid
    \bm{z}_i)}{\sum_{j=1}^K\widehat{\pi}_{j}^{(l)}f_{j,\widehat{\theta}^{(l)}}(\bm{y}_i\mid
    \bm{z}_i) }
\end{align}
In the M-step we obtain the new class probabilities
\begin{align}
  \widehat{\pi}_{j}^{(l+1)} = \frac{1}{n}\sum_{i=1}^n \widehat{\pi}_{ji}^{(l)}
\end{align}
and $\widehat{\theta}^{(l+1)}$ is derived by solving
\begin{align}
  \argmax_\theta Q(\theta; \widehat{\theta}^{(l)}) = \sum_{i=1}^n\sum_{j=1}^K
  \widehat{\pi}_{ji}^{(l)} \log\left(f_{j,\theta}(\bm{y}_i\mid
    \bm{z}_i)\right)
\end{align}
which for a general LLVM mixture is optimized iteratively (NR
pr. default).  In principle a model relating class probabilities to
covariates could also be included in this setup leading to a M-step
where we instead of the simple expression (\ref{eq:pihat}) would have
to maximize a weighted multinomial logit model.

To analyze a mixture model in \pkg{lava} the plugin package
\pkg{lava.mixture} needs to be loaded. The function \code{mixture}
fits the mixture model to a list of \code{lvm} objects implicitly
defining the number of mixture components of the model. Via the
\code{control} argument the parameters of the EM algorithm can be
adjusted, such as the starting values (\code{start}), number of random
starting points (\code{nstart}), convergence tolerance (\code{tol},
change in log-likelihood), etc.
\begin{Schunk}
\begin{Sinput}
> library(lava.mixture)
> mixture(list(m1,m2,...),data=mydata,control,...)
\end{Sinput}  
\end{Schunk}
Instead of a list, a single \code{lvm} object can be given as argument
with the argument \code{k} specifying the number of mixture
components. 

\begin{figure}[htbp]\centering
\includegraphics{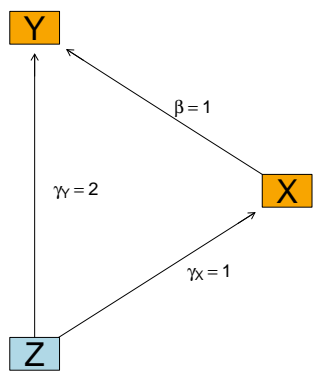}
\caption{Model \code{mix0} with unmeasured confounder $z$.}
\label{fig:mix0}
\end{figure}

As an example we will simulate data from a simple model (see Figure
\ref{fig:mix0}), where we have a single dichotomous unmeasured
confounder $z$ ($P(z=1)=0.5$), which have a direct linear effect on
the outcome of interest $Y$
\begin{align}
 Y = \mu_Y + \beta X  + \gamma_Y Z  + \epsilon_Y
\end{align}
and on the predictor $X$, which we assume is conditionally normally
distributed given $Z$
\begin{align}
  X = \mu_X + \gamma_X Z  + \epsilon_X
\end{align}
In our simulation we will let all intercepts be zero, residual variances 1,
and slopes as defined by Figure \ref{fig:mix0}:
\begin{Schunk}
\begin{Sinput}
> mix0 <- lvm(list(Y ~ X + Z, X ~ Z))
> distribution(mix0, ~Z) <- binomial.lvm()
> d0 <- sim(mix0, p = c(`Y<-Z` = 2), n = 500)
\end{Sinput}
\end{Schunk}
\relsize{-2}
\origfigsize
Next, we will define the mixture regression that takes into account
that we have unobserved heterogeneity caused by $Z$. The base model is
the simple linear regression model of $Y$ given $X$, however, with a
\code{covariance} call we define $X$ to be endogenous and let all
parameters except for the intercepts of $Y$ and $X$ be fixed between the classes
\begin{Schunk}
\begin{Sinput}
> mix1 <- lvm(Y ~ X)
> covariance(mix1, ~X) <- "v"
> mix1 <- baptize(mix1)
> intercept(mix1, ~Y + X) <- NA
\end{Sinput}
\end{Schunk}
defining the model
\begin{gather}
  Y = \mu_{y,c} + \beta X + \epsilon, \\
  \begin{pmatrix}
    \epsilon \\
    X 
  \end{pmatrix} \sim \mathcal{N}\left(
  \begin{pmatrix}
    0 \\
    \mu_{x,c}
  \end{pmatrix}
  ,
  \begin{pmatrix}
    \sigma_y^2 & 0 \\
    0 & \sigma_x^2
  \end{pmatrix}\right)
\end{gather}
with the index $c$ denoting the class. The model with two classes is
fitted with the call
\begin{Schunk}
\begin{Sinput}
> M <- mixture(mix1, d0, k = 2)
> (s <- summary(M))
\end{Sinput}
\end{Schunk}
\begin{Schunk}
\begin{Soutput}
Cluster 1 (n=231, Prior=0.4687):
--------------------------------------------------
                    Estimate Std. Error Z value  Pr(>|z|)
Regressions:                                             
   Y<-X              0.95114  0.07881   12.06853   <1e-12
Intercepts:                                              
   Y                 2.11558  0.15240   13.88137   <1e-12
   X                 1.11484  0.11280    9.88306   <1e-12
Residual Variances:                                      
   Y                 1.14553  0.12566    9.11583         
   X                 1.05875  0.11837    8.94435         

Cluster 2 (n=269, Prior=0.5313):
--------------------------------------------------
                    Estimate Std. Error Z value  Pr(>|z|)
Regressions:                                             
   Y<-X              0.95114  0.07881   12.06853   <1e-12
Intercepts:                                              
   Y                 0.15417  0.11353    1.35802 0.1745  
   X                -0.10255  0.10257   -0.99982 0.3174  
Residual Variances:                                      
   Y                 1.14553  0.12566    9.11583         
   X                 1.05875  0.11837    8.94435         
--------------------------------------------------
AIC= 3325.163 
||score||^2= 0.0001193449 
\end{Soutput}
\end{Schunk}
We see that the mixture model gives a regression coefficient of
0.951 which is close to the true value.
This should be compared to the biased OLS estimate of $\beta$:
\begin{Schunk}
\begin{Sinput}
> coef(summary(lm(Y ~ X, d0)))
\end{Sinput}
\begin{Soutput}
             Estimate Std. Error  t value     Pr(>|t|)
(Intercept) 0.8785888 0.06556336 13.40061 3.472204e-35
X           1.3675927 0.05108919 26.76873 1.804800e-98
\end{Soutput}
\end{Schunk}
In fact for the given set of parameters the bias is 
\begin{align}
  \bias(\widehat{\beta}_{\text{OLS}}) = 
  \beta - \frac{(\beta\gamma_X^2+\gamma_Y\gamma_X)\var(Z)+\beta\var(\epsilon_X)}{\gamma_X^2\var(Z)
    + \var(\epsilon_X)} = \tfrac{2}{5}
\end{align}
and replicating the above simulation 1000 times we obtain the
following statistics for the $\beta$ parameter
\begin{center}
  \begin{tabular}{cccc}
    \hline
    Estimator & Bias & Std.Err & MSE \\
    \hline
    OLS & 0.403 & 0.0510 & 0.165 \\
    Mixture & 0.002 & 0.0822 &  0.068 \\
    \hline 
  \end{tabular}
\end{center}
which clearly shows the advantage of the mixture regression model.

The example generalizes to several binary confounders ($2^k$ classes
with $k$ confounders, and the continuous case could likewise be
approximated by a finite number of mixtures) and if we were suspecting
an interaction effect between $X$ and $Z$ we could also have allowed
the slope parameter ($\beta$) to vary freely in the two 
classes. The difficult task of choosing an optimal number of
components in the mixture could be based on cross-validation, but
currently this is not implemented in \pkg{lava.mixture} (nor is the
problem of adjusting the standard errors for this
model selection which could be based on a bootstrap
resampling). Still we believe that a mixture analysis as in the
previous example could serve as an important sensitivity analysis in
many applications.

Two variants of the EM-algorithm are also implemented in the
\code{mixture} function (via the argument \code{type}): CEM
(Classification EM) where each observation in the E-step is assigned
to the class with the highest maximum posterior probability
(\ref{eq:pihat}), and StEM (Stochastic EM) where each observation is
assigned randomly to a class based on a draw from a multinomial
distribution with the posterior probabilities as parameter. In both
cases, the M-step reduces to the maximization of a simple multigroup
LLVM (see Section \ref{sec:multigroup}).  The latter approach leads to
a time-homogeneous Markov chain of the parameters, which under certain
regularity conditions is ergodic with a normal stationary distribution
with a mean that is a consistent estimate of the mixture parameters
\citep{celeux93:_asymp_em}. In some situations, the StEM has been
observed to have faster convergence and being more likely to converge
to the global maximum of the log-likelihood as the stochastic
variation introduced can push the "EM"-algorithm away from a
saddle-point or local maximum \citep{dieboltip96}. However, assessment
of convergence of the Markov chain remains a challenging problem. In
any case, the CEM or StEM algorithm might be useful for finding good
starting values for the EM algorithm \citep{Biernacki2003561}.

In models with an unstructured mean and covariance in each class, the
function \code{mvnmix} can be used because it exploits the fact, that
the M-step has a closed-form solution (in fact the likelihood
is unbounded, however, this is of more technical than
practical interest as we typically can disregard the obviously wrong
solutions to the score equations). As an example we will fit a
two-component Gaussian mixture model to the waiting times between
eruptions and the durations of the eruptions for the Old Faithful
geyser in Yellowstone National Park:
\begin{Schunk}
\begin{Sinput}
> data(faithful)
> mixff <- mvnmix(faithful, k = 2)
> (s <- summary(mixff))
\end{Sinput}
\end{Schunk}
\begin{Schunk}
\begin{Soutput}
Cluster 1 (n=97, Prior=0.3559):
--------------------------------------------------
                     Estimate Std. Error Z value  Pr(>|z|)
Intercepts:                                               
   eruptions          2.03639  0.03495   58.27283   <1e-12
   waiting           54.47852  0.62846   86.68510   <1e-12
Residual Variances:                                       
   eruptions          0.06917  0.01074    6.43961         
   eruptions,waiting  0.43517  0.17984    2.41979 0.01553 
   waiting           33.69728  5.77754    5.83246         

Cluster 2 (n=175, Prior=0.6441):
--------------------------------------------------
                     Estimate  Std. Error Z value   Pr(>|z|) 
Intercepts:                                                  
   eruptions           4.28966   0.03349  128.06900   <1e-12 
   waiting            79.96812   0.47131  169.67075   <1e-12 
Residual Variances:                                          
   eruptions           0.16997   0.02112    8.04628          
   eruptions,waiting   0.94061   0.18679    5.03562 4.763e-07
   waiting            36.04621   4.09090    8.81132          
--------------------------------------------------
AIC= 2282.528 
||score||^2= 8.611289e-16 
\end{Soutput}
\end{Schunk}

\relsize{-2}
\begin{figure}[htbp]\centering
\includegraphics{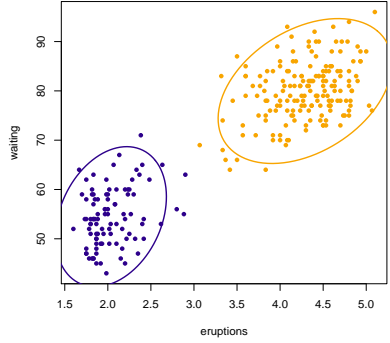}
\caption{\code{plot(mixff,col="darkblue","orange")}: 2-component
  Gaussian Mixture Model fit of Old Faithful data (95\% prediction
  regions for each mixture distribution and each observation where
  assigned the color of the mixture with the highest posterior
  probability)}
\label{fig:faithful}
\end{figure}
\origfigsize

\subsection{Binary data}
In epidemiology binary data are among the most common types of
endpoints and often correlated binary data are collected.  Several
methods have been proposed to deal with this sort of data,
e.g. marginal models \citep{liang86:_longit}, conditional maximum
likelihood estimation \citep{andersen1971} or numerical integration to
obtain the marginal likelihood of the observed data
\citep{pinheirochao06}.  Via the package
\pkg{lava.tobit} \citep{lavatobit} we can estimate LLVMs where a subset of the
endogenous variables are binary using a probit-link (only
subtle differences with a logit-link resulting in a scaling of roughly
1.7 of the parameters).

As an example we will look at a simple factor analysis model with a covariate
\begin{gather}
  \pr(Y_{ij}=1|\eta_i,X_i) = \Phi\left(\mu_j +
    \lambda_j\eta_i\right), \\
  \eta_i = \gamma X_i + \zeta_i,
\end{gather}
where $\zeta_i\sim\mathcal{N}(0,\sigma^2)$, and $\Phi$ is the
standard normal cumulative distribution function.

There are several ways to specify this model in \code{lava}, but here
we will use the \code{binary} function:
\begin{Schunk}
\begin{Sinput}
> mprobit <- lvm(list(c(y1, y2, y3) ~ eta, eta ~ x))
> latent(mprobit) <- ~eta
> binary(mprobit) <- endogenous(mprobit)
> set.seed(1)
> dprobit <- sim(mprobit, 500)
\end{Sinput}
\end{Schunk}
With the \code{binary} call the endogenous variables, $Y_{ij}$, are
changed from being continuous to being dichotomous, where we assume
that there exists a latent conditionally normally distributed variable,
$Y_{ij}^*$, such that
\begin{align}\label{eq:thres1}
  Y_{ij} =
  \begin{cases}
    1,& Y_{ij}^*>0 \\
    0,& Y_{ij}^*\leq 0.
  \end{cases}
\end{align}
The MLE is obtained as usual
\begin{Schunk}
\begin{Sinput}
> lava.options(param = "hybrid", trace = 1, method = "NR")
> (eprobit <- estimate(mprobit, dprobit))
\end{Sinput}
\end{Schunk}
\begin{Schunk}
\begin{Soutput}
                    Estimate Std. Error  Z value  Pr(>|z|)
Measurements:                                             
   y2<-eta           0.78017    0.15378  5.07336 3.909e-07
   y3<-eta           0.92687    0.18950  4.89114 1.003e-06
Regressions:                                              
   eta<-x            1.09228    0.15999  6.82737 8.649e-12
Intercepts:                                               
   y1                0.12358    0.09374  1.31829    0.1874
   y2               -0.06024    0.08104 -0.74330    0.4573
   y3                0.04414    0.08930  0.49432    0.6211
Residual Variances:                                       
   eta               1.32052    0.40634  3.24979          
\end{Soutput}
\end{Schunk}

Another interesting example is the logit-probit-normal model
\citep{caffo2006linkprobitnormal} which is a conditional model with
marginal fixed effects on a logit scale. E.g. a random intercept model
with a single covariate $X_{ij}$ can be formulated as
\begin{align}
  \pr(Y_{ij}=1\mid \eta_i, X_{ij} ) = \Phi\left(\Phi^{-1}\left\{\frac{1}{1+\exp(-\mu-\beta X_{ij})}\right\}\sqrt{1+\sigma^2} + \eta_i\right),
\end{align}
where $\sigma^2$ is the variance of the random effect $\eta_i$. The
nonlinear parameterization ensures the condition
\begin{align}
  \logit \pr(Y_{ij}=1\mid X_{ij}) = \mu + \beta X_{ij}.
\end{align}

Specification is straightforward using the \code{constrain} method,
and standard likelihood theory can be applied on this model in
contrast to the generalized estimating equation framework,
e.g. likelihood ratio testing, profile likelihood confidence limits,
and analysis with data missing at random as described in Section
\ref{sec:multigroup}.

As an example we simulate observations from a simple random intercept
probit model
\begin{Schunk}
\begin{Sinput}
> margm <- lvm(c(y1, y2) ~ x)
> regression(margm, c(y1, y2) ~ u) <- "gamma"
> intercept(margm, endogenous(margm)) <- "mu"
> binary(margm) <- endogenous(margm)
> covariance(margm, ~u) <- 1
> latent(margm) <- ~u
> dmarg <- sim(margm, 100)
\end{Sinput}
\end{Schunk}

The logit-probit-normal model is then specified as
\begin{Schunk}
\begin{Sinput}
> regression(margm, c(y1, y2) ~ x) <- 0
> constrain(margm, mu ~ x + alpha + beta + gamma) <- function(z) qnorm(1/(1 + 
+     exp(-z[2] - z[3] * z[1]))) * sqrt(1 + z[4]^2)
\end{Sinput}
\end{Schunk}
and estimates are obtained in the usual way
\begin{Schunk}
\begin{Sinput}
> (emargm <- estimate(margm, dmarg))
\end{Sinput}
\end{Schunk}
\begin{Schunk}
\begin{Soutput}
              Estimate Std. Error Z value  Pr(>|z|)
Measurements:                                      
   y1<-u       2.08070    0.57583 3.61339 0.0003022
Intercepts:                                        
   alpha       0.04207    0.20986 0.20049    0.8411
   beta        1.35056    0.25772 5.24039 1.602e-07
\end{Soutput}
\end{Schunk}

\subsection{Censored data}
Censoring is a complication that is often encountered in cohort
studies but can also be seen in experimental studies where
thresholding of measurements may occur due to limiting precision of
laboratory equipment.

We will assume that the data-generating mechanism is defined by
(\ref{eq:measurement}) and (\ref{eq:structural}) and that a subset of
the endogenous variables $Y_{ij}^*$, $j\in J$ are censored such that for
given censoring times $C_j$, $j\in J$ we only observe
\begin{align}\label{eq:tobitright}
  {Y}_{ij} =
  \begin{cases}
    Y_{ij}^*, & Y_{ij}^*< C_{ij}\\
    C_{ij}, & Y_{ij}^*\geq C_{ij}
  \end{cases}
\end{align}

As an example we will simulate data from a regression model (see Figure
\ref{fig:mix0}), where we have a single dichotomous mediator
$Z$ ($P(Z=1)=0.5$), which have a direct linear effect on
the outcome of interest $Y$
\begin{align}
 Y = \mu_Y + \beta X  + \gamma_Y Z  + \epsilon_Y
\end{align}
and on the predictor $X$, which we assume is conditionally normally
distributed given $Z$
\begin{align}
  X = \mu_X + \gamma_X Z  + \epsilon_X
\end{align}
In our simulation we will let all intercepts be zero, residual variances 1,
and slopes as defined by Figure \ref{fig:mix0}:
\begin{Schunk}
\begin{Sinput}
> med0 <- lvm(list(Y ~ X + Z, X ~ Z))
> distribution(med0, ~Z) <- binomial.lvm()
> d0 <- sim(med0, p = c(`Y<-Z` = 2), n = 500)
\end{Sinput}
\end{Schunk}

We further assume we only observe the
thresholded versions of $Y$ and $X$:
\begin{Schunk}
\begin{Sinput}
> dtobit <- transform(d0, X = as.factor((X > 0) * 1), Y = Surv(pmin(Y, 
+     2), Y < 2))
\end{Sinput}
\end{Schunk}

As \code{X} is coded as a factor and \code{Y} is coded as a right-censored
\code{Surv}-object (combinations of left and right censoring are
allowed) the \code{estimate} method automatically applies a Probit and
Tobit model respectively
\begin{Schunk}
\begin{Sinput}
> (etobit <- estimate(med0, dtobit))
\end{Sinput}
\end{Schunk}
\begin{Schunk}
\begin{Soutput}
                    Estimate Std. Error  Z value Pr(>|z|)
Regressions:                                             
   Y<-X              0.94429    0.07790 12.12215   <1e-12
   Y<-Z              2.04975    0.13527 15.15247   <1e-12
    X<-Z             1.00076    0.12154  8.23419   <1e-12
Intercepts:                                              
   Y                -0.02887    0.07733 -0.37334   0.7089
   X                -0.07362    0.07994 -0.92096   0.3571
Residual Variances:                                      
   Y                 0.97204    0.10225  9.50687         
\end{Soutput}
\end{Schunk}

The Probit/Tobit model framework also has important applications in
the causal modeling framework where it allows us to elegantly define
direct and indirect effects for binary and censored data in complex
path analyses \citep{ditlevsen05:_mediat_propor}. The interpretation
in this setup is linked to the assumption that the observations are
generated by an unobserved continuous variable following a conditional
normal distribution, and the indirect effects can therefore be
quantified via the \code{effects} method:

\begin{Schunk}
\begin{Sinput}
> effects(etobit, Y ~ Z)
\end{Sinput}
\begin{Soutput}
Total effect of 'Z' on 'Y':
		2.994759 (Approx. Std.Err = 0.1509939)
Direct effect of 'Z' on 'Y':
		2.04975 (Approx. Std.Err = 0.135275)
Indirect effects:
	Effect of 'Z' via Z->X->Y:
		0.9450087 (Approx. Std.Err = 0.1358391)
\end{Soutput}
\end{Schunk}

\newcommand{\VV}{\bm{\Omega}_\theta}
\newcommand{\mm}{\bm{\xi}_\theta}
\newcommand{\mmi}{\bm{\xi}_{\theta,i}}
\newcommand{\Ss}{\bm{\Sigma}_\theta}
\newcommand{\uu}{\bm{\mu}_\theta}

\subsubsection{Inverse probability weights}
For models with complex sampling (survey data) and as an alternative
method to deal with censored or missing data it is convenient to
introduce Inverse Probability Weights in the estimation procedure
\citep{horvitzthompson1952,rotnitzkyrobbins1995}. The
estimating equations in this situation becomes
\begin{align}
  \begin{split}\label{eq:w}
  \mathcal{U}_i^{\bm{\mathcal{W}}}(\theta; \bm{Y}_i,\bm{Z}_i) &= 
  -\frac{1}{2}\Big\{
  \Big(\frac{\partial\mvec\VV}{\partial\bm{\theta}'}\Big)'\Big(\mvec\Big[(\VV^{-1}
    \\
    &\qquad - \VV^{-1}(\bm{Y}_i-\mmi)(\bm{Y}_i-\mmi)'\VV^{-1}\Big]\bm{\mathcal{W}}_i\Big)
\\
&\qquad+
  2\left(\frac{\partial\mvec{\mmi}}{\partial\bm{\theta}'}\right)'\VV^{-1}\bm{\mathcal{W}}_i(\bm{Y}_i-\mmi)
  \Big\} = 0
  \end{split}
\end{align}
where $\bm{\mathcal{W}}_i$ is the weight-matrix for the $i$th
observation, and $\mmi$ and $\VV$ are the model-specific mean and
covariance matrix of $\bm{Y}_i$ given covariates $\bm{Z}_i = (\bm{X}_i',\bm{V}_i',\bm{W}_i')'$.


In \pkg{lava}, equation (\ref{eq:w}) can be solved with a diagonal
weight matrix using the estimator \code{weighted} and using the
argument \code{weight} with the \code{estimate} method. The
\code{weight} argument should be either a matrix with the weights of
the endogenous variables of the model (or a named matrix with a subset
of the variables) or alternatively a character vector with the names in
the data.frame that corresponds to weights\footnote{For the
  \code{lava.tobit} package the \code{weight} argument is already
  reserved and the \code{weight2} argument should be used instead and
  further the estimator should not be changed from the default.}  (the
weights are then assigned to the variables in the models in the order
they appear, see e.g the \code{vars} function). For multigroup models
a list of matrices or character-vectors is expected.

\subsection{Instrumental variables}
In econometrics \emph{Instrumental Variable} (IV) estimators are popular
tools for dealing with the problem of covariates that are correlated
with the residual term of the response variable. In this situation ordinary linear
regression analysis will yield biased estimates.
The idea is to identify an IV, which is a variable fulfilling
the conditions
\begin{enumerate}
\item Correlated with the problematic covariate, $X$, given all other covariates
\item Uncorrelated with the residual error, $\epsilon$, of the
  response, $Y$.
\end{enumerate}
The estimator can be then be formulated as Two-Stage Least-Squares
(2SLS) approach. In the first step, regress $X$ on the IV(s) and obtain
the predicted covariate, $\widehat{X}$. In the second stage, regress $Y$ on
$\widehat{X}$ (and other covariates). Consistency follows under
very weak assumptions \citep{greene2002,angrist01:iv}.

The method can also be applied to estimate parameters in SEMs \citep{bollen1996}.
In the following we will assume that the model of interest is a SEM (no
random slopes, single group) without linear or non-linear parameter constraints
\begin{align}\label{eq:semms}
  \bm{Y} = \bm{\nu} + \bm{\Lambda}\bm{\eta} + \bm{K}\bm{X} +
  \bm{\epsilon},
\end{align}
\begin{align}\label{eq:semst}
  \bm{\eta} = \bm{\alpha} + \bm{B}\bm{\eta} + 
  \bm{\Gamma}\bm{X} + \bm{\zeta},
\end{align}
with a zero diagonal of $\bm{B}$. We further assume that we have at
least one indicator for each latent variable. This means that there exists a
subset of endogenous variables $\widetilde{\bm{Y}}$ of $\bm{Y}$, with
no other predictors than the latent variable. From these we can
identify the latent variables
\begin{align}\label{eq:etay}
  \bm{\eta} = \widetilde{\bm{Y}} - \widetilde{\bm{\epsilon}}.
\end{align}
Substituting (\ref{eq:etay}) into
(\ref{eq:semms}) and (\ref{eq:semst}) we obtain the following equation
for one of the endogenous variables, $Y_j$ (not part of $\widetilde{\bm{Y}}$):
\begin{align}\label{eq:osem1}
  Y_j = \nu_j +
  \bm{\Lambda}_j\widetilde{\bm{Y}} +
  \bm{K}_j\bm{X} + \underbrace{\epsilon_j-\bm{\Lambda}_j\widetilde{\bm{\epsilon}}}_{u_j},
\end{align}
and similarly for the surrogate
\begin{align}\label{eq:osem2}
  \widetilde{Y}_r = \alpha_r + \bm{B}_r\widetilde{\bm{Y}} +
  \bm{\Gamma}_r\bm{X} +
  \underbrace{\zeta_r+\widetilde{\epsilon}_r-\bm{B}_r\widetilde{\epsilon}}_{u_r}.
\end{align}
Equations (\ref{eq:osem1}) and (\ref{eq:osem2}) only includes observed
predictors but parameters cannot be estimated using OLS because of
obvious correlations between predictors and residuals.
Instruments have to be identified that are uncorrelated with the
residual terms ($u_r$ or $u_j$) in 
(\ref{eq:osem1})-(\ref{eq:osem2}), while at
the same time being correlated with the part of $\widetilde{\bm{Y}}$
that are entering the equation as predictors. In \pkg{lava} these
conditions are checked automatically via the model-implied covariance
structure, selecting the largest possible set of instrument variables
for each equation. To implicitly select the instruments for the
different regression equations, covariance between specific residual
terms should be added to the model structure via the \code{covariance}
method (thereby indirectly disqualifying a variable as an 
instrument candidate). As all equations are solved simultaneously
\citep{bollenfest} the covariance matrix of the estimates are
available (i.e. \code{vcov}) and Wald tests via the \code{compare}
method are possible. Consistent estimates of the variance parameters,
are obtained via MLE in the model with all other parameters fixed at
the IV estimates (can be disable with the control parameter
\code{variance=FALSE}).

\begin{figure}[htbp]\centering
\includegraphics{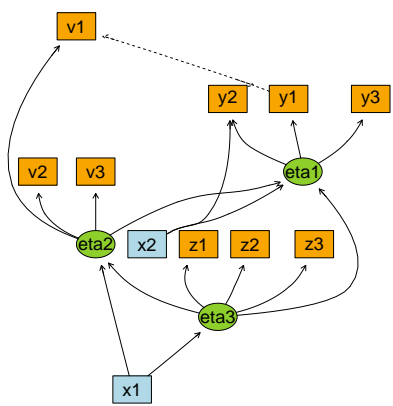}
\caption{\code{plot(mIV)}}
\label{fig:mIV}
\end{figure}

To illustrate the method we simulate data from a complex latent model
with three measurement models (see Figure \ref{fig:mIV}):
\begin{Schunk}
\begin{Sinput}
> mIV <- lvm()
> regression(mIV) <- c(y1, y2, y3) ~ eta1
> regression(mIV) <- c(v1, v2, v3) ~ eta2
> regression(mIV) <- c(z1, z2, z3) ~ eta3
> latent(mIV) <- ~eta1 + eta2 + eta3
> regression(mIV) <- eta1 ~ eta2 + eta3 + x2
> regression(mIV) <- eta2 ~ eta3 + x1
> regression(mIV) <- eta3 ~ x1
> covariance(mIV) <- y1 ~ v1
> regression(mIV) <- y2 ~ x2
> regression(mIV, c(y1[0], v1[0], z1[0]) ~ eta1 + eta2 + eta3) <- c(1, 
+     1, 1)
> dIV <- sim(mIV, 1000)[, manifest(mIV)]
\end{Sinput}
\end{Schunk}
\relsize{-2}
\origfigsize In this model we explicitly defined \code{y1}, \code{v1} and \code{z1} as the
indicators in (\ref{eq:etay}) setting the factor loadings to one and
intercepts to zero. To apply the
instrumental variable estimator using the model-implied
instruments, we simply add the argument \code{estimator="IV"} to the
\code{estimate} function, and as we have already chosen the
indicator-set we set $\code{fix=FALSE}$:
\begin{Schunk}
\begin{Sinput}
> summary(eIV <- estimate(mIV, dIV, estimator = "IV", fix = FALSE))
\end{Sinput}
\end{Schunk}
\begin{Schunk}
\begin{Soutput}
Latent variables: eta1 eta2 eta3 
Number of rows in data=1000
--------------------------------------------------
                    Estimate Std. Error  Z value Pr(>|z|)   std.xy
Measurements:                                                     
   y1<-eta1          1.00000                               0.96794
   y2<-eta1          0.99358    0.01199 82.89643   <1e-12  0.89359
   y3<-eta1          0.98599    0.01175 83.90153   <1e-12  0.96603
    v1<-eta2         1.00000                               0.91585
    v2<-eta2         0.99870    0.01972 50.64819   <1e-12  0.92661
    v3<-eta2         1.02062    0.02023 50.45033   <1e-12  0.92472
   z1<-eta3          1.00000                               0.81644
   z2<-eta3          0.94467    0.03249 29.07141   <1e-12  0.80369
   z3<-eta3          0.98653    0.03382 29.17123   <1e-12  0.81647
Regressions:                                                      
   y2<-x2            0.98771    0.04606 21.44170   <1e-12  0.22739
    eta1<-eta2       1.05111    0.06144 17.10780   <1e-12  0.64088
    eta1<-eta3       0.89998    0.10608  8.48410   <1e-12  0.32826
    eta1<-x2         0.97850    0.05490 17.82364   <1e-12  0.25048
   eta2<-eta3        0.90539    0.06139 14.74821   <1e-12  0.54161
   eta2<-x1          1.05768    0.07972 13.26707   <1e-12  0.44770
    eta3<-x1         0.97241    0.04543 21.40496   <1e-12  0.68806
Intercepts:                                                       
   y1                0.00000                               0.00000
   y2               -0.03743    0.04464 -0.83837   0.4018 -0.00861
   y3               -0.07527    0.04534 -1.66018  0.09688 -0.01886
   eta1              0.01279    0.05512  0.23202   0.8165  0.00327
   v1                0.00000                               0.00000
   v2               -0.03401    0.04561 -0.74560   0.4559 -0.01323
   v3               -0.02803    0.04679 -0.59910   0.5491 -0.01065
   eta2              0.09849    0.05343  1.84351  0.06526  0.04131
   z1                0.00000                               0.00000
   z2               -0.00774    0.04259 -0.18173   0.8558 -0.00462
   z3                0.02848    0.04488  0.63450   0.5258  0.01652
   eta3              0.05345    0.04585  1.16578   0.2437  0.03747
Residual Variances:                                               
   y1                1.02973                               0.06310
   y1,v1             0.53963                               0.05131
   y2                0.98617                               0.05217
   y3                1.06375                               0.06678
   eta1              1.01587                               0.06644
   v1                1.09261                               0.16123
   v2                0.93365                               0.14140
   v3                1.00333                               0.14490
   eta2              0.98078                               0.17254
   z1                1.01747                               0.33342
   z2                0.99507                               0.35408
   z3                0.99004                               0.33338
   eta3              1.07110                               0.52657
--------------------------------------------------
Estimator: IV 
--------------------------------------------------
                                   
Latent variables     eta1,eta2,eta3
Surrogate variables: y1,v1,z1      
                                   
 Response Instruments               
 y2       y3,v2,v3,z1,z2,z3,x2,x1   
 y3       y2,v2,v3,z1,z2,z3,x2,x1   
 eta1     v2,v3,z2,z3,x2,x1         
 v2       y2,y3,v3,z1,z2,z3,x2,x1   
 v3       y2,y3,v2,z1,z2,z3,x2,x1   
 eta2     z2,z3,x2,x1               
 z2       y1,y2,y3,v1,v2,v3,z3,x2,x1
 z3       y1,y2,y3,v1,v2,v3,z2,x2,x1
 eta3     x1,x2                     
--------------------------------------------------
\end{Soutput}
\end{Schunk}

The IV estimator has the advantage of being a non-iterative procedure
and requires weaker assumptions than MLE. There are also indications
that IV estimators are performing well in low sample-sizes and are
generally more robust to structural model mis-specification than MLE
\citep{bolleniv2007}. Where applicable an IV analysis is therefore a
good choice of method for analyzing the model fit of a
structural equation model fitted with MLE (e.g. based on a Hausman type test
statistics).

\section{Graphics}\label{sec:graphics}
\begin{table}[hb]
  \centering
  \begin{tabular}{ll}
    \textbf{Function} & \textbf{Task} \\ \hline
    \code{plot} & Plots path diagram of model \\
    \code{labels} & Defines labels for variables \\ 
    \code{edgelabels} & Defines labels and style for edges of the graph \\ 
    \code{nodecolor} & Defines color and style of nodes/variables \\ 
    \code{Graph} & Extracts graph (\code{graphNEL} object)\\ 
    \hline
  \end{tabular}
  \caption{Graphics functions.}\label{tab:graphics}
\end{table}

A \code{plot} method is available for both the \code{lvm},
\code{multigroup} and \code{lvmfit} classes.
Plotting a \code{lvmfit} object shows the user whether \pkg{lava} has linked
the model to the data as intended. A common mistake is that a variable
name in the model specification does not appear in the data. In this
case \pkg{lava} consider the corresponding variable to be latent which
is easily identified from the plot.

Layout and rendering of the graphs are achieved via \pkg{Rgraphviz}
\citep{rgraphvizR}, which in combination with the \pkg{tikzDevice}
\citep{tikzdev} makes it possible to produce publication quality path
diagrams.

In the following we will use model \code{m1} defined in Section
\ref{sec:modelspec} as the example. To enhance the graph we will add some
labels to the nodes using the function \code{labels}
\footnote{for more information on mathematical annotation in \code{R}
  we refer to the \code{plotmath} help-page.}
\begin{Schunk}
\begin{Sinput}
> labels(m1) <- c(u1 = expression(eta[1]), u2 = expression(eta[2]))
\end{Sinput}
\end{Schunk}
Similarly, subscripted versions of the observed variables could be
defined but we will keep them as they are for now. The labels of an
object can be examined with \code{labels(m)}.

Labels of edges can be defined with the \code{edgelabels}-function, e.g.
to define new labels, colors and line width for the edges from $x_1$ to
$u_1$ and from $u_1$ to $y_{11}$ we call
\begin{Schunk}
  \begin{Sinput}
> m1 <- m0
> edgelabels(m1, u1~x1, lwd=3, col="blue") <- expression(beta[1])
> edgelabels(m1, y11~u1, lwd=3, col="red",
+           labcol="red") <- expression(beta[2])       
    
  \end{Sinput}
\end{Schunk}
In addition the argument \code{labels=TRUE} can be parsed to the
\code{plot} method to add parameter names (named
\code{p1},\code{p2},... if no names were previously given using,
e.g., \code{regression}) to the edges of the plot. This will override (but
not delete) previously defined edge label attributes.

The color of the nodes is automatically added to the graph. To disable
this functionality the argument \code{addcolor=FALSE} should be
parsed to the \code{plot}-method. New coloring and style can be added via
\code{nodecolor}:
\begin{Schunk}
  \begin{Sinput}
nodecolor(m1) <- "indianred"
nodecolor(m1, ~y11+y12+y13, 
+          labcol="white", lwd=c(3,1,1)) <- "lightblue"
nodecolor(m1, ~x1+x2, labcol="red", 
+          border=c("black","white")) <- "white"    
  \end{Sinput}
\end{Schunk}

\subsection{Graph Attributes}

\begin{figure}[!ht]
  \centering
\includegraphics{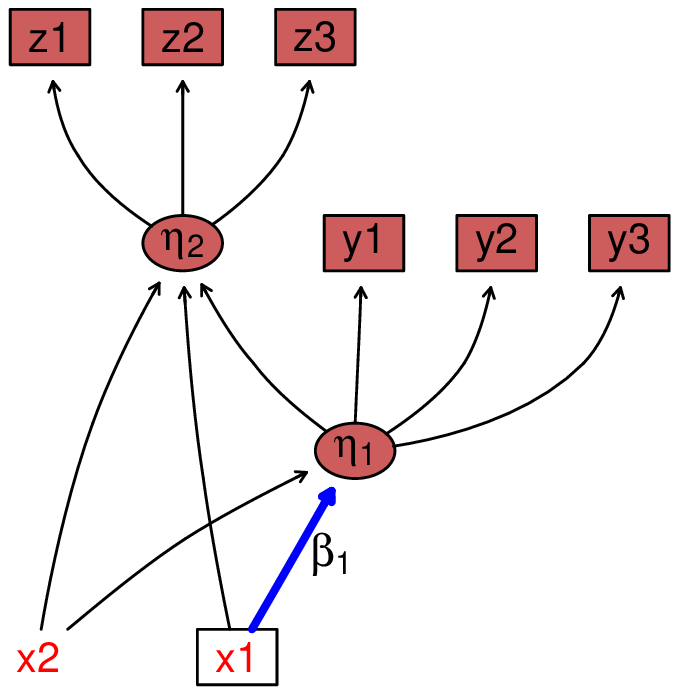}
\caption{\code{plot(m1)}}
\label{fig:m1feature}
\end{figure}

Specific attributes of the graph such as font size, line width
etc. can also be controlled via the \code{nodeRenderInfo} and
\code{edgeRenderInfo} functions called on the \code{graph} object in a
\code{lvm}-object accessible via the \code{Graph} function (also
available for \code{lvmfit}-objects). E.g. to set the font size to 2
of all edge-labels (see Figure \ref{fig:m1feature}) we would write
\begin{Schunk}
\begin{Sinput}
> edgeRenderInfo(Graph(m1))$cex <- 1.5
\end{Sinput}
\end{Schunk}
\relsize{-2}
\origfigsize

\begin{figure}[!ht]
  \centering
\includegraphics{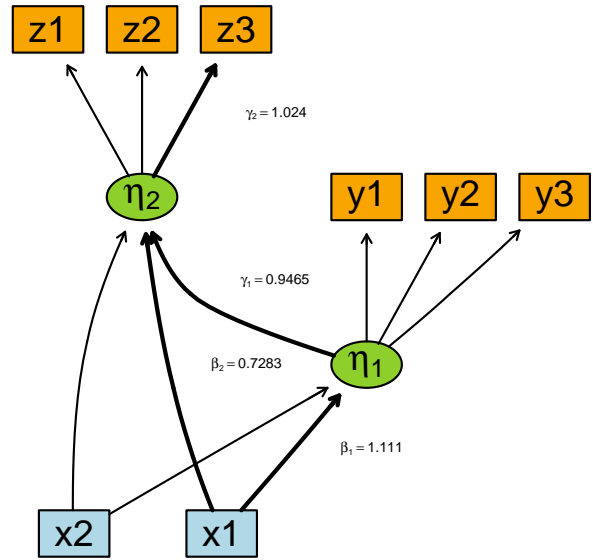}
\caption{Selected estimates from model \code{m1} (\code{plot(e,diag=FALSE)}))}
\label{fig:e2}
\end{figure}

Methods for visualizing \code{lvmfit} are also available. As an
example we will extract the pathways from \code{x1} to \code{z3} with
the \code{path} method and highlight and label the corresponding
arrows in the path-diagram (see Figure \ref{fig:e2}):
\begin{Schunk}
  \begin{Sinput}
Graph(e) <- Graph(e, add=TRUE)
labels(Graph(e)) <- c(u1=expression(eta[1]), u2=expression(eta[2]))
mypath <- path(e, z3~x1)
edgeRenderInfo(Graph(e))$lwd[unlist(mypath$edges)] <- 2
edgeRenderInfo(Graph(e))$label <- NA
edgelabels(e, edges=unique(unlist(mypath$edges)), cex=1.1) <- 
+           c("beta[1]", "gamma[1]", "gamma[2]", "beta[2]")    
  \end{Sinput}
\end{Schunk}
\origfigsize

\subsection{Graph Layout}

\begin{figure}[!ht]
  \centering
\includegraphics{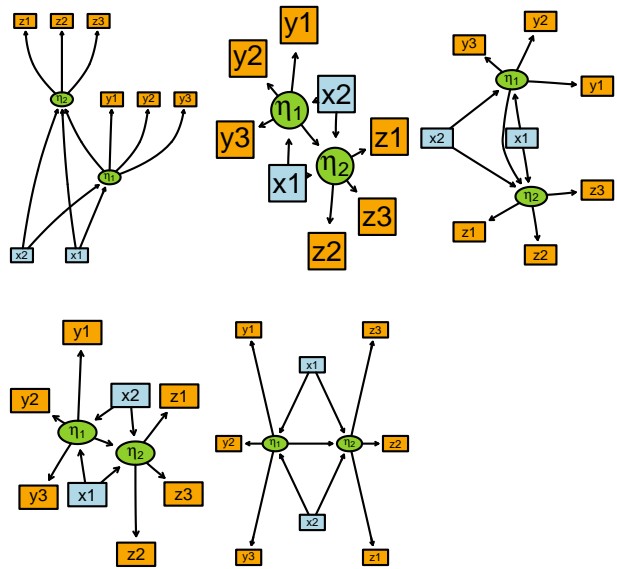}
  \caption{Different graph layout algorithms from top left to bottom right: dot, neato, twopi, fdp, circo.}
  \label{fig:layout}  
\end{figure}

Several automatic graph layout algorithms are available through the
\code{layoutType} argument (see Figure \ref{fig:layout}). Additionally, the graph (obtainable via
the \code{Graph} method) can be saved with the \code{doDot}
function and be processed in an external program supporting the
\code{dot}-format (graphviz).

\clearpage

\section{Application: Brain serotonin transporter imaging data}
\label{sec:pet}
\begin{figure}[htbp]
  \centering
  \hspace*{-2cm}
  \mbox{
    \includegraphics[width=0.58\textwidth,keepaspectratio]{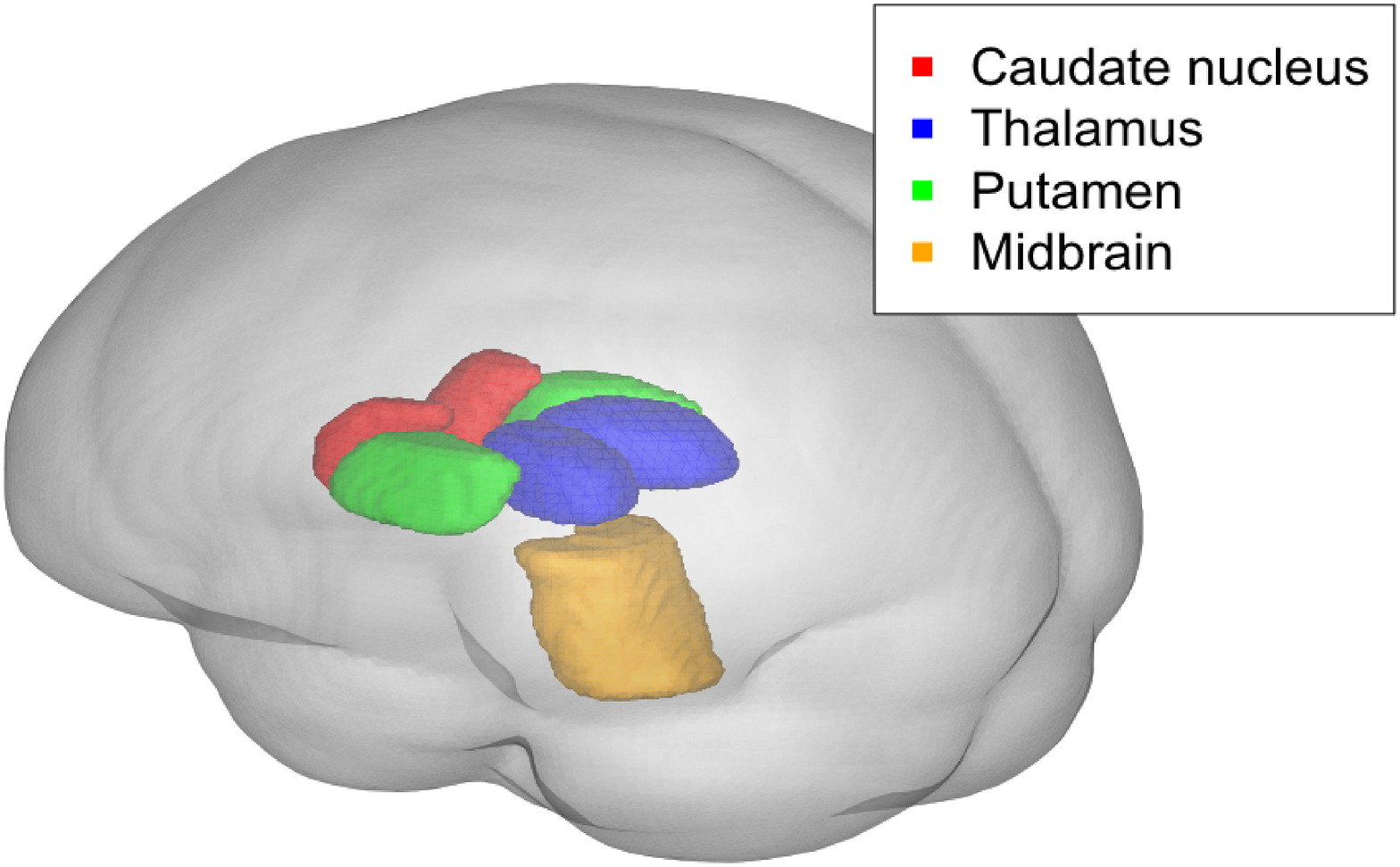}
    \includegraphics[height=0.4\textwidth,keepaspectratio=TRUE]{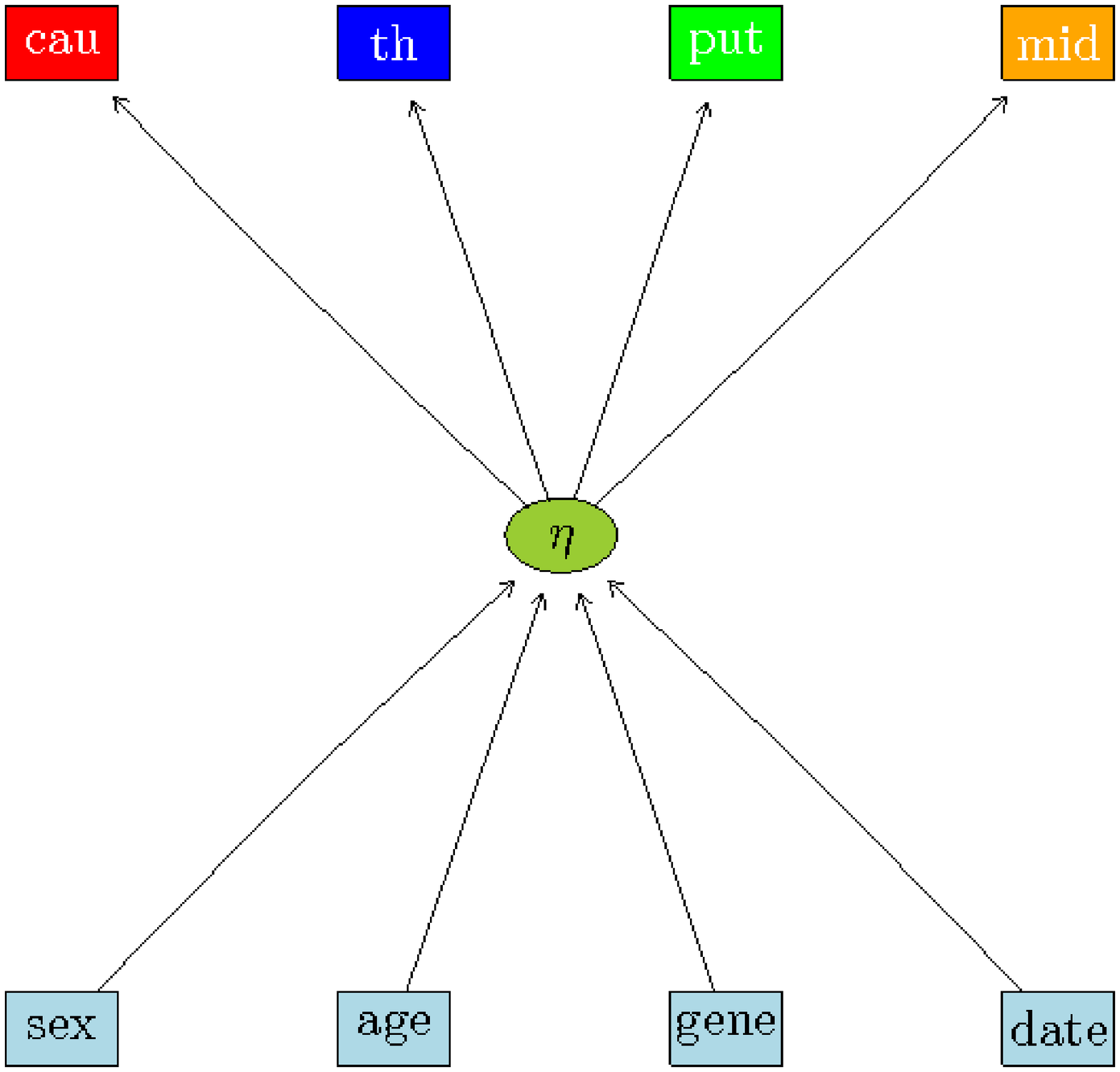}
  }
  \caption{Brain regions of interests and initial \htt
    model.}
  \label{fig:rois}
\end{figure}

We consider 54
observations from \citep{Kalbitzer2010} with measurements of the
serotonin transporter (\htt) in the human brain as measured by
\emph{Positron Emission Tomography} (PET) techniques. The outcome of
interest, \htt, was quantified as binding potential of the specific
tracer binding (\bpnd) in four regions of interest, which a priori were
identified as high-binding and reliable measurements of \htt. The
serotonergic system has been suggested to have a strong impact on mood
and as a potential predictor of the development of seasonal affective
disorder (SAD). The aim of the original study was to explore the
association between levels of \htt and the interaction between
seasonality and a repeat polymorphism in the promoter region of the
serotonin transporter gene (5-HTTLPR). This was achieved by linear
regression on each regional outcome. However, these data are
characterized by high inter-regional correlation, and in the following
we will supplement the original analysis with a multivariate analysis
taking this aspect into account.  It has been demonstrated that \htt
levels respond to chronic changes in brain serotonin (5-HT) levels as
suggested by studies of the effect of SSRI treatment and in animal
studies. It has therefore been suggested that the common regulator of
\htt is this underlying 5-HT level \citep{DavidErritzoe03032010}
which, however, cannot be measured directly in vivo. This biological
model can be captured by a structural equation model, with a simple
measurement model describing the four regions of interest, and 
a structural model in which exogenous variables affects the \htt measurements
indirectly through the intermediate latent variable as shown on Figure
\ref{fig:rois}.


We chose to model the seasonal effect using a harmonic function with
a period of one year, described by the amplitude, $A$, and the
translation parameter, $\delta$ (time of peak):
\begin{align}\label{eq:harmonic}
  A\cos\left(\frac{2\pi(t-\delta)}{365}\right) =
  \underbrace{A\cos\left(\frac{2\pi\delta}{365}\right)}_{\beta_1}
  \cos\left(\frac{2\pi t}{365}\right) +
  \underbrace{A\sin\left(\frac{2\pi\delta}{365}\right)}_{\beta_2}\sin\left(\frac{2\pi
      t}{365}\right),
\end{align}
where $t$ is the date of the scan.
This parameterization approximately embeds a simpler model, where the seasonal
effect is described as a linear function of the amount of daylight
minutes on the date of the PET scan. However, the harmonic curve adds
the flexibility of letting the time of peak be a free parameter which
allows a possible delayed seasonal effects on the serotonin transporter
to be taken into account. In \pkg{lava} the seasonal effect could be
modeled directly (via \code{constrain}) as the left-hand-side of
(\ref{eq:harmonic}), thus directly quantifying $A$ and $\delta$. Here
we use the linear parameterization on the right-hand-side using the
cosine and sine transformed time variables as predictors. As in the
original study, we also adjust for possible main effects of age, gender
and the 5-HTTLPR polymorphism (dichotomized as carriers of the (short)
s-variation vs. non-carriers (long-long alleles))

\begin{Schunk}
\begin{Sinput}
> httmod <- lvm(c(cau,th,put,mid) ~ eta)
> regression(httmod) <- eta ~ f(cos,b1)+f(sin,b2)+age+sex+gene
\end{Sinput}
\end{Schunk}

\begin{figure}[htbp]
  \centering
    \includegraphics[height=7cm, keepaspectratio=TRUE]{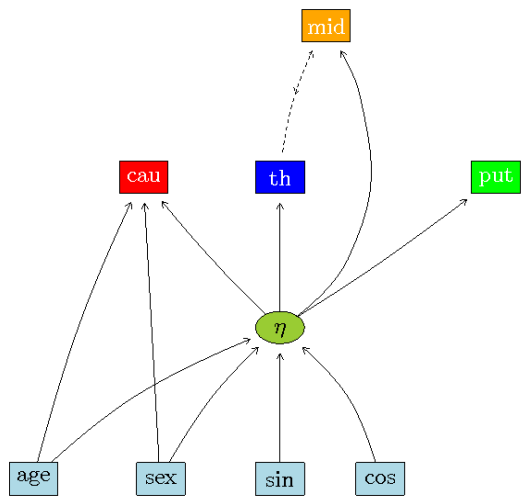}
    \caption{  Final \htt-seasonality model.}
  \label{fig:rois2}
\end{figure}

As described in Section \ref{sec:inference}, we assessed the model fit via
residual plots, score tests and the global $\chi^2$-test, and
concluded that the local independence between the midbrain and
thalamus region was not plausible.  Significantly different age and
gender effects on caudate nucleus were also identified in this process
and therefore added to the model.
\begin{Schunk}
\begin{Sinput}
> regression(httmod) <- cau~sex+age
> covariance(httmod) <- th~mid
\end{Sinput}
\end{Schunk}
No additional evidence against the model was identified, thus leading
to a final model as illustrated in Figure \ref{fig:rois2}:
\begin{SaveVerbatim}{Vplotcode} 
        > tikz(file="sertsem1.tex",8,8,standAlone=TRUE)
        > m <- lvm(~cau+th+put+mid)
        > regression(m) <- c(cau,mid,th,put)~eta
        > latent(m) <- ~eta; labels(m) <- c(eta="$\\eta$")
        > regression(m) <- eta~sex+age+date
        > nodecolor(m,vars(m)) <- 0
        > nodecolor(m,endogenous(m),labcol="white") <- 
        +     c("red","blue","green","orange")
        > plot(m)
        > dev.off()
\end{SaveVerbatim}
\begin{Schunk}
\begin{Sinput}
> tikz(file="sertsem1.tex",8,8,standAlone=TRUE)
> m <- lvm(~cau+th+put+mid)
> regression(m) <- c(cau,mid,th,put)~eta
> latent(m) <- ~eta; labels(m) <- c(eta="$\\eta$")
> regression(m) <- eta~sex+age+date
> nodecolor(m,vars(m)) <- 0
> nodecolor(m,endogenous(m),labcol="white") <- 
+     c("red","blue","green","orange")
> plot(m)
> dev.off()
\end{Sinput}
\end{Schunk}

Parameter estimates are obtained by maximum likelihood using a
Newton-Raphson algorithm:
\begin{Schunk}
\begin{Sinput}
> lava.options(method="NR",trace=1,tol=1e-12,param="relative")
> httmod.fit <- estimate(httmod,data=dasb)
\end{Sinput}
\end{Schunk}

One of the aims of the analysis is to quantify the seasonal effect, therefore
we will also estimate the translation and amplitude (unsigned) 
\begin{align}
  \widehat{\delta} =
  \arctan(\widehat{\beta}_2/\widehat{\beta}_1)365/(2\pi)
 \end{align}
 \begin{align}
   |\widehat{A}| = \sqrt{\widehat{\beta}_1^2+\widehat{\beta}_2^2}
 \end{align} 
\begin{Schunk}
\begin{Sinput}
> constrain(httmod.fit,delta~b1+b2) <- function(x) atan(x[2]/x[1])*365/(2*pi)
> constrain(httmod.fit,A~b1+b2) <- function(x) sqrt(x[1]^2+x[2]^2)
> summary(httmod.fit, std=NULL)
\end{Sinput}
\end{Schunk}
\begin{Schunk}
\begin{Soutput}
||score||^2= 1.171417e-11 
Latent variables: eta 
Number of rows in data= 54
--------------------------------------------------
                       Estimate  Std. Error     Z value    Pr(>|z|)
Measurements:                                                      
   cau<-eta          1.0000e+00                                    
   th<-eta           9.6834e-01  3.0972e-01  3.1264e+00  1.7693e-03
   put<-eta          1.2119e+00  3.5519e-01  3.4120e+00  6.4499e-04
   mid<-eta          1.1317e+00  7.5972e-01  1.4896e+00  1.3633e-01
Regressions:                                                       
   cau<-age          6.7822e-03  1.9574e-03  3.4648e+00  5.3054e-04
   cau<-sex         -3.2907e-01  7.4087e-02 -4.4416e+00  8.9284e-06
    b1:eta<-cos      1.0722e-01  3.9546e-02  2.7112e+00  6.7043e-03
    b2:eta<-sin     -3.7658e-02  4.1421e-02 -9.0915e-01  3.6327e-01
    eta<-age        -2.9471e-03  1.4856e-03 -1.9838e+00  4.7282e-02
    eta<-sex        -7.8173e-03  4.9608e-02 -1.5758e-01  8.7479e-01
    eta<-GnonLL     -4.5062e-02  4.7688e-02 -9.4494e-01  3.4469e-01
Intercepts:                                                        
   cau               0.0000e+00                                    
   th                2.8698e-01  5.9078e-01  4.8577e-01  6.2713e-01
   put              -4.1726e-01  6.8945e-01 -6.0521e-01  5.4504e-01
   mid               2.0487e+00  1.4175e+00  1.4453e+00  1.4838e-01
   eta               1.9647e+00  8.7535e-02  2.2445e+01      <1e-16
Residual Variances:                                                
   cau               5.4901e-02  1.2519e-02  4.3855e+00            
   th                6.9573e-02  1.4911e-02  4.6658e+00            
   th,mid            9.6725e-02  3.4453e-02  2.8074e+00  4.9940e-03
   put               1.0455e-02  9.2801e-03  1.1266e+00            
   mid               6.9218e-01  1.3472e-01  5.1377e+00            
   eta               2.0335e-02  9.8548e-03  2.0635e+00            

Non-linear constraints:
         Estimate  Std. Error     Z value    Pr(>|z|)        2.5%   97.5%
delta -19.6215810  19.5444140  -1.0039483   0.3154035 -57.9279285 18.6848
A       0.1136378   0.0426363   2.6652834   0.0076923   0.0300722  0.1972
--------------------------------------------------
Estimator: gaussian 
--------------------------------------------------
Number of observations = 54 
 Log-Likelihood = -57.88782 
 BIC = 223.2812 
 AIC = 155.7756 
 log-Likelihood of model = -57.88782 
 log-Likelihood of saturated model = -46.94523 
 Chi-squared statistic: Q = 21.88517 , df = 14 , P(Q>q) = 0.08100473 
--------------------------------------------------
\end{Soutput}
\end{Schunk}

As an alternative to the Wald test we can conduct a LRT to test the
significance of the seasonal parameters:
\begin{Schunk}
\begin{Sinput}
> httmod0 <- httmod; kill(httmod0) <- ~cos+sin
> httmod0.fit <- estimate(httmod0,dd,control=list(start=coef(httmod.fit)))
> compare(httmod0.fit,httmod.fit)
\end{Sinput}
\end{Schunk}
\begin{Schunk}
\begin{Soutput}
	Likelihood ratio test

data:  
chisq = 12.3978, df = 2, p-value = 0.002032
sample estimates:
log likelihood (model 1) log likelihood (model 2) 
               -64.53133                -58.33241 
\end{Soutput}
\end{Schunk}

Thus, we find a highly significant seasonal effect ($p=0.002$). 
The estimated time of peak and 95\% Wald confidence limits can be quantified as
\begin{Schunk}
\begin{Sinput}
> format(as.Date("2010-1-1")+constraints(httmod.fit)["delta",c(1,5:6)],
+  "%d%b")
\end{Sinput}
\begin{Soutput}
Estimate     2.5%    97.5% 
 "12Dec"  "04Nov"  "19Jan" 
\end{Soutput}
\end{Schunk}
and with an estimate around middle of December this is in reasonable
agreement with a suggested peak around winter solstice (about December 21).
The parameter estimates of the initial model (Figure \ref{fig:rois})
yielded very similar seasonal parameter estimates and confidence limits.

Next, we will examine the interaction between season and the 5-HTTLPR
polymorphism. This is done via a multigroup analysis and hence we
divide the data into two groups defined by the genetic variable
\begin{Schunk}
\begin{Sinput}
> d1 <- subset(dd, G=="LL")
> d2 <- subset(dd, G=="nonLL")
\end{Sinput}
\end{Schunk}
A standard multigroup analysis could be conducted where the parameters
$\beta_1$ and $\beta_2$ would be allowed to vary freely in the two groups.
However, a more biological plausible model is to fix the translation
parameter $\delta$ in the two groups to be the same and let the
amplitude $A$ be
free. This can be implemented via the left-hand-side of
(\ref{eq:harmonic}) and the \code{constrain}-method
\begin{Schunk}
\begin{Sinput}
> m <- baptize(kill(httmod,~G+cos+sin))
> intercept(m,~cau+eta) <- list(0,"mu1")
> regression(m,cau~eta) <- 1
> regression(m,eta~Day) <- 0
> m2 <- m
> intercept(m2,~eta) <- "mu2"
> mycos <- function(x) x[2]+x[3]*cos(2*pi*(x[1]-x[4])/365)
> constrain(m,mu~Day+nu1+A1+delta) <- mycos
> constrain(m2,mu2~Day+nu2+A2+delta) <- mycos
\end{Sinput}
\end{Schunk}
Here we explicitly chose caudate nucleus as our reference region by
fixing the factor loading to 1 and intercept to 0. The intercept
parameters \code{nu1} and \code{nu2} describes the main effect of the
genotype. The significance of the interaction can then be examined
with a LRT against the first model
\begin{Schunk}
\begin{Sinput}
> httmod.2 <- estimate(list(m,m2),list(d1,d2))
> compare(httmod.2,httmod.fit)
\end{Sinput}
\begin{Soutput}
	Likelihood ratio test

data:  
chisq = 4.7267, df = 1, p-value = 0.0297
sample estimates:
log likelihood (model 1) log likelihood (model 2) 
               -55.52446                -57.88782 
\end{Soutput}
\end{Schunk}
The estimated amplitude parameters with 95\% confidence limits are
\begin{Schunk}
\begin{Sinput}
> confint(httmod.2)[c("A1","A2"),c(1,3,4)]
\end{Sinput}
\begin{Soutput}
     Estimate    Pr(>|z|)        2.5%
A1 0.01672275 0.691577803 -0.06589577
A2 0.14670814 0.002547141  0.05142227
\end{Soutput}
\end{Schunk}
and the difference in amplitude can be quantified as
\begin{Schunk}
\begin{Sinput}
> constrain(httmod.2, dA~A1+A2) <- diff
> constraints(httmod.2)["dA",]
\end{Sinput}
\begin{Soutput}
   Estimate  Std. Error     Z value    Pr(>|z|)        2.5%       97.5% 
0.129985393 0.061819428 2.102662508 0.035495282 0.008821541 0.251149245 
\end{Soutput}
\end{Schunk}
with peak time around January 1:
\begin{Schunk}
\begin{Sinput}
coef(httmod.2)[[1]]["delta",]
\end{Sinput}
\begin{Soutput}
   Estimate  Std. Error     Z value    Pr(>|z|) 
 0.24805873 17.55228131  0.01413256  0.98872422 
\end{Soutput}
\end{Schunk}
Hence we see a statistical significant difference in amplitude between
the two genotypes ($p=0.03$, 95\% confidence limits $[0.01;
0.25]$), with carriers of the short 5-HTTLPR allele having on average
a higher seasonal variation in SERT binding. 

\begin{figure}[htp!]
  \centering
    \includegraphics[width=12cm,height=8cm,keepaspectratio=true]{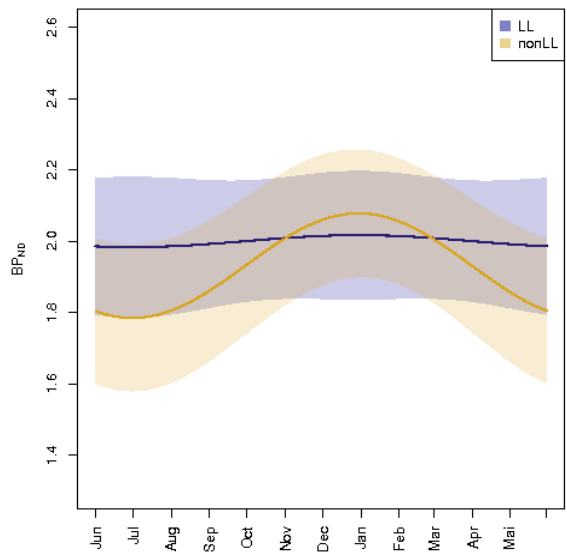}
  \caption{Estimated harmonic curve of nonLL and LL group with 95\%
    point-wise confidence limits.}
  \label{fig:harmonic}
\end{figure}

To visualize the harmonic curves of the two genotypes we need to
predict the parameters \code{mu1} and \code{mu2}. This can also be
achieved with \code{constraints} where we supply the argument
\code{idx} indicating which non-linear parameter to extract. For a
\code{multigroup} model we also need to specify which group to extract
this parameter from via the argument \code{k}. For instance predicting
the two means on each day of the year with 95\% confidence limits we can
write
\begin{Schunk}
\begin{Sinput}
> mu1 <- constraints(e,k=1,idx="mu",data=data.frame(Day=1:365),level=0.95)
> mu2 <- constraints(e,k=2,idx="mu2",data=data.frame(Day=1:365),level=0.95)
\end{Sinput}
\end{Schunk}
See Figure \ref{fig:harmonic} for a plot of these two curves.

This study shows that S-allele carriers (nonLL) on average have a much
more varying \htt binding level which could suggest a decreased
serotonin concentration in the winter in this group. This could be
caused by the decreasing amount of daylight in this period in the
study country (Denmark) and supplements previous studies showing that
S-allele carriers have higher risk of developing SAD. A detailed
discussion can be found in the original paper \citep{Kalbitzer2010}.



\section{Conclusion} 
A package \pkg{lava} has been developed which covers the
classical covariance structure analysis and which hopefully can
serve as a platform for methodological development in the field
of structural equation models and related models.

The key features of the package is 
\begin{enumerate}
\item Easy interactive specification and visualization of complex models
\item Simulation routines (for a broad class of models beyond
  the LLVM)
\item Extensions to binary and censored data via \pkg{lava.tobit}
\item Multigroup analyses 
\item MLE with data missing at random
\item Non-linear parameter constraints and covariate effects
\item Asymptotically correct standard errors for clustered correlated data
\end{enumerate}

Further, the program is built up around a series of modules
(optimizers, estimators, plot hooks, simulations hooks, pre and post
estimation hooks) which should ensure that future extensions
can be written quite easily. This has been one of the main aims
during the development of the \pkg{lava} package.

Additional extensions of the package is currently under preparation
including non-linear random effects and MLE for a broader class of the
exponential family with estimation based on adaptive quadrature
rules. In this process, we are preparing to export loop-intensive parts
of the program to \proglang{C++}, which also should give a considerable
computational and memory (via call-by-reference) improvement for some of
the closed-form likelihood models (e.g. random slope models) and
models with large number of variables.

If you use \pkg{lava} please cite this paper and the R software
in publications.



\section{Acknowledgments}
This work was supported by The Danish Agency for Science, Technology
and Innovation.

\appendix

\section{Some zero-one matrices}\label{sec:zeroone}
In this section we will define a few matrix-operators in order to
define various conditional moments.  Let $\bm{B}\in\R^{n\times m}$ be
a matrix, and define the indices $\bm{x} =
\{x_{1},\ldots,x_{k}\}\in\{1,\ldots,n\}$, and $\bm{y} =
\{y_{1},\ldots,y_{l}\}\in\{1,\ldots,m\}$.  We define
$\bm{J}_{n,\bm{x}} = \bm{J}_{\bm{x}} \in\R^{(n-k)\times n}$ as the
$n\times n$ identify matrix with rows $\bm{x}$ removed. E.g.
\begin{align*}
  \bm{J}_{6,(3,4)} =
  \left(\begin{array}{cccccc}
      1 & 0 & 0 & 0 & 0 & 0 \\
      0 & 1 & 0 & 0 & 0 & 0 \\ 
      0 & 0 & 0 & 0 & 1 & 0 \\
      0 & 0 & 0 & 0 & 0 & 1 \\
    \end{array}\right).
\end{align*}
To remove rows $\bm{x}$ from $\bm{B}$ we simply multiply from the left
with $\bm{J}_{n,\bm{x}}$. If we in addition want to remove columns $\bm{y}$ we
multiply with the transpose of $\bm{J}_{n,\bm{y}}$ from the right:
\begin{gather}
  \bm{J}_{n,\bm{x}}\bm{B}\bm{J}_{n,\bm{y}}'.
\end{gather}
We will use the notation $\bm{J}$ to denote the matrix that removes
all latent variables ($\bm{\eta}$) from the vector of all variable,
$\bm{U}$ as defined in (\ref{eq:Udef}). We denote $\bm{J}_{\bm{Y}}$
the matrix that only keeps endogenous variables ($\bm{Y}$).

We also need an operator that cancels out rows or
columns of a matrix/vector.  Define the square matrix
$\bm{p}_{n,\bm{x}}\in\R^{n\times n}$ as the identity-matrix with
diagonal elements at position $\bm{x}$ canceled out:
\begin{align}
  \bm{p}_{n,\bm{x}}(i,j) =
  \begin{cases}
    1, & i=j, \ i\not\in\bm{x}, \\
    0, & \text{else}.
  \end{cases}
\end{align}
E.g.
\begin{align*}
  \bm{p}_{6,(3,4)} =
  \left(\begin{array}{cccccc}
      1 & 0 & 0 & 0 & 0 & 0 \\
      0 & 1 & 0 & 0 & 0 & 0 \\ 
      0 & 0 & 0 & 0 & 0 & 0 \\ 
      0 & 0 & 0 & 0 & 0 & 0 \\
      0 & 0 & 0 & 0 & 1 & 0 \\
      0 & 0 & 0 & 0 & 0 & 1 \\
    \end{array}\right).
\end{align*}
To cancel out rows $\bm{x}$ and columns $\bm{y}$ of the matrix
$\bm{B}\in\R^{n\times m}$ we calculate
\begin{align*}
  \bm{p}_{n,\bm{x}}\bm{B}\bm{p}_{m,\bm{y}}'.
\end{align*}
We will use $\bm{p}_{\bm{X}}$ and $\bm{p}_{\complement\bm{X}}$ as the matrix that
cancels out the rows corresponding to the index of the exogenous
variables ($\bm{X}$) respectively the matrix that cancels out all rows
but the ones corresponding to $\bm{X}$.

\section{The Score Function and Information}
\label{sec:scorehessian}
In this section we will calculate the analytical derivatives of the
log-likelihood.
In order to obtain these results we will first introduce the notation
of some common matrix operations. Let $\bm{A}\in\R^{m\times n}$, then
we define the column-stacking operation:
\begin{align*}
  \mvec(\bm{A}) =
  \begin{pmatrix}
    a_1 \\
    \vdots \\
    a_n
  \end{pmatrix},
\end{align*}
where $a_i$ denotes the $i$th column of $\bm{A}$. The unique
\emph{commutation matrix}, $\R^{mn\times mn}$ is defined by
\begin{align}
  \bm{K}^{(m,n)}\mvec(\bm{A}) = \mvec(\bm{A}').
\end{align}
Letting $\bm{H}^{(i,j)}$ be the $m\times n$-matrix with one at
position $(i,j)$ and zero elsewhere, then
\begin{align*}
  \bm{K}^{(m,n)} = \sum_{i=1}^m\sum_{j=1}^n (\bm{H}^{(i,j)}\otimes\bm{H}^{(i,j)}{}'),
\end{align*}
e.g.
\begin{align*}
  \bm{K}^{(2,3)} =
  \left(
    \begin{array}{cc:cc:cc}
      1 & 0 & 0 & 0 & 0 & 0 \\
      0 & 0 & 1 & 0 & 0 & 0 \\
      0 & 0 & 0 & 0 & 1 & 0 \\ \hdashline
      0 & 1 & 0 & 0 & 0 & 0 \\
      0 & 0 & 0 & 1 & 0 & 0 \\
      0 & 0 & 0 & 0 & 0 & 1 \\
    \end{array}
  \right).
\end{align*}
It should be noted that product with a commutation matrix can be
implemented very efficiently instead of relying on a direct
implementation of the above mathematical definition.

Let $\bm{A}\in\R^{m\times n}$ and $\bm{B}\in\R^{p\times q}$ then the
\emph{Kronecker product} is the $mp\times nq$-matrix:
\begin{align*}
  \bm{A}\otimes\bm{B} =
  \begin{pmatrix}
    a_{11}\bm{B} & \cdots & a_{1n}\bm{B} \\
    \vdots & & \vdots \\
    a_{m1}\bm{B} & \cdots & a_{mn}\bm{B} \\    
  \end{pmatrix}
\end{align*}
We will calculate the derivatives of (\ref{eq:loglik1}) by means of
matrix differential calculus.
The \emph{Jacobian matrix} of a matrix-function $F\colon
\R^n\to\R^{m\times p}$ is the $mp\times n$ matrix defined by
\begin{align*}
  DF(\bm{\theta}) = \frac{\partial \mvec F(\bm{\theta})}{\partial\bm{\theta}'}.
\end{align*}
Letting $\dif$ denote the differential operator (see \cite{MR940471}),
the first identification rule states that $\dif\mvec F(\bm{\theta}) =
A(\bm{\theta})\dif \bm{\theta} \Rightarrow DF(\bm{\theta}) =
A(\bm{\theta})$.

\subsubsection{Score function}
Using the identities $\dif\log\abs{\bm{X}} = \tr(\bm{X}^{-1}\dif \bm{X})$ and
$\dif \bm{X}^{-1} = -\bm{X}^{-1}(\dif\bm{X})\bm{X}^{-1}$, and applying
the product rule we get
\begin{align}
  \dif \fun{\ell}{\bm{\theta}} &= -\frac{n}{2}\fun{\tr}{\SigmaF^{-1}\dif\SigmaF}
  - \frac{n}{2}\fun{\tr}{\dif\{\wS\SigmaF^{-1}\}} \\
  &= - \frac{n}{2}\fun{\tr}{\SigmaF^{-1}\dif\SigmaF} +
  \frac{n}{2}\fun{\tr}{\wS\SigmaF^{-1}[\dif\SigmaF]\SigmaF^{-1}},
\end{align}
where
\begin{align}
  \dif\SigmaF &=
  \left\{\dif\IA\right\}\PT\IA' +
  \IA\left\{\dif\PT\IA'\right\} \\ 
  &=   \left\{\dif\IA\right\}\PT\IA' + 
  \IA\left\{\dif\PT\right\}\IA'
  +  \IA\PT\left\{\dif\IA\right\}' \\
  &= \left\{\dif\IA\right\}\PT\IA' +
  [\left\{\dif\IA\right\}\PT\IA']' +
  \IA\left\{\dif\PT\right\}\IA',
\end{align}
and 
\begin{align}
  \dif\IA = \bm{J}(\one_m-\AT)^{-1}\left\{\dif\AT\right\}(\one_m-\AT)^{-1}.
\end{align}
Taking $\mvec$'s it follows that
\begin{align}
  \dif\mvec\IA = 
  \left[((\one_m-\AT)^{-1})'\otimes \IA\right]\dif\mvec\ftb{A}\dif\bm{\theta},
\end{align}
hence by the first identification rule
\begin{align}\label{eq:dG}
  \frac{\partial\mvec\IA}{\partial\bm{\theta}'} = 
  \left[((\one_m-\AT)^{-1})'\otimes \IA\right] \frac{\partial\mvec\ftb{A}}{\partial\bm{\theta}'},
\end{align}
and similarly
\begin{align}\label{eq:dSigma}
  \frac{\partial\mvec\SigmaF}{\partial\bm{\theta}'} &= (\one_{k^2} + \bm{K}^{(k,k)})
  \left[\IA\ftb{P}\otimes\one_k\right]\frac{\partial\mvec\IA}{\partial\bm{\theta}'}
  + 
  \left[\IA\otimes\IA\right]\frac{\partial\mvec\ftb{P}}{\partial\bm{\theta}'},
\end{align}
and finally (exploiting the symmetry of $\SigmaF$ and commutative
property under the trace operator) we obtain the gradient
\begin{align}
  \frac{\partial \ell(\bm{\theta})}{\partial\bm{\theta}}
  =  
  \frac{n}{2}\left(\frac{\partial\mvec\SigmaF}{\partial\bm{\theta}'}
  \right)'\mvec\left[\SigmaF^{-1}\wS\SigmaF^{-1}\right] 
  -\frac{n}{2}\left(\frac{\partial\mvec\SigmaF}{\partial\bm{\theta}'}
  \right)'\mvec\left[\SigmaF^{-1}\right].
\end{align}
Next we examine the model including a mean structure (\ref{eq:loglik1}).
W.r.t. to the first differential we observe that
\begin{align}
  \dif
  \tr\left\{\ftb{T}\SigmaF^{-1}\right\} 
  &= 
  -\tr\left\{\ftb{T}\SigmaF^{-1}(\dif\SigmaF)\SigmaF^{-1}\right\} 
  + \tr\left\{(\dif\ftb{T})\SigmaF^{-1}\right\}.
\end{align}
Hence
\begin{align}
  \begin{split}
    \frac{\partial \ell(\bm{\theta})}{\partial\bm{\theta}} &=
    \frac{n}{2}\left(\frac{\partial\mvec\SigmaF}{\partial\bm{\theta}'}
    \right)'\mvec\left[\SigmaF^{-1}\ftb{T}\SigmaF^{-1}\right]
    -\frac{n}{2}\left(\frac{\partial\mvec\SigmaF}{\partial\bm{\theta}'}
    \right)'\mvec\left[\SigmaF^{-1}\right]
    \\
    &\qquad -
    \frac{n}{2}\left(\frac{\partial\mvec\ftb{T}}{\partial\bm{\theta}'}\right)'\mvec\left(\SigmaF^{-1}\right).
  \end{split}\label{eq:scoremean}
\end{align}
Further by the chain-rule 
\begin{align}\label{dT}
    \frac{\partial\mvec\ftb{T}}{\partial\bm{\theta}'} = 
    \frac{\partial (\wmu-\ftb{v})}{\partial\bm{\theta}'}
    \frac{\partial\mvec\ftb{T}}{\partial(\wmu-\ftb{\xi})'} =
    - \left[\one_k\otimes(\wmu-\ftb{\xi})
    + (\wmu-\ftb{\xi})\otimes\one_k\right]\frac{\partial\ftb{\xi}}{\partial\bm{\theta}'},
\end{align}
and
\begin{align}
  \dif \ftb{\xi} = (\dif\IA)\ftb{v} + \IA(\dif\ftb{v}).
\end{align}
Taking $\mvec$ ($\IA\ftb{v} = \one\IA\ftb{v}$):
\begin{align}
  \frac{\partial\ftb{\xi}}{\partial\bm{\theta}'} = 
  (\ftb{v}'\otimes \one_k)\frac{\partial\mvec\IA}{\partial\bm{\theta}'} + 
\IA\frac{\partial\ftb{v}}{\partial\bm{\theta}'}.
\end{align}
We have calculated the full score but in some situations it will be
useful to evaluate the score in a single point.
The contribution of a single observation to the log-likelihood is
\begin{align}
  \ell(\bm{\theta}\mid \bm{z}_i) \propto \frac{1}{2}\log\abs{\SigmaF} + \frac{1}{2}(\bm{z}_i-\ftb{\xi})'\SigmaF^{-1}(\bm{z}_i-\ftb{\xi}),
\end{align}
or as in (\ref{eq:loglik1}) where we simply exchange $\ftb{T}$ with
$\bm{T}_{\bm{z}_i,\bm{\theta}} =
(\bm{z}_i-\ftb{\xi})(\bm{z}_i-\ftb{\xi})'$, hence the score is as
in (\ref{eq:scoremean}) where (\ref{dT}) is calculated with $\bm{z}_i$
instead of $\widehat{\bm{\mu}}$.
Alternatively, letting $\bm{z_i}-\ftb{\xi} = \ftb{u} = \ftb{u}(i)$:
\begin{align}
  \begin{split}
    \dif(\ftb{u}'\SigmaF^{-1}\ftb{u}) &=
    \ftb{u}'\left[(2\SigmaF^{-1})\dif\ftb{u} +
      (\dif\SigmaF^{-1})\ftb{u}\right] \\
    &= - \ftb{u}'\left[(2\SigmaF^{-1})\dif\ftb{\xi} +
      \SigmaF^{-1}(\dif\SigmaF)\SigmaF^{-1}\ftb{u}\right],
  \end{split}
\end{align}
where we used that for constant symmetric $\bm{A}$ the differential of a quadratic
form is
\begin{align}
  \dif(\bm{u}'\bm{A}\bm{u}) =  2\bm{u}'(\bm{A})\dif\bm{u}. 
\end{align}
Hence the contribution to the score function of the $i$th observation is
\begin{align}
  \begin{split}
    \mathcal{S}_i(\bm{\theta}) &=
    -\frac{1}{2}\Big\{
    \mvec(\SigmaF^{-1})\frac{\partial\mvec\SigmaF}{\partial\bm{\theta}'}
    -2\ftb{u}'\SigmaF^{-1}\frac{\partial\mvec\ftb{\xi}}{\partial\bm{\theta}'}
    \\
    &\qquad -
    (\ftb{u}'\SigmaF^{-1}\otimes\ftb{u}'\SigmaF^{-1})\frac{\partial\mvec\SigmaF}{\partial\bm{\theta}'}
    \Big\}
    \\
    &= -\frac{1}{2}\left\{ \left(\mvec(\SigmaF^{-1}) -
        \mvec(\SigmaF^{-1}\ftb{u}\ftb{u}'\SigmaF^{-1})\right)\frac{\partial\mvec\SigmaF}{\partial\bm{\theta}'}
      -2\ftb{u}'\SigmaF^{-1}\frac{\partial\mvec\ftb{\xi}}{\partial\bm{\theta}'}
    \right\},
  \end{split}
\end{align}
where the score-function evaluated in $\bm{\theta}$ is
$\mathcal{S}(\bm{\theta}) = \sum_{i=1}^n \mathcal{S}_i(\bm{\theta})$.

\subsubsection{The Information matrix}
The second order partial derivative is given by
\begin{align}\label{myd2l}
  \frac{\partial\ell(\bm{\theta})}{\partial\theta_i\theta_j}
  = -\frac{1}{2}\frac{\partial}{\partial\theta_i}\left[
   \left\{\mvec(\SigmaF^{-1}) - \mvec(\SigmaF^{-1}\ftb{u}\ftb{u}'\SigmaF^{-1})
    \right\}
\frac{\partial\mvec\SigmaF}{\partial\theta_j}
- 2\ftb{u}'\SigmaF^{-1}\frac{\partial\ftb{\xi}}{\partial\theta_j}
 \right].
\end{align}
Taking negative expectation with respect to the true parameter
$\bm{\theta}_0$ we obtain the \emph{expected information}
\citep{MR940471}, which get rid of all second order derivatives
\begin{align}\label{eq:I}
  \mathcal{I}(\bm{\theta}_0) &= \frac{1}{2}\left(\left.
    \frac{\partial\mvec\SigmaF}{\partial\bm{\theta}'}\right|_{\bm{\theta}=\bm{\theta}_0}
  \right)'(\bm{\Omega}_{\bm{\theta}_0}^{-1}\otimes\bm{\Omega}_{\bm{\theta}_0}^{-1})
  \left(\left.
      \frac{\partial\mvec\SigmaF}{\partial\bm{\theta}'}\right|_{\bm{\theta}=\bm{\theta}_0}
  \right) \\
  &\quad + 
  \left(\left.
      \frac{\partial\ftb{\xi}}{\partial\bm{\theta}'}\right|_{\bm{\theta}=\bm{\theta}_0}
  \right)'
  \bm{\Omega}_{\bm{\theta}_0}^{-1}
  \left(\left.
      \frac{\partial\ftb{\xi}}{\partial\bm{\theta}'}\right|_{\bm{\theta}=\bm{\theta}_0}
  \right).
\end{align}
We will further derive the observed information in the case where the second derivatives vanishes in the case of the matrix functions
$\ftb{A}$,$\ftb{P}$ and $\ftb{v}$. Now
\begin{align}
  \dif^2\ftb{G} = \dif\left[\bm{J}\IAi(\dif\ftb{A})\IAi\right].
\end{align}
Hence
\begin{align}\label{eq:d2G}
  \frac{\partial^2 \ftb{G}}{\partial\theta_i\partial\theta_j} =
  \ftb{G}\left[\frac{\partial\ftb{A}}{\partial\theta_i}\IAi\frac{\partial\ftb{A}}{\partial\theta_j}
    + \frac{\partial\ftb{A}}{\partial\theta_j}\IAi\frac{\partial\ftb{A}}{\partial\theta_i}\right]\IAi.
\end{align}

Next we will find the derivative of (\ref{eq:dSigma}). We let $m$
denote the number of variables, $p$ the number of parameters, and $k$
the number of observed variable (e.g. $\ftb{G}\in\R^{k\times m}$ and
the number of columns in the derivatives are $p$).
We have $\ftb{G}\ftb{P}\in\R^{k\times m}$ and using rules for
evaluating the differential of Kronecker-product (see \cite{MR940471}
pp. 184)
we obtain
\begin{align}\label{eq:part1}
  \begin{split}
    \frac{\partial\mvec}{\partial\bm{\theta}'}(
    \ftb{G}\ftb{P}\otimes\one_{k}) &= \left(\one_{m}\otimes
      \bm{K}^{(k,k)}\otimes\one_k\right)\left(\one_{km}\otimes\mvec\one_k\right)\frac{\partial\mvec\ftb{G}\ftb{P}}{\partial\bm{\theta}'}
    \\
    &= \left(\one_{m}\otimes
      \bm{K}^{(k,k)}\otimes\one_k\right)\left(\one_{km}\otimes\mvec\one_k\right)
    \times \\
    &\qquad \left[
      (\ftb{P}\otimes\one_k)\frac{\partial\mvec\ftb{G}}{\partial\bm{\theta}'}
      +
      (\one_m\otimes\ftb{G})\frac{\partial\mvec\ftb{P}}{\partial\bm{\theta}'}
    \right].
  \end{split}
\end{align}
And
\begin{align}\label{eq:part2}
  \begin{split}
    \frac{\partial\mvec}{\partial\bm{\theta}}\left[(\ftb{G}\otimes\ftb{G})\frac{\partial\mvec\ftb{P}}{\partial\bm{\theta}'}\right]
    &=
    \left[\frac{\partial\mvec\ftb{P}}{\partial\bm{\theta}'}'\otimes\one_{k^2}\right]\frac{\partial\mvec\ftb{G}\otimes\ftb{G}}{\partial\bm{\theta}'}
    \\
    &= 
    \left[\frac{\partial\mvec\ftb{P}}{\partial\bm{\theta}'}'\otimes\one_{k^2}\right]
    \left(\one_m\otimes \bm{K}^{(m,k)}\otimes\one_k\right)\times \\
    &\qquad
    \left(\one_{km}\otimes\mvec\ftb{G}+
      \mvec\ftb{G}\otimes\one_{km}\right)
      \frac{\partial\mvec\ftb{G}}{\partial\bm{\theta}'}.
  \end{split}
\end{align}
Hence from (\ref{eq:part1}) and (\ref{eq:part2}) and using rules for
applying the $\mvec$ operator on products of matrices we obtain
\begin{align}
  \begin{split}\label{eqq1}
    \frac{\partial^2\mvec\SigmaF}{\partial\bm{\theta}\partial\bm{\theta}'}
    &=
    \left[\left(\frac{\partial\mvec\ftb{G}}{\partial\bm{\theta}'}\right)'\otimes\left(
        \one_{k^2}+\bm{K}^{(k,k)}
      \right)\right]
    \left(\one_{m}\otimes
      \bm{K}^{k+k}\otimes
      \one_k\right)\left(\one_{km}\otimes\mvec\one_k\right)
    \times \\
    &\qquad \left[
      (\ftb{P}\otimes\one_k)\frac{\partial\mvec\ftb{G}}{\partial\bm{\theta}'}
      +
      (\one_m\otimes\ftb{G})\frac{\partial\mvec\ftb{P}}{\partial\bm{\theta}'}
    \right] + \\
    &\qquad  
    \Big(\one_p \otimes (\ftb{G}\ftb{P}\otimes\one_k)\Big)
    \frac{\partial^2\mvec\ftb{G}}{\partial\bm{\theta}\partial\bm{\theta}'}
    + \\
    &\qquad
    \left(\left(\frac{\partial\mvec\ftb{P}}{\partial\bm{\theta}'}\right)'\otimes\one_{k^2}\right)
    \left(\one_m\otimes \bm{K}^{(m,k)}\otimes\one_k\right)\times \\
    &\qquad
    \Big(\one_{km}\otimes\mvec\ftb{G}+
      \mvec\ftb{G}\otimes\one_{km}\Big)
      \frac{\partial\mvec\ftb{G}}{\partial\bm{\theta}'},
  \end{split}
\end{align}
with the expressions for the derivatives and second derivatives of
$\ftb{G}$ given in (\ref{eq:dG}) and (\ref{eq:d2G}). 
Further
\begin{align}
  \begin{split}\label{eqq2}
    \frac{\partial^2\ftb{\xi}}{\partial{\bm{\theta}}\partial\bm{\theta}'} &=
    \frac{\partial\mvec}{\partial\bm{\theta}'}(\ftb{v}'\otimes
    \one_k)\frac{\partial\mvec\IA}{\partial\bm{\theta}'} +
    \frac{\partial\mvec}{\partial\bm{\theta}'}\IA\frac{\partial\ftb{v}}{\partial\bm{\theta}'} \\
    &=
    \left(\left(\frac{\partial\mvec\ftb{G}}{\partial\bm{\theta}'}\right)'\otimes\one_k\right)\Big(\one_m\otimes\mvec\one_k\Big)\frac{\partial\mvec\ftb{v}}{\partial\bm{\theta}'} \\
    &\qquad+ (\one_p\otimes(\ftb{v}'\otimes\one_k))
    \frac{\partial^2\mvec\ftb{G}}{\partial\bm{\theta}\partial\bm{\theta}'} \\
    &\qquad+ 
    \left(\frac{\partial\mvec\ftb{G}}{\partial\bm{\theta}'}\right)'\frac{\partial\ftb{v}}{\partial\bm{\theta}'},
  \end{split}
\end{align}
and
\begin{align}\label{eqq3}
  \frac{\partial\mvec}{\partial\bm{\theta}'}\SigmaF^{-1} =
  -(\SigmaF^{-1}\otimes\SigmaF^{-1})\frac{\partial\mvec\SigmaF}{\partial\bm{\theta}'},
\end{align}
and
\begin{align}
  \dif\left(\SigmaF^{-1}\ftb{u}\ftb{u}'\SigmaF^{-1}\right) &= 
  -\SigmaF^{-1}(\dif\SigmaF)\SigmaF^{-1}\ftb{u}\ftb{u}'\SigmaF
  + \SigmaF^{-1}(\dif\ftb{u})\ftb{\mu}'\SigmaF^{-1} \\
  &\quad +
  \SigmaF^{-1}\ftb{u}(\dif\ftb{u}')\SigmaF^{-1}
  - \SigmaF^{-1}\ftb{u}\ftb{u}'\SigmaF(\dif\SigmaF)\SigmaF^{-1}.
\end{align}
By using the identity $\mvec(ABC) = (C'\otimes A)\mvec(B)$ several times we obtain
\begin{align}\label{eqq4}
  \frac{\partial\mvec}{\partial\bm{\theta}'}\SigmaF^{-1}\ftb{u}\ftb{u}'\SigmaF^{-1}
  &=
  -\left([\SigmaF^{-1}\ftb{u}\ftb{u}'\SigmaF^{-1}]'\otimes\SigmaF^{-1}\right)\frac{\partial\mvec\SigmaF}{\partial\bm{\theta}'} \\
  &\quad- \left([\ftb{u}'\SigmaF^{-1}]'\otimes\SigmaF^{-1}\right)\frac{\partial\mvec\ftb{\xi}}{\partial\bm{\theta}'} \\
  &\quad- \left(\SigmaF^{-1}\otimes[\SigmaF^{-1}\ftb{u}]\right)\frac{\partial\mvec\ftb{\xi}}{\partial\bm{\theta}'} \\
  &\quad- \left(\SigmaF^{-1}\otimes[\SigmaF^{-1}\ftb{u}\ftb{u}'\SigmaF^{-1}]\right)\frac{\partial\mvec\SigmaF}{\partial\bm{\theta}'},
\end{align}
and the second order derivative of the log-likelihood (\ref{myd2l})
now follows from applying the product rule with (\ref{eqq1}),
(\ref{eqq2}), (\ref{eqq3}) and (\ref{eqq4}).




\bibliographystyle{elsarticle-harv}


\end{document}